\documentclass{aa}  

\usepackage{graphicx}
\usepackage{txfonts}
\usepackage{color}
\begin{document}

   \title{What can be learnt from UHECR anisotropies observations?}

   \subtitle{Paper I : dipole modulation}

\subtitle{Paper I : dipole moment of the arrival direction distribution}
\subtitle{Paper I : dipole moment of the distribution of arrival directions}

\subtitle{Paper I : large-scale anisotropies and composition features}

   \author{D. Allard
          \inst{1},
          J. Aublin, %\fnmsep\thanks{Grand vainqueur de la belette de Winchester}, 
          B. Baret
         \and E. Parizot
          }

   \institute{Université de Paris, CNRS, Astroparticule et Cosmologie, F-75013 Paris, France\\
             \email{allard@apc.in2p3.fr}
             }

   \date{Received ... ; accepted ... }

% \abstract{}{}{}{}{} 
% 5 {} token are mandatory
 
  \abstract
  % context heading (optional)
  % {} leave it empty if necessary  
   { In the recent years, evidences for an anisotropic distribution of ultra-high-energy cosmic rays (UHECRs) have been claimed, notably a dipole modulation in right ascension  has been reported by the Auger collaboration above the $5\sigma$ significance threshold.
   %   Ultra-high-energy cosmic rays have long been a center of great interest in astroparticle physics, due to the possibility they offer to constrain the physical processes and astrophysical parameters in the most energetic sources of the universe. In the recent years various  evidences of the presence of anisotropies in the UHECR sky have been claimed by several experiments, in particular a right ascension dipole modultion passing the $5\sigma$ significance threshold has been reported by the Pierre Auger observatory. These observations bring back to the front of the scene the eventuality of constraining the nature of UHECR sources despite the evidences that the  composition is becoming heavier at highest energies.
}
  % aims heading (mandatory)
   {
   We investigate the implications of the current data regarding large scale anisotropies, including higher order multipoles, and examine to what extent they can be used to shed some light on the origin of UHECRs, and constrain the astrophysical and/or physical parameters of the source scenarios. We investigate the possibility of observing an associated anisotropy of the UHECR composition and discuss the potential benefit of a good determination of the composition and of the separation of the different nuclear components. We also discuss the interest and relevance of observing the UHECR sky with larger exposure future observatories.
%   We investigate the potential constraints which can be brought on the nature of UHECR sources by the various recently reported observations related to  UHECR large scale  anisotropies, in particular the above mentioned observation of a dipole modulation as well as of the absence of significant signal for higher order multipoles. We  investigate the possibility of observing an associated anisotropy of the UHECR composition and discuss, the relevance of a good determination of the composition and of the separation of the different nuclear components of the UHECR dataset in the context of anisotropy studies. We also discuss the interest and relevance of observing the UHECR sky with larger exposure future observatories.
   }
  % methods heading (mandatory)
   {
   We simulate realistic UHECR sky maps for a wide range of astrophysical scenarios satisfying the current observational constraints, taking into account the energy losses and the photo-dissociation of the UHE protons and nuclei, as well as their deflections by intervening magnetic fields. We investigate scenarios in which the UHECR source distribution follows that of the galaxies in the Universe (with possible biases), varying the UHECR source composition and spectrum, as well as the source density and the magnetic field models. For each of them, we simulate 300 realisations of independent datasets corresponding to various assumptions for the statistics and sky coverage, and apply similar analyses as those used by the Auger collaboration for the search of large scale anisotropies.
%   We simulated realistic UHECR sky maps for a wide range of possible astrophysical scenarios motivated by the current observational constraints, taking the energy losses and photo-dissociation of the UHE protons and nuclei into account, as well as their deflections by intervening magnetic fields. Datasets were built assuming various statistics and sky coverages, in particular that of the Pierre Auger observatory at the time of the discovery of the dipole modulation as a reference. The simulated datasets were analysed with methods similar to those used by the Pierre Auger collaboration for the search of large scale anisotropies. A statistical study of the resulting anisotropies was performed for each astrophysical scenario, varying the UHECR source composition and spectrum and the source density and exploring a set of three hundred independent realizations for each choice of a parameter set. We brought a particular attention to scenarios assuming UHECR sources spatial distribution follows that of the galaxies in the Universe.
   }
  % results heading (mandatory)
   {We find that: i) reproducing the first-order (dipole) anisotropy observed in the Auger data, as well as its evolution as a function of energy, is relatively easy within our general assumptions ; ii) this general agreement can be obtained with different sets of assumptions on the astrophysical and physical parameters, and thus cannot be used, at the present stage, to derive strong constraints on the UHECR source scenarios and draw model-independent constraints on the various parameters individually ; iii) the actual direction of the dipole modulation reconstructed from the Auger data, in the energy bin where the signal is most significant, appears highly non natural in essentially all scenarios investigated, and calls for a reconsideration of their main assumptions, either regarding the source distribution itself or the assumed magnetic field configuration, especially in the Galaxy ; iv) the energy evolution of the reconstructed dipole direction contains potentially important information, which may become constraining for specific source models when larger statistics is collected ; v) for such high-statistics datasets most of our investigated scenarios predict a significant quadrupolar modulation, especially if a the light component of UHECRs can be extracted from the all-particle dataset ; vi) except for protons, the energy range in which the GZK horizon strongly reduces is a key target for anisotropy searches for each given nuclear species ; vii) although a difference in the average composition of the UHECRs in regions having a different count rate is naturally expected in our models, it is unlikely that the composition anisotropy recently reported by Auger can be explained by this effect, unless the reported amplitude is a strong positive statistical fluctuation of an intrinsically weaker signal.
   }
  % conclusions heading (optional), leave it empty if necessary 
   {}


   \maketitle
%
%-------------------------------------------------------------------

\section{Introduction}

The nature of ultra-high-energy cosmic-rays (UHECRs) astrophysical sources and of the mechanism(s) at play to produce them are among the most fascinating and long standing puzzles of high-energy astrophysics, despite decades of intense efforts on experimental developments and theoretical modeling. The latest generation of detectors, mainly the Pierre Auger Observatory (\citet{AugerObs}; hereafter Auger) in the southern hemisphere (Argentina), as well as Telescope Array (\citet{TAObs}; hereafter TA) in the northern hemisphere (USA) have collected data with unprecedented precision and statistics, and provided very valuable information regarding all of the three main characteristics of UHECRs: their energy spectrum, their composition, and the distribution of their arrival directions.

Among the most important results obtained in the recent years, the Auger measurements indicate that the composition of UHECRs is mixed and predominantly light around the so-called ankle in the cosmic-ray energy spectrum (a few 10$^{18}$~eV) and becomes progressively heavier as the energy increases \citep{AugerCompoa, AugerCompob, AugerMix2016}. The most natural interpretation of this composition trend is that it is a signature of a charge-dependent maximum energy of the nuclei accelerated at the dominant sources of UHECRs (as is generically expected for acceleration processes mostly sensitive to the rigidity of the particles), with a relatively low maximum rigidity, say $\rm E_{max}/Z < 10^{19}$~eV (see {\it e.g} \citet{Allard2012} and references therein). Together with detailed measurements of the UHECR spectrum \citep{AugerICRC2019, AugerSpec2020, TASpec2020}, the Auger data on composition bring important constraints on the physics of UHECR accelerators. It cannot, however, lead to an unambiguous identification of the nature of the sources, since: {\it i)} while the global evolution of the UHECR nuclear abundances towards a heavier composition at the highest energies seems now well established, its translation in terms of actual relative abundances of the different nuclear species is much more uncertain and {\it ii)} there are, in any case, no robust theoretical predictions of unambiguous composition signatures of the many different types of UHECR candidate sources. This makes it all the more important to complete the set of constraints on the identity of these elusive sources by including the observation of anisotropies in the UHECR arrival directions.

After several encouraging hints followed by disappointed hopes ({\it e.g} \citet{Uchiori2000, FW2004, AugerAGN2007, AugerAGN2010}), for the first time the observation of an anisotropy  passing the 5$\sigma$ discovery threshold as been reported by Auger \citep{AugerDip2017,AugerDipLong2018}. More precisely this study establishes the existence of a dipole modulation in right ascension in the arrival direction of UHECR events observed above $8\,10^{18}~$eV. Moreover, some evidence of intermediate scale anisotropies is claimed at the $\sim$3 to 4.5 $\sigma$ post-trial level, both by Auger \citep{AugerSFG2018, AugerICRC2019} and TA \citep{TASpot2014}, using blind searches based on top-hat analyses to detect significant event clustering as well as searching for positive correlations with candidate sources or catalogs of astrophysical objects.

These signals indicate with a high confidence level that the UHECR sky is genuinely anisotropic and revive the prospect of revealing the position and the nature of UHECR in a foreseeable future. However, given the above-mentioned evidence for a progressive extinction of the light (low-charge) component of UHECRs and the increasing abundance of intermediate-mass and heavy nuclei at higher energy, and resting on our current knowledge of the strength and structure of the Galactic magnetic field (GMF), it is very likely that the angular deflections of UHECRs remain large even at the highest energies, leading not only to a spread over the sky of the UHECRs coming from a given source, but also to a significant shift of their average arrival direction away from the source direction (except for very special locations in the sky, which moreover depend on the adopted GMF model). Thus, using the observed anisotropies to deduce the positions and nature of the UHECR sources appears particularly difficult, and in any case very uncertain at this stage.

Admittedly, these anisotropy observations appear to support the general idea that UHECRs are of extragalactic origin (as already accepted, almost universally, based on simple theoretical considerations). Yet, whether they hold unambiguous signatures of a correlation of UHECR sources with the distribution of ordinary matter in the universe is not entirely settled. And even if such a correlation holds (as is also expected from simple astrophysical considerations), it is not clear whether there is any hint or evidence of biases in this correlation that would for instance be the signature of either a preferred or a disfavoured presence of UHECR sources in some specific astrophysical environments (such as the nearby richest galaxy clusters, starburst galaxies, radio-loud galaxies, etc.). Nor whether a given potential source ({\it e.g.} Centaurus~A, or M87, etc.) or a specific class of sources is required to account for the data, or even simply favoured, in some quantifiable way. Anisotropy signals have only been obtained so far against the (astrophysically very improbable) assumption of a perfectly isotropic distribution, and do not allow to significantly discriminate between a large number of possible source scenarios.

%To the best of our knowledge, the tentative conclusions that have been drawn so far from various studies of UHECR phenomenology have all been reached under rather restrictive assumptions regarding either the UHECR source model or the GMF model, but would not necessarily hold in general, since these models are only possibilities among others, which are also compatible with the current observational constraints, and which would often lead to different conclusions. Moreover, despite several attempts to account for \emph{a subset of} the available data in terms of some specific astrophysical scenarios {\color{red}[ref.?]}, so far none of them have proven able to account for the complete set of observational constraints.

In this series of papers, our main purpose is to examine what can and what cannot be inferred from the set of available data, what would be needed in priority to improve the situation, and to what extent future experimental developments can help reaching the first goal of UHECR astrophysics: identifying at least the class of sources responsible for their very existence. This includes considerations of what would be offered by a better separation of the light and heavy nuclear components of the UHECR spectrum, which is an important motivation for the upgrade program of Auger \citep{AugerUpgrade2020}, as well as of the benefits to be expected from larger statistics and/or consistent full-sky observations, either from the ground \citep{GRAND2020, GCOS2021} or from space \citep{Bertaina2019, POEMMA2021}.

Our strategy is to investigate a range of astrophysical scenarios, varying the density of UHECR sources, their 3D distribution in space around our Galaxy and their effective composition and energy spectrum at their injection site. In each case, we build simulated sky maps taking into account the propagation of the UHECRs in the cosmological photon backgrounds and in (models of) the extragalactic and Galactic magnetic fields. These sky maps are built assuming various statistics (total numbers of events) and various exposures corresponding to different observatory locations, using Auger and its statistics at the time of the  publication of the dipole signal as a benchmark. For each astrophysical scenario, we investigate both the so-called “cosmic variance” (by producing many realisations of an actual source distribution satisfying the same general hypotheses) and the “statistical variance” (by generating many datasets with a given statistics for each particular realisation). The obtained skymaps are then analyzed from the point of view of different anisotropy observables. We examine the sensitivity of our predictions to the various categories of astrophysical hypotheses, and discuss the importance of the statistical and cosmic variances. We assess the compatibility of the predictions for each scenario with the current observations, as well as the potential benefits from future datasets that would correspond to higher statistics and/or uniform full-sky coverage and/or improved discrimination between nuclear species.

In the present paper, we concentrate on the dipolar and multipolar modulation of the arrival directions of UHECRs, at different energies. In Sect. 2, we describe the various astrophysical scenarios explored and underline their main ingredients. In Sects.~\ref{sec:transport}, \ref{sec:data} and \ref{sec:content}, we describe the simulations themselves and give more details about the production of representative UHECR skymaps and their content. In Sect.~\ref{sec:results}, we present and discuss our main results, comparing the expected anisotropies for each scenario with the actual measurements of Auger. This includes dipolar and higher order anisotropies, both in terms of amplitude and position on the sky. We also discuss composition related issues, such as the potential interest of an accurate separation of the nuclear components as well as the possible existence of a difference in the average composition of the UHECRs in regions having different event count rates. Finally, a summary and general discussion is given in Sect.~\ref{sec:conclusion}. The general study of possible significant excesses in the flux of UHECRs in specific regions of the sky (through top-hat and Li\&Ma analyses), as well as of the correlation of the UHECR arrival directions with known sources and source catalogs, is left for a companion paper.

\section{Source models and astrophysical parameters}

\subsection{UHECR source distribution in 3D space}
\label{sec:sourceDistrib}

The spatial distribution of the UHECR sources is obviously one of the critical ingredients of any UHECR model, when considering the distribution of their arrival directions on Earth. In the absence of a clear prescription for the nature of the sources, a simple and natural choice is to assume a distribution similar to that of ordinary matter. Therefore, the starting point for the 3D distribution of sources in our simulations is provided by the catalog of galaxies of the 2MASS Redshift Survey catalog (2MRS, \citet{Huchra2012}), used in a similar way as described in detail in (\cite{BRDO2014}), with some improvements indicated below. The catalog, linked with the Extragalactic Distance Database (EDD, \citet{EDD2009}) to obtain a distance estimate of the nearby galaxies (which cannot be estimated accurately using the Hubble law), benefits from the recent addition of the Cosmicflows-3 distance catalogue \citep{CF32016}, as well as the updated nearby galaxy catalog of \citet{Kar2013}. Moreover the membership of individual galaxies (within $\rm3500\,km/s$) to larger groups or associations of galaxies can be assessed thanks to the database derived from the Kourkchi-Tully groups catalog \citep{Kourk2017}. As in \citet{BRDO2014}, the catalog is subsequently enriched to account for missing sources in the Galactic plane according to the method proposed in \citet{Crook2007}.

For our purpose, i.e. the simulation of UHECR skymaps associated with specific astrophysical scenarios, we need to specify the position of all the sources (of a given class) likely to contribute to the observed UHECR flux. Since the sources are not known, we must use a proxy for their location, making the assumption that they are distributed in a similar way as galaxies in general, or some classes of galaxies. It would be best to use a volume-limited catalog of galaxies, completing it beyond the maximum radius of completeness according to some specific prescription. However, the 2MASS catalog is a magnitude-limited catalog, resulting from a survey in the near-infrared K-band that includes galaxies with apparent magnitude $K_s \leq 11.25$. It is thus affected by radial-selection effects, and it is not necessarily reliable to deduce the distribution of certain types of galaxies (say of lower luminosity, thus excluded from the catalog beyond a certain distance) from that of the most luminous galaxies, for which the catalog is in principle complete up to larger distances.
%Indeed, at each distance $D$ only galaxies brighter than an absolute magnitude $M_0(D)$ (or the corresponding luminosity $L_0(D)$) are observed with an apparent magnitude $K_s$.
We have no other choice, however, than deducing the likely spatial distribution of sources from the available galaxies in the catalog at hand.

In the following, we use two different approaches for the choice of sources, as detailed in the next paragraphs: {\it i)} a {\it volume-limited catalog approach}, which allows us to have a fixed distribution of sources and study the influence of various physical parameters on the expected UHECR anisotropies, with all other parameters fixed, and {\it ii)} a {\it mother catalog approach}, with which we can specifically study the cosmic variance of the anisotropy features under study.

\subsubsection{Volume-limited catalog approach}

To build a volume-limited catalog containing a complete sample of galaxies closer than a given distance $D_{\max}$, one must set a lower limit on the absolute luminosity of the considered galaxies, such that any such galaxy would indeed be seen up to distance $D_{\max}$. Thus, for each choice of $D_{\max}$, one obtains the corresponding absolute luminosity $L_0(D_{\max})$ beyond which the catalog is complete (up to $D_{\max}$), and that limit results in a specific value of the source density, namely the density of the sources (here galaxies) with absolute luminosities larger than $L_0(D_{\max})$, simply obtained by dividing the number of such galaxies by the volume of the sphere of radius $D_{\max}$ (see Sect~3.3 of \citet{BRDO2014} and the corresponding figures for details).

We produced in this way four volume-limited catalogs, complete up to distances $D_{\max} = 40$, 104, 144 and 176~Mpc, chosen to provide source densities of interest for the present UHECR studies.%\footnote{Note that for the highest density volume limited catalog, we manually remove source with distance lower than 1 Mpc, and in particular M31 (which would otherwise pass the K-band luminosity threshold $L_0$), since this very nearby galaxy, if present in the catalog, would systematically contribute significantly to the diffuse UHECR spectrum, dominate anisotropies patterns and produce features obviously at odd with the data, at least in the range of astrophysical assumptions we consider in the following.}
The corresponding values are given in Table~\ref{table:1}. Of course, each of these volume-limited catalogs becomes increasingly incomplete as the distance increases above $D_{\max}$, since sources with similar luminosities ($L > L_0(D_{\max})$) eventually pass below the apparent magnitude threshold, $K_\mathrm{s}$, of the 2MASS catalog. These sources, however, cannot be ignored when computing their contribution to the UHECR flux observed on Earth. Therefore, the above volume-limited catalogs must be completed by the addition of an increasing numbers of missing sources as the distance above $D_{\max}$ increases.

To estimate the number of these missing sources in a given distance bin, we use the results of large-scale structure simulations (\citet{LSSS2018}; hereafter LSSS), constrained by the Cosmicflows2 peculiar velocities catalog \citep{CF22014}. These simulations provide a 3D spatial distribution of the matter density in the universe around us, which in turn provides a local over-density factor in the considered distance bin (compared to the average), by which we rescale the selected source density associated with our initial choice of $D_{\max}$. Then, for the direction of these missing sources, we also draw it randomly according to the spatial distribution in the LSSS at that distance. Note that these constrained LSSS, which were used for the first time in \citet{No2019} to model UHECR anisotropies, are used up to a distance $\sim800$ Mpc. Beyond that distance, we draw the missing sources randomly assuming a homogeneous and isotropic distribution. Finally, we ascribe an absolute luminosity to each missing source by drawing it randomly according to the actual 2MASS luminosity function in the K-band, but only between the limiting luminosity, $L_0(D_{\max})$, corresponding to the source density under consideration, and the luminosity above which any source would be seen at the considered distance, and thus already included in the 2MRS catalog.

It is important to note that the galaxies closer than $D_{\max}$ selected for these volume-limited catalogs are chosen solely on the basis of their K-band luminosity, which is unlikely to be a particularly relevant criterion to select UHECR source candidates. We shall nevertheless use these galaxies as proxies for the position of UHECR sources, as they still capture some important properties of the local matter distribution and allow us to specify a definite source distribution, from which we can study the separate influence of other astrophysical parameters. 

%It is important to note that for these volume-limited catalogs, the selected galaxies closer than $D_{max}$ which will serve as inferred UHECR sources in the following are chosen on the sole criterion of their K-band luminosity, which is however unlikely to be a particularly relevant criterion to select UHECR source candidate. This will however allow us to study the importance of various physical parameter such a the extra-galactic or galactic magnetic fields, the composition at the source, the effect of a partial sky coverage or of the limited statistics of current and future data-sets on a precisely fixed (at least within $D_{max}$) spatial distribution of sources. 

\begin{table}
\caption{Characteristics of the volume-limited catalogs. $L_0$ represents the minimum luminosity of the selected galaxies in the K-band, in solar units.}             % title of Table
\label{table:1}      % is used to refer this table in the text
\centering                          % used for centering table
\begin{tabular}{c c c c}        % centered columns (4 columns)
\hline\hline                 % inserts double horizontal lines
$D_{max}$ (Mpc) & $\rho$ ($\rm Mpc^{-3}$) & $L_0$  ($L_\sun$ units)\\    % table heading 
\hline                        % inserts single horizontal line
   40 & $7.6\times10^{-3}$ & $1\times10^{10}$  \\      % inserting body of the table
   104 & $1.4\times10^{-3}$ & $7\times10^{10}$     \\
   144 & $3.7\times10^{-4}$ & $1.3\times10^{11}$    \\
   176 & $1.1\times10^{-4}$ & $2\times10^{11}$     \\
   
\hline   
%inserts single line
\end{tabular}
%\tablefoot{$L_0$ represents the minimum luminosity of the selected galaxies in the K-band (in solar units).}
\end{table}

\subsubsection{Mother catalog  approach}
However, we also need to study the importance of the cosmic variance, i.e. how the actual, contingent distribution of sources that happen to be currently active around us may impact the typical observations that can be expected to be made. In other words, what dispersion in the various theoretical predictions would result from different realisations of the same general astrophysical scenario? In the ignorance of the definite sources of the observed UHECRs, this is important to determine how much can be inferred about the actual astrophysical processes at work from the present or future datasets.

To this end, we will use the volume-limited catalog with the largest source density, namely the one corresponding to $D_{max}=40$ Mpc, as a {\it mother catalogue} from which many different sub-catalogs can be obtained, with lower source densities, by random sampling. In particular, we study the cosmic variance associated with scenarios in which the UHECR sources are randomly chosen within the mother catalogue with source densities $10^{-3},\,10^{-3.5},\,10^{-4},$ and $10^{-5}\,\rm Mpc^{-3}$. As indicated in Table~\ref{table:1}, the complete set of galaxies in the mother catalog corresponds to a source density of $\sim7.6\times10^{-3}$. Thus, to explore a UHECR scenario with a source density of $10^{-4}\,\rm Mpc^{-3}$, say, we simply select randomly 1 galaxy out of 76 in the mother catalog. We repeat this procedure 300 times for each density to obtain 300 different realisations of the source distribution for the same astrophysical scenario.

\begin{figure}[t!]
   \centering
   \includegraphics[width=\hsize]{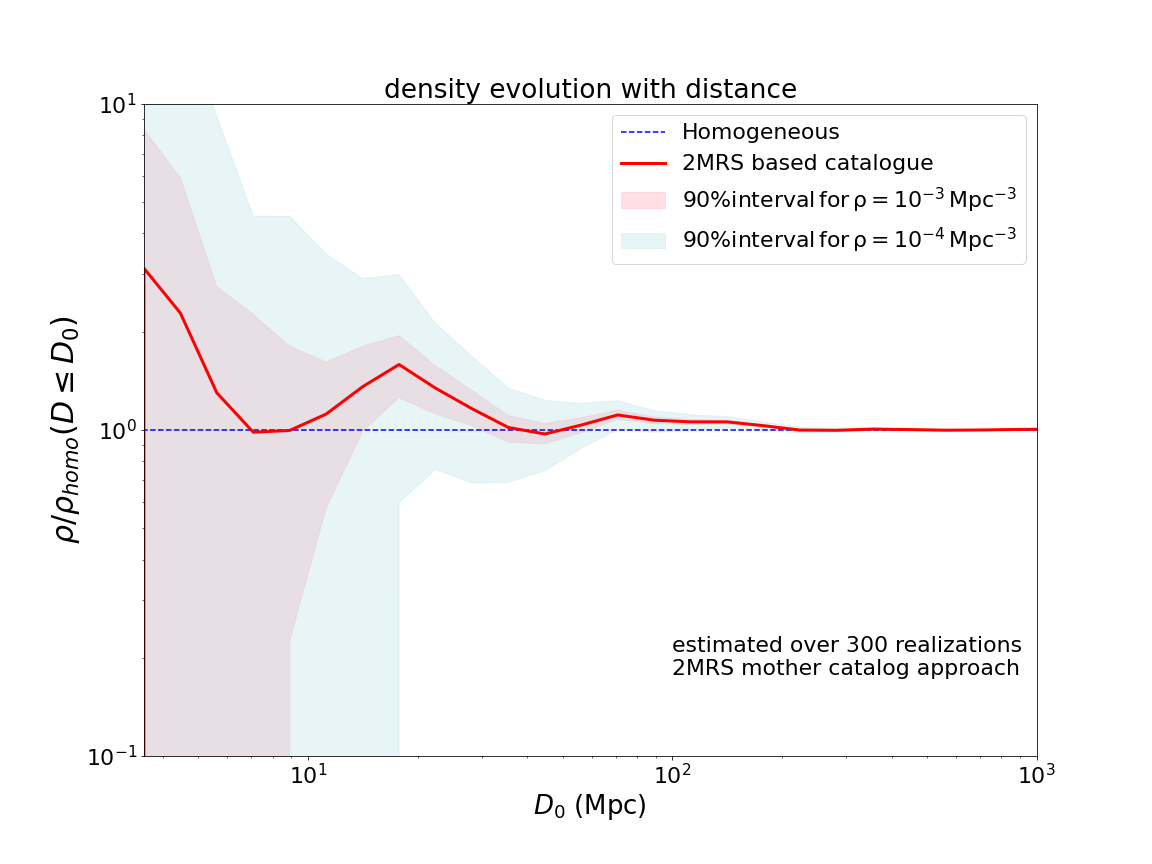}
      \caption{Average density of UHECR sources in a sphere of radius $D_0$ normalised to the case of a homogeneous universe, as function of that radius, for 300 different realisations of a UHECR source distribution randomly drawn from the mother catalog (see text) with an overall density of $10^{-3}\,\rm Mpc^{-3}$ (red) or $10^{-4}\,\rm Mpc^{-3}$ (blue). The thick red line shows the average overdensity for 300 realisations (thus close to that of the mother catalog itself), while the shaded area shows the spread around this average, and contains 90\% of the realisations (excluding the 5\% most overdense and the 5\% most underdense realisations).}
         \label{Fig2MRSDens}
   \end{figure}

Figure~\ref{Fig2MRSDens} shows the overdensity of the mother catalog (integrated up to a given distance, as a function of that distance), compared to a homogeneous distribution, as well as the range of variation of the source distribution as a function of distance, for the different realisations. The red and blue shaded areas contain 90\% of the 300 realisations, respectively for a source density of $10^{-3}\,\rm Mpc^{-3}$ and $10^{-4}\,\rm Mpc^{-3}$ (excluding the 5\% most overdense and most underdense).

\subsubsection{Additional comments on source catalogs}

The above-mentioned volume-limited catalogs are nothing but one particular realisation of a catalog of sources with the corresponding source density, which has no particular reason to be preferred to any other obtained by sub-sampling the mother catalog, except in the very specific (and unlikely) assumption that only the galaxies with an intrinsic luminosity larger that the corresponding cut are actual UHECR sources. However, these are the ones we will use, as one example among others, when we will need a fixed distribution of sources to study the specific influence of the various physical parameters on the different predictions, as well as their statistical variance. NB: the only modification that we will make on the volume-limited catalogs will be to manually remove source with distance lower than 1 Mpc, and in particular M31 (which would otherwise pass the K-band luminosity threshold $L_0$ in the case of the largest density volume-limited catalogue), since this neighbour galaxy, if present in the catalog, would systematically contribute significantly to the diffuse UHECR spectrum, dominate anisotropies patterns and produce features obviously at odds with the data, at least in the range of astrophysical assumptions that we consider in the following.

Note that the volume-limited catalog with a source density $\rho_{\mathrm{s}}=1.4\times 10^{-3}\,\rm Mpc^{-3}$ corresponds to the most stringent luminosity cut allowing to keep most of the often cited local candidate sources, such as Centaurus~A, M81/82 or NGC253 (together with higher luminosity and more distant galaxies such as NGC1068, M87 or Fornax~A). We  consider it as our baseline volume-limited catalog in the following. It is worth noting, however, that this particular realisation of the 2MRS catalog exhibits an overdensity in the inner 20~Mpc radius, which is larger than the average value shown in red in Fig.~\ref{Fig2MRSDens}. This is an example of cosmic variance, which in this particular case tends to give rise to somewhat higher anisotropy levels than an average random realisation of the catalog with similar density (see below).

%In the following studies, we will use the above two types of catalogs: i) the “volume-limited catalog approach” to study the statistical variance of the predictions, as well as the importance of the various physical parameter on the predictions, assuming a fixed distribution of sources, and ii) the “seed catalog approach” to study and quantify the cosmic variance. 

Our study also includes “biased” versions of the 2MRS catalog, in which we either suppress or enhance the injection of UHECRs from potential sources located in the richest nearby cluster and associations of galaxies, as could be suggested by some specific UHECR source models, e.g. involving cluster-scale accretion shocks (effective enhancement) or starburst galaxies (potential suppression, due to the smaller amount of gas in galaxies in the environment of large clusters, see {\it e.g} \citet{Guglielmo2015}). Another type of biased catalog will be used, where we impose the presence of a given astrophysical source (such as CenA, M81/M82 or NGC253) when subsampling the mother catalog, in order to study specific scenarios.

Finally, we consider the case of a random distribution of sources, drawn with various source densities from an underlying homogeneous and isotropic distribution. This allows to examine to what extent the study of dipolar and multipolar anisotropies can discriminate between a scenario where the UHECR sources follow the local structure of matter in the universe and a purely random configuration of discrete sources.

\subsection{General characteristics of the sources}

Simple models for the propagation of UHECRs from their sources to the Earth, taking into account the energy losses, nuclear photodissociation and deflections in the intervening magnetic fields, allow to reproduce the overall energy spectrum with a limited set of parameters, under the simplifying assumption that all the UHECR sources are identical, i.e. {\it i)} they have the same intrinsic power, {\it ii)} they inject UHECRs in the extragalactic medium continuously, {\it iii)} with the same energy spectrum consisting of a power-law with some rigidity-dependent (usually exponential) cut-off, and {\it iv)} with the same nuclear abundance ratios (hereafter “composition”). The main free parameters of such models are the global source density, the logarithmic slope of the source spectrum, the energy (or rigidity) scale of the cutoff in the UHECRs injection spectrum, and the relative abundances of the various nuclei at the source. From the phenomenological point of view, these parameters are in principle independent from one another (at least until an actual acceleration model in a given astrophysical environment is proposed), but they must be chosen consistently to reproduce the observed spectrum and composition. Only certain combinations can lead to simulated dataset compatible with the existing constraints.

Although the above simplifying assumptions appear natural in the current stage of investigation, in reality it is likely that individual sources differ from one another regarding their spectral index, composition and maximum energy (or rigidity). However, even though introducing such source-to-source variability would necessarily increase the variance of the different model predictions, the above assumptions still allow to explore a wide range of models whose predictions, when varying their input parameters, should cover the range of patterns that may be expected as far as anisotropies are concerned (which is the main focus of this paper). Relaxing these assumptions would require the introduction of additional free parameters on which there are no observational constraints at the moment, and would thus be justified only in the framework of a specific theoretical source model, which is not our goal here.

%The most standard propagation studies that take into account the energy losses and angular deflections of the UHECRs allow to reproduce the whole sky spectrum with a limited set of parameters, under the simplifying assumption that all UHECR sources are essentially identical, i.e. {\it i)} they have the same intrinsic power, {\it ii)} they inject UHECRs in the extragalactic medium continuously, {\it iii)} with the same power-law energy spectrum, and {\it iv)} with the same composition. The remaining free parameters are the source density, the logarithmic slope of the source spectrum, the maximum energy of the UHECRs and the relative abundances of the various nuclei at the source. These parameters are not independent, and must be chosen so as to reproduce correctly the observed spectrum and composition. In reality, it is likely that individual sources show some variability in terms of their cosmic-ray output (that is, their spectral shape composition and/or maximum energy/rigidity). However, the above assumptions allow to explore a  set of models which should be representative of the range of patterns one may expect from the point of view of anisotropies. Relaxing them would introduce more free parameters on which there are no observational constraints at the moment, and this would be justified only in the framework of a specific theoretical source model which is not our goal here. 

The only assumption that we will relax is that of an identical power for all sources, which will be replaced in the case when we use the above-mentioned volume-limited catalogs by the assumption that the UHECR sources have the same intrinsic luminosity distribution as the galaxies in the catalog. By contrast, when exploring the cosmic variance with the mother catalog, we will mostly assume that UHECR sources are standard candles. The case of a power-law or broken power-law luminosity distribution will also be briefly discussed.
We shall not consider in this study the case of short-lived or transient sources (which were recently discussed for instance in \citet{GAPLP2017}). However, we do not expect that the main conclusions of our study would be significantly different if we considered this class of sources as well.

\begin{table*}[t!]
\caption{Characteristics of the four generic source models: A, B, C and D (see text), giving the spectral index, the relative normalisation of the rigidity spectrum for different nuclear components, and the maximum rigidity (common to all species). The last column indicates whether the model assumes a cosmological evolution of the source luminosity (as function of redshift/time). When it is the case, the evolution is assumed to be similar to that of GRBs, as estimated in \cite{Wand2010}.}             % title of Table
\label{table:2}      % is used to refer this table in the text
\centering                          % used for centering table
\begin{tabular}{c c c c c c c c c c}        % centered columns (8 columns)
\hline\hline                 % inserts double horizontal lines
Model & $\beta$ & $\alpha(H)$  &  $\alpha(He)$  &  $\alpha(C)$  &  $\alpha(O)$  & $\alpha(Si)$  & $\alpha(Fe)$  &  $R_{max}$ (EV) & evolution \\    % table heading 
%\hline                        % inserts single horizontal line
%   A & 0.45 & 20.75 & 49.45 & 10.62 & 16.12 & 2.72 & 1.33 & 1.6  & yes \\      % inserting body of the table
%   B & -0.85 & 7.72 & 12.86 & 34.37 & 38.00 & 4.56 & 2.47 & 4.5 & no    \\
%   C & 0.6 & 17.03 & 47.35 & 11.98 & 18.98 & 2.78 & 1.87 & 1.4 & yes   \\
\hline                        % inserts single horizontal line
   A & -0.45 & 20. & 49. & 10. & 16. & 2.7 & 1.3 & 1.6  & yes \\      % inserting body of the table
   B & 0.85 & 7.7 & 13. & 34. & 38. & 4.6 & 2.5 & 4.5 & no    \\
   C & -0.6 & 17. & 47. & 12. & 19. & 2.8 & 1.9 & 1.4 & yes   \\
\end{tabular}
\begin{tabular}{c c} 
   \hline

   D$\,\,\,\,\,\,\,\,\,\,\,\,\,\,\,\,\,\,\,\,\,\,\,\,\,\,\,\,\,\,\,\,\,\,\,\,\,\,\,\,\,$ adapted and  modified from \citet{GAMP2015}  $\,\,\,\,\,$  $\,\,\,\,\,$ $\,\,\,\,\,$ $\,\,\,\,\,$& $\,\,\,$yes \\
   
\hline   
%inserts single line
\end{tabular}
%\tablefoot{The column "evolution" refers to the cosmological evolution of the luminosity of the sources. For the models with evolution we chose the example of GRBs source evolution as estimated in \cite{Wand2010}.}
\end{table*}

\subsection{Source spectrum and composition}
\label{sec:sourceSpectrumAndComposition}

The UHECR source composition, i.e. the relative abundance of the various nuclei injected as UHECRs by the source, has a strong influence on the level of anisotropy that one can expect to measure. It had long been hoped that increasing the statistics at the highest energies would give rise to the observation of very strong anisotropies at small angular scales, in the form of tight multiplets of events within a few degrees from each other. Under reasonable assumptions about the strength of the Galactic and extragalactic magnetic fields, such an expectation was judged reasonable under the implicit assumption that most of the highest energy cosmic rays would be protons, with only a few sources contributing to the observed flux. However, such multiplet observations did not occur so far, and it is clear that the evolution of the UHECR composition observed by Auger above the ankle of the cosmic-ray spectrum makes the situation much less favorable, as far as small angular scale anisotropies are concerned.

All the models considered here belong to the category of the so-called “low Emax models”, which refers to mixed-composition models in which the protons do not reach the highest energies. In these models, protons are accelerated by the sources only up to a maximum energy that is lower than the energy range of the GZK suppression. As a consequence, the source composition above the ankle is gradually becoming heavier, by successive extinction of the lighter components. For the purpose of our study, we select extragalactic source spectra and compositions that allow to reproduce correctly the observed UHECR spectrum and composition above the ankle, and examine the associated properties of the arrival direction distribution of these cosmic-rays above 4 EeV.

We focus on four different source models, noted A, B, C and D, whose characteristics are reported in Tab.~\ref{table:2}. In models A, B and C, the energy spectrum at the sources is assumed to be a power-law with the same logarithmic index, $-\beta$, for all nuclear species, ended by an exponential cutoff at a rigidity scale, $R_{max}$, which is also supposed to be the same for all species. As a consequence, the maximum energy scale, $E_{max,i}$, for each nucleus $i$ is proportional to its charge, $Z_i$ : $E_{max,i} = Z_i \times  R_{max}$. The rigidity spectrum is thus given, for nuclei $i$, by: $${\mathrm{d}N_i/\mathrm{d}R} = \alpha_i\times K R^{-\beta} \exp(-R/R_{max}),$$ where $\alpha_i$ is an abundance coefficient, and $K$ is an overall normalisation factor adjusted to reproduce the observed UHECR flux.
%$\frac{\mathrm{d}N_i}{\mathrm{d}R}(R) = \alpha_i R^{\beta} \exp(-\frac{R}{R_{max}})$.
In addition to the spectral index, maximum rigidity and relative abundances (at a given rigidity), Tab.~\ref{table:2} indicates whether a cosmological evolution of the intrinsic source power is assumed.

Although these models can be very different from the point of view of the source phenomenology, they result in propagated spectra and compositions which are compatible with the current data. We adopt model A as our baseline. Model B shows very different abundance ratios at the source, but leads to a similar evolution with energy of the mean nuclear mass, characterised by the quantity $\langle \ln{A}\rangle$ (where $A$ is the atomic number), up to $\sim2\,10^{19}~$eV. It then leads to a lighter composition than model A at the highest energies, which makes it interesting to explore an alternative composition trend, still compatible with the observations (see below). %, where the highest energies have a lower mean $\ln{A}$ towards the bottom of the favoured range by the Auger data.
As for model C, it has a similar evolution of $\langle \ln{A}\rangle$ with energy, but is globally heavier than model A (thereby leading to generally less anisotropic sky maps), towards the top of the favoured range for $\ln{A}$. Finally, model D is a modified version of the effective source spectrum derived from the study by some of us of the acceleration of UHECRs at the internal shocks of gamma-ray bursts (GRBs), reported in \citet{GAMP2015}. For this revised model, we modified the assumed GRB luminosity function so that the effective source spectrum leads to a better agreement with the measured spectrum after propagation (the spectrum obtained with the initially assumed luminosity function would be slightly too soft, when using our updated photonuclear cross sections spectrum: see below). As a result, the evolution of the composition, from light to heavy, is slightly more pronounced than in the original version of the model. Most importantly, it is quite different from the other three models (notably with a significantly larger proton abundance around the ankle), which is why we discuss it here as well.

Note that we did not choose the parameters of these models by means of a minimization or other statistical methods (see e.g. \citet{AugerFit2017} for a thorough discussion), but rather on the basis of a visual evaluation of their global agreement with the data. This is because: {\it i)} the analysis of the UHECR extensive air showers shows that the current hadronic interaction models are not able to fully explain the observations. As a result, and although the general trend of the evolution of the composition appears to be better and better established, a detailed determination of the UHECR composition in terms of the relative abundances of the different nuclei or groups of nuclei is currently out of reach (not mentioning the systematic and statistical uncertainties in the measurements themselves). As a consequence, a sophisticated optimisation of the models to match the composition estimated from the data does not appear sufficiently reliable, nor useful in the context of our studies. Moreover, {\it ii)} when one considers the cosmic variance associated with the different astrophysical scenarios, that is the uncertainty on the actual position of the sources for a given hypothesis on their global distribution, one finds that the realisation-to-realisation variation of the UHECR spectrum after propagation from the sources to the Earth can be much larger than what would result from even a large change in the source composition, especially when the assumed source density reaches values as low as, say, $10^{-4}\rm Mpc^{-3}$. Therefore, deriving an optimised source composition on the basis of a specific source model and source distribution, and then sticking to it when varying the model parameters, is not appropriate. In principle, one would have to repeat the minimization procedure for each realization of the source distribution randomly sub-sampled from the mother catalog. Yet, fixing the composition based on a reasonable agreement with the general observed phenomenology is interesting in the framework of our study in that it allows to examine the specific effect of varying the different parameters individually.

Finally, it is important to keep in mind that the parameters given in Tab.~\ref{table:2}, under the simplifying assumptions of standard candles with identical power-law spectra and exponential cutoffs, are nothing but effective parameters accounting for the assumed extragalactic UHECRs above 4 EeV, and are relevant for the comparison of the model predictions with the data in that energy range only. In particular, even though the very hard source spectral indices would imply, if taken at face value, a negligible contribution of this extragalactic component below the ankle (see {\it e.g.} \cite{Aloi2014}), the actual UHECR component modelled by models A, B and C above the ankle does not have to have this property. As a matter of fact, models involving a soft ($\beta\ga 2$) spectral index for protons and much harder spectra for heavier nuclei, can very well have similar cosmic-ray outputs above 4 EeV. This is notably the case of the above-mentioned GRB acceleration model, as a result of the competition between photo-interactions and escape in the accelerator, as we described in \citet{GAMP2015} and discussed in the context of the Galactic-to-extragalactic transition in \citet{GAP2015}. Thus, the model parameters are effective in that precise sense that, when propagated from the corresponding sources, the UHECRs that arrive at Earth above 4 EeV have on average (over source realisations and datasets) a similar spectrum and composition as the one actually observed.

To illustrate this point we performed a fit of the source spectrum of the different nuclei in the case of model A above 5 EeV, leaving the spectral index free for the different species and assuming a squared exponential cut-off rather than a simple exponential cut-off. A best fit spectral index of $\sim 2$ and a value of $E_{max}\simeq 5$~EeV were found for the proton component, while indices of the order of $\sim 0.6$ were found for heavier nuclei. These values are quite different from those given in Tab.~\ref{table:2}, simply because a squared exponential was assumed instead of a regular one, even though the fitted spectra and abundances are the very same. Now, with this second set of values, the proton component would indeed be much softer than that of the other nuclei, and would thus result in very different conclusions if extrapolated below 4~EeV. This clearly shows that the effective model parameters that we use in the framework of this study should not be over-interpreted, and that, in particular, the discussion of the observed anisotropies above 4 EeV is largely independent from a discussion of the transition from the Galactic to the extragalactic cosmic-ray component.

\subsection{Magnetic field models}
\label{sec:magneticFieldModels}

The last astrophysical ingredient of the computations, especially crucial when studying the distribution of the UHECR arrival directions and associated anisotropies, is the model of the magnetic fields intervening between the source and the detector, both outside and inside the Galaxy.

\subsubsection{Extragalactic magnetic fields}
\label{sec:EGMF}

The extragalactic magnetic field (EGMF) is poorly known: not only its origin, but all its aspects influencing the propagation of UHECRs, namely its spatial distribution, intensity, coherence length and time evolution, are currently uncertain. Observations imply the presence of $\mu$G fields in the core of large galaxy clusters. However, the spatial extension of these large field regions and their volume filling factor in the universe are difficult to evaluate. Efforts have been made, in the past two decades (see {\it e.g} pioneering works by \citet{Dolag2002, Sigl2002, Armengaud2005}) to model local magnetic fields using simulations of large-scale structure formation that include an MHD treatment of the magnetic field evolution. These simulations rely on different assumptions regarding the origin of the fields (see the recent discussion in \citet{Hack2018}) and the mechanisms involved in their growth. The outcome of the different simulations can strongly differ, notably concerning the volume filling factors for strong fields (above 1 nG, say), which is particularly important for UHECR transport, and varies significantly from one simulation to the other. Thus, given the lack of observational constraints, it is not clear to what extent they capture even the main characteristics of the EGMF. Note also that in order to give an actual value for the magnetic field, their outcome must be normalized \emph{a posteriori} to the values observed at the present epoch in the central regions of the galaxy clusters.

In view of these uncertainties, we chose to restrict our study to the general effect of different amounts of deflections of the UHECRs in the intergalactic medium, i.e. before they enter the Galaxy, by assuming purely turbulent, homogeneous magnetic fields of different levels, ranging from $10^{-2}$~nG (which leads to quasi-negligible deflections) to 10~nG (which is already above the commonly accepted upper limit for non negligible filling factors, see {\it e.g.} \citet{Blasi1999}). This simplified approach leaves enough flexibility for us to investigate the impact of a magnetic blurring of the UHECRs before they interact with the Galactic magnetic fields, whose effects are dominant, and to test a wide range of possible deflections for the astrophysical scenarios under consideration, while keeping the amount of needed computational resources reasonable. For the same reason, we fix the coherence length of the assumed isotropic Kolmogorov-like turbulent field to a value of $\lambda_{\mathrm{c}}$=200~kpc, noting that for this topology of the EGMF, there is a degeneracy between the field intensity and the coherence length, as far as the particle deflections are concerned.
 More sophisticated EGMF models can be implemented as in \cite{Witt2018} or \citet{Hack2018}. However in that case, the entire 3D extragalactic propagation must be simulated again each time one changes the source distribution and/or density (because of the inhomogeneity which implies that different sources at the same distance will have different contributions. Since the investigation of the cosmic variance is an important aspect of our study, this approach proved impossible in our case. Besides, the knowledge of the EGMF is currently too uncertain for one to confidently relie on a definite EGMF configuration. Note that intermediate approaches, interesting simple alternative to complex hydrodynamical or magneto-hydrodynamical simulations, offering more freedom to test different models of magnetic field evolution with matter density and inhomogeneous EGMFs, have also been proposed in \citet{Kotera2008a, Kotera2008b}.

%Our description of the structure and spatial distribution of the EGMF is arguably less sophisticated and realistic than that of \cite{Witt2018} or \citet{Hack2018}. It however allow us more flexibility to test the influence of a large number of astrophysical parameters and of the cosmic variance. Let us note that intermediate approaches, interesting simple alternative to complex hydrodynamical or magneto-hydrodynamical simulations, offering more freedom to test different models of magnetic field evolution with matter density and inhomogeneous EGMFs, have also been proposed in \citet{Kotera2008a, Kotera2008b}.}}

\subsubsection{The Galactic magnetic field}
\label{sec:GMF}

To simulate the transport of the UHECRs in the Galaxy, we implemented different models of the Galactic magnetic field (GMF). For our baseline model, we followed the modeling of \citet{JF2012a, JF2012b} for the coherent, turbulent and striated components of the GMF (see {\it e.g.}, \citet{Jaffe2010, Jaffe2011}). One of the salient characteristics of this model was to include for the first time a so-called X-field which is motivated by observations of X-shaped field structures in external galaxies (see {\it e.g.} \citet{Beck2009}). As an alternative GMF model, we consider the simpler “ASS+RING” model  proposed by \citet{Sun2008, Sun2010}. For both models, we consider the parameters as updated after the comparison of their predicted polarized synchrotron and dust emissions with that measured by the Planck satellite mission, as reported in \citet{GMFPlanck2016}. We refer to the former model as the “JF12+Planck” model, and to the latter as the “Sun+Planck” model.

In all cases, we follow the numerical modeling procedure of \citet{Giac1999} for the random turbulent component of the GMF and assume a Kolmogorov-like turbulence assuming various values for the coherence length, namely $\lambda_{\mathrm{c}} = 20$, 50, 200, 500, or 2000 pc, bracketing the favored value of $\sim 200$~pc according to \citet{Beck2016}. NB: when simulating the isotropic random field, we include between 3.5 and 5 decades of the turbulence wave-numbers, depending on the assumed value of $\lambda_{\mathrm{c}}$.

%\section{UHECR propagation, datasets and skymaps production}

\subsection{Energy losses and photo-nuclear cross-sections}

For each of the explored astrophysical scenarios, i.e. each set of assumptions regarding the parameters and physical ingredients presented above, we simulate the propagation of the UHECRs from their sources to the Earth, using the Monte-Carlo code presented in \citet{Allard2005}, taking into account the relevant energy loss processes, the possible change of nuclear species through photonuclear interactions with the various photon backgrounds, and the deflections of the particles in the extragalactic and Galactic magnetic fields. We refer the reader to the detailed description of the relevant processes in the above reference, and simply note here that we have updated the Giant Dipole Resonance (GDR) cross-sections and reactions branching ratios of the ~380 isotopes considered between $^2$H and $^{56}$Fe, using the cross-sections predicted by the latest version of the TALYS code \citep{Kon2004, Gor2008}, namely TALYS-1.95.

The recent improvements implemented in this framework (which will be presented in more detail elsewhere) include: {\it i)}, updated Lorentzian-based parameterizations of the E1-strength of the different isotopes \citep{Plu2018}, {\it ii)} a phenomenological treatment of {\it isospin forbidden transitions} (which affects in particular the branching ratios of $(\gamma, x\alpha)$ reactions), {\it iii)} a description of nuclear level densities better adapted to light and intermediate nuclei \citep{Kie2020}, and {\it iv)} the replacement of Lorentzian-based parameterizations of the total E1-strength by experimental data for light and intermediate nuclei isotopes ({\it i.e.} $\sim$Be to Si), whenever available (see {\it e.g.} the compilation of experimental data by \citet{Ish2002}). Note that, while giving a somewhat robuster treatment of the GDR interactions of the UHECRs than in our previous works, the use of these updated cross-sections does not have any strong impact on the conclusions of the present study. The use of our previous set of cross-sections would essentially lead to different parameters for the source spectra and compositions providing a satisfying description of the data (see above), and only slight numerical differences in the results presented below.

\section{UHECR transport in 3D space}
\label{sec:transport}
\subsection{Two-step process}

The transport of UHECRs in 3D space is treated numerically in two separate steps. In the first step, we propagate the UHECR protons and nuclei in the extragalactic medium, and derive their characteristics as they enter our Galaxy, whose limit is arbitrarily defined for the present purpose as a sphere of radius 50~kpc around the Galactic center. We thus obtain the energy, mass and arrival direction (on that sphere) of individual UHECRs injected by a given source located at a distance $D$ and Galactic coordinates $l$ and $b$, at a redshift/time $z$, and then add the contributions of all sources in the particular realisation of the source model under consideration.

For this first step, we use the fast integration method described in detail in \citet{GAP2008} to compute the 3D trajectories as influenced by the EGMF (see in particular Sect.~4 of this reference, where a comparison with a full numerical integration is given). This allows us to keep track of the time (i.e. redshift) dependence of the energy losses, without having to assume rectilinear transport.

In the second step, we propagate the UHECRs arriving at the (fictitious) border of the Galaxy through the GMF to the Earth, to determine their actual arrival direction at the detector location, and thus build the final simulated dataset. This is done by using a “transfer function” in angular space, obtained by inverting the function that maps the different directions on the celestial sphere to the corresponding directions at the entrance of the Galaxy, itself obtained by back-propagating negatively charged nuclei from the Earth outwards through the GMF, as explained below.

\begin{figure*}[ht!]
   \centering
  
   \includegraphics[width=8cm]{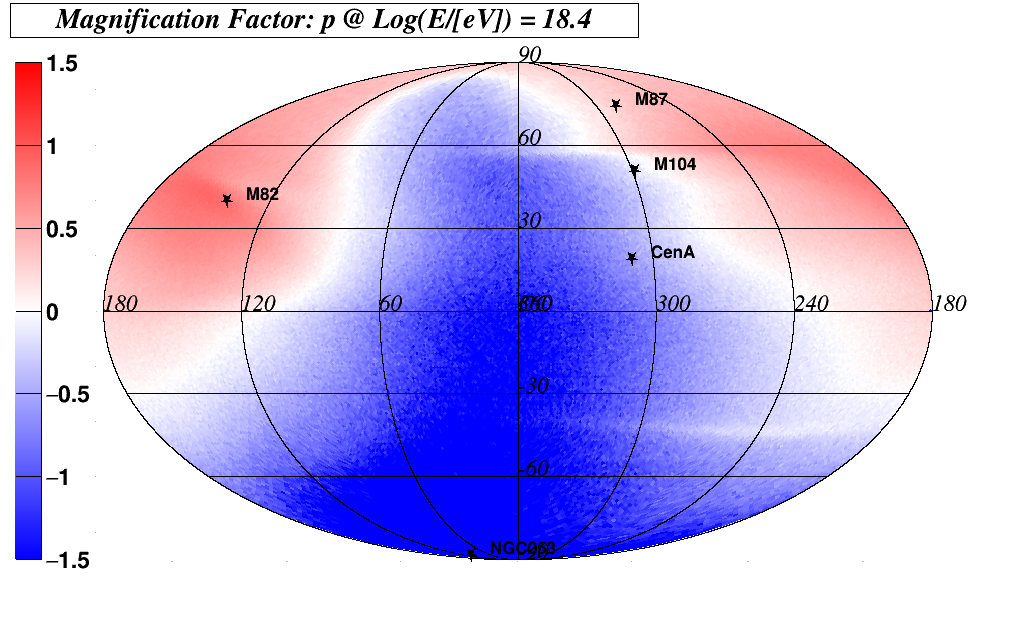}
    \includegraphics[width=8cm]{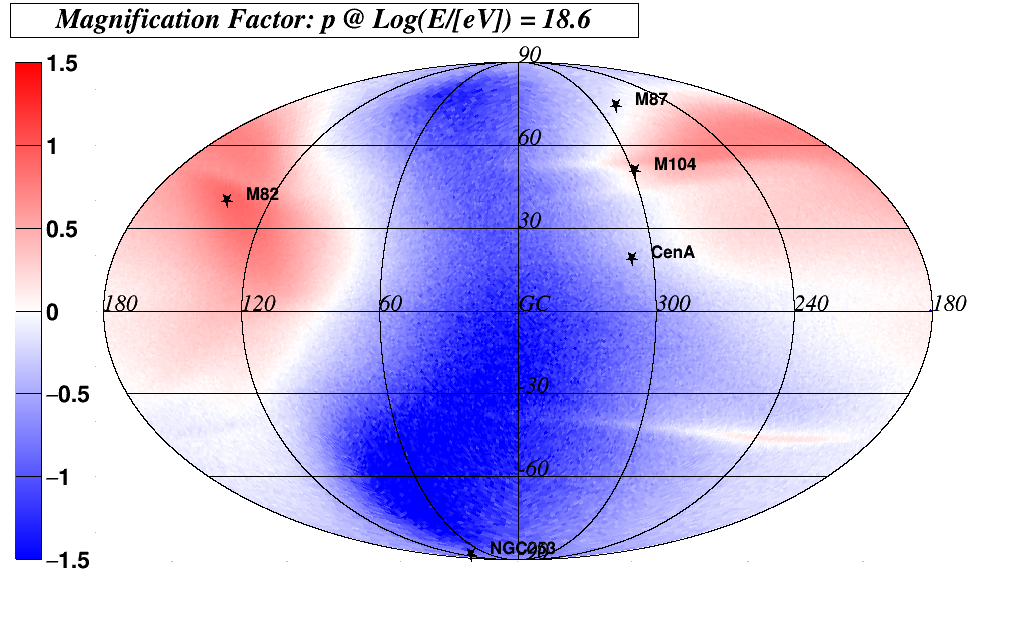}
    \includegraphics[width=8cm]{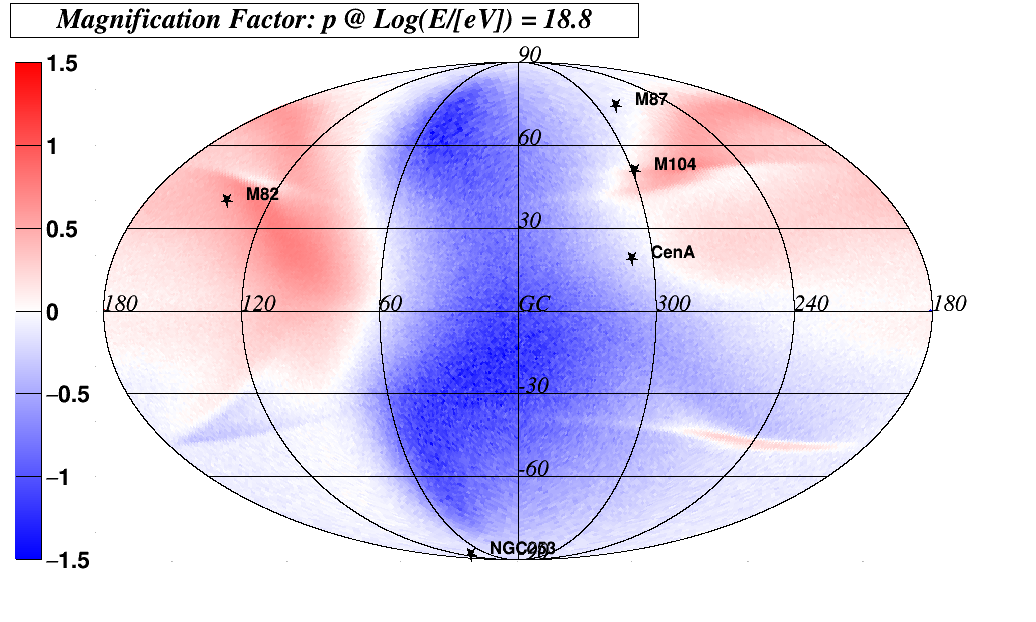}
     \includegraphics[width=8cm]{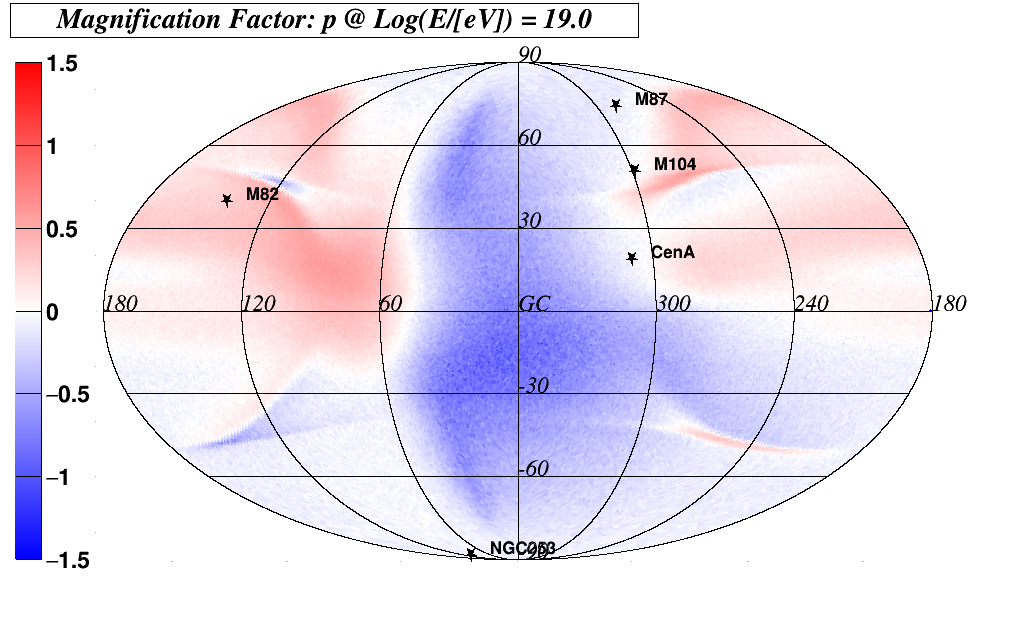}
      \caption{Magnification maps in Galactic coordinates for rigidities from $10^{18.4} \rm\, to\, 10^{19}$~V assuming the "JF12+Planck" GMF model with a coherence length of the turbulent component $\lambda_{\mathrm{c}}=200$~pc. The color scale indicates the logarithm of the magnification factor (in base 10).
              }
         \label{FigMagFac1}
   \end{figure*}

\begin{figure}
   \centering
   \includegraphics[width=\hsize]{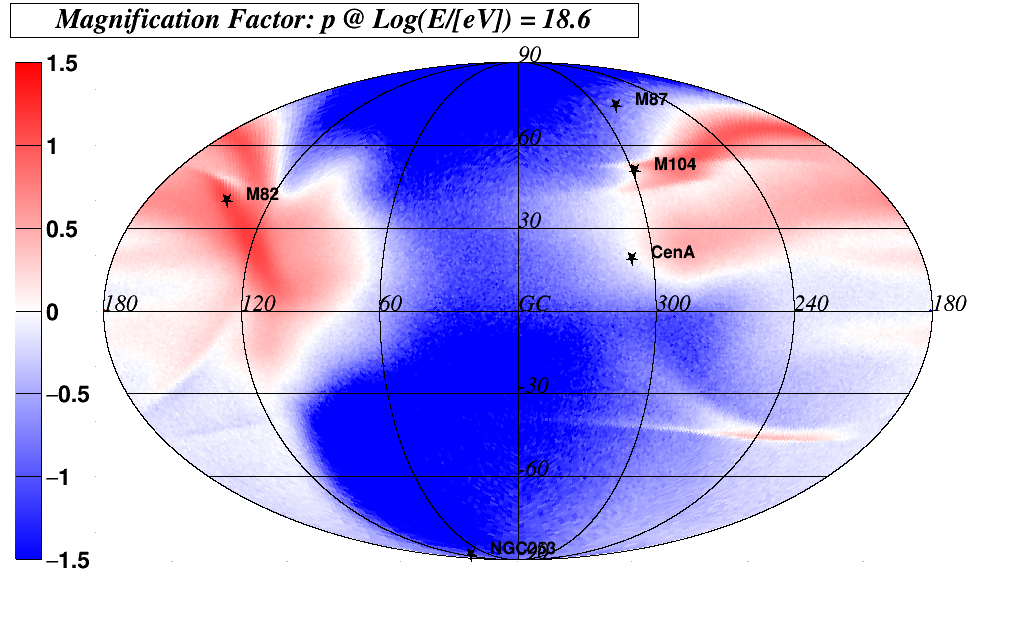}
   \includegraphics[width=\hsize]{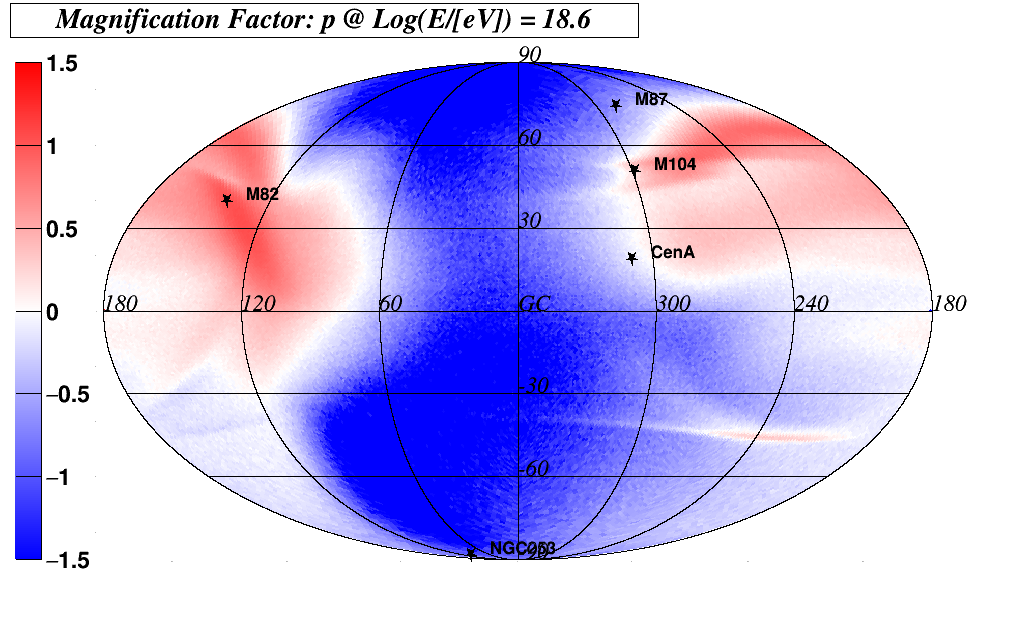}
    \includegraphics[width=\hsize]{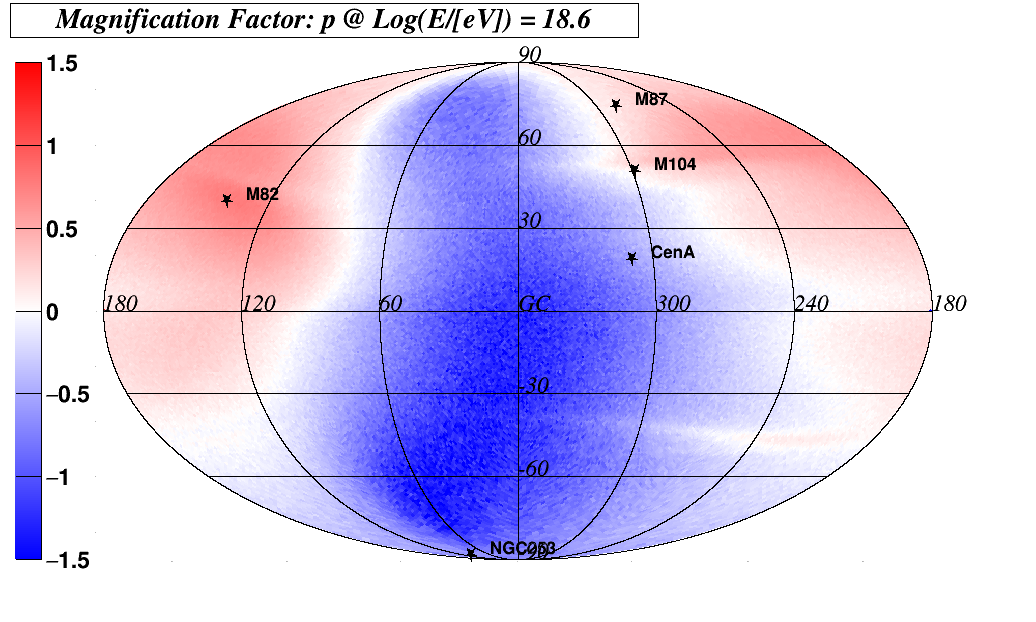}
    \includegraphics[width=\hsize]{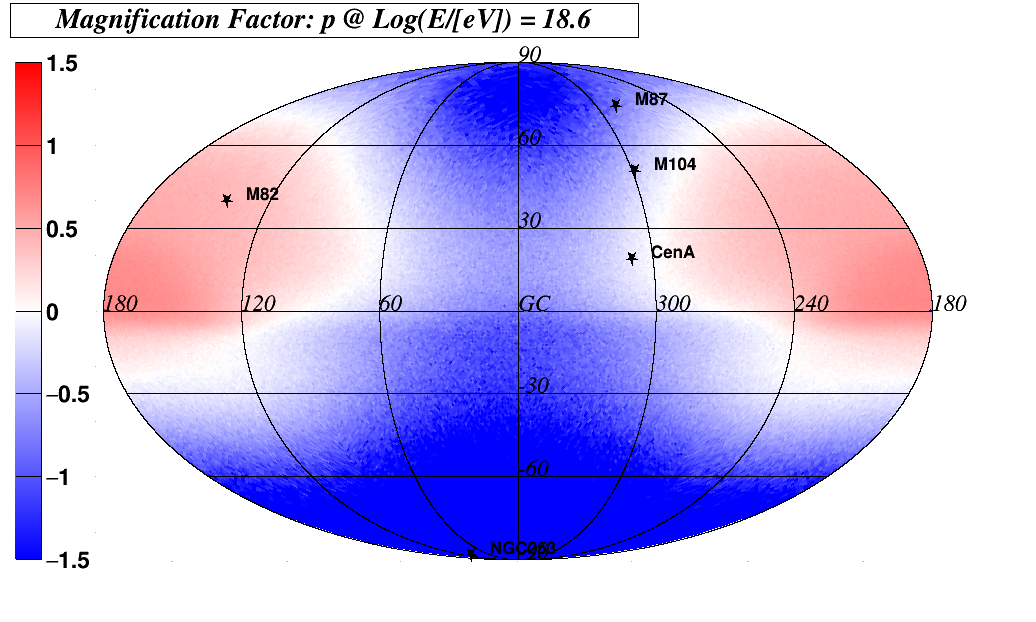}
      \caption{Same as Fig.~\ref{FigMagFac1} for particles of rigidity $R = 10^{18.6}$~V, for different GMF models and coherence lengths of the turbulent component. From top to bottom : "JF12+Planck" with $\lambda_{\mathrm{c}}=20$~pc, "JF12+Planck" with $\lambda_{\mathrm{c}}=50$~pc, "JF12+Planck" with $\lambda_{\mathrm{c}}=500$~pc, "Sun+Planck" with $\lambda_{\mathrm{c}}=200$~pc.
              }
         \label{FigMagFac2}
\end{figure}

\begin{figure}
   \centering
   \includegraphics[width=\hsize]{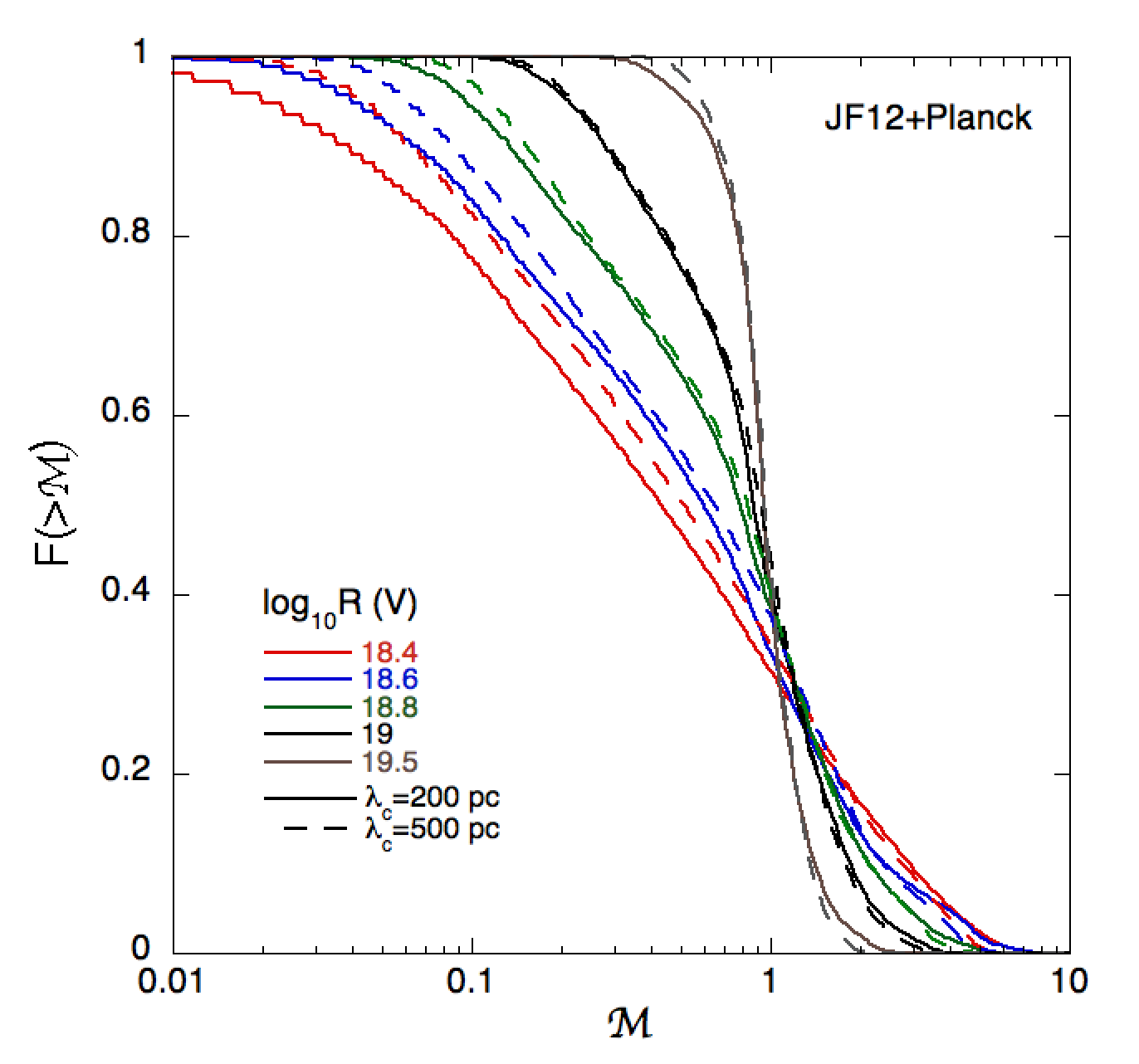}
      \caption{Fraction of the sky $F(>\mathcal{M})$ filled with magnification factors larger than $\mathcal{M}$ as a function of $\mathcal{M}$. Various rigidities are considered (see legend), the GMF model studied is "JF12+Planck" and the considered coherence lengths are $\lambda_{\mathrm{c}}=200$~pc (lines) and $\lambda_{\mathrm{c}}=500$~pc (dashed lines)
              }
         \label{FigMagFac3}
\end{figure}
\subsection{Angular transfer function of the GMF}

To connect the arrival direction of the particles into the Galaxy (as resulting from the extragalactic propagation) to the direction in which they are eventually observed on Earth, we apply the following procedure (see \cite{BRDO2014} for more details).

First, we “back-propagate” a very large number of protons away from the Earth, until they reach a sphere of radius 50 kpc centered on the Galactic center, beyond which the influence of the GMF is negligible (according to the GMF models described in Sect.~\ref{sec:magneticFieldModels}). More specifically, we propagate anti-protons with fixed energies between $10^{17}$~eV and $10^{20}$~eV, by steps of $\Delta \log(E) = 0.1$, starting from the Earth in different directions. For each energy, the starting directions are regularly distributed over the celestial sphere using an HEALPix grid \citep{Gors2005} with resolution parameter Nside = 1024, which corresponds to 12,582,912 directions, or a pixel size of $\sim$3.5 arcmin. The spatial transport of the particles is then computed by simply integrating the equation of the trajectory governed by the Lorentz force (see {\it e.g}, \citet{Stanev1997, Harari1999} for early discussions). The back-propagation gives us, for each rigidity, a one-to-one relation between the $\sim$12.6 million starting directions on the celestial sphere and the direction in which the corresponding (anti-)UHECR leaves the Galaxy, which we may call “conjugate directions”.

We then use a coarser sampling of the celestial sphere, choosing a HEALPix resolution parameter Nside = 64, which defines 49,152 pixels of slightly less than 1~deg$^2$ equally distributed over the sky, each of which contains 256 of the original directions on the thinner grid. Thus each direction on the sky (with a resolution of $\sim$1 deg) is now linked with 256 directions at the boundary of the Galaxy, which are in effect the arrival directions of UHECRs with the rigidity under consideration that would be observed from Earth in the particular direction of the broader “sky pixel” (for a given GMF model).

Finally, the angular transfer function that is needed to build a simulated sky map is obtained by inverting the relation between the conjugate pixels, i.e. by ascribing to each arrival direction on the Galactic border (i.e. each UHECR entering the Galaxy) a given direction on the celestial sphere, where the UHECR will be recorded in the dataset. To this end, we simply keep track of all the “Earth pixels” (on the finest angular grid) that conjugate with a given “Galactic pixel” (coarsest angular grid), and randomly draw one of them (with equal probability) each time that Galactic pixel is hit by a UHECR obtained from the first (extragalactic) step of the transport calculation.

\subsection{Magnification factors}
\label{sec:magFactor}

On average, of course, each Galactic pixel is associated with 256 conjugate Earth pixels. However, the specific pattern of deflections induced by the chosen GMF model results in unevenly distributed relations between Earth pixels and Galactic pixels. Some Galactic pixels turn out to be associated with fewer directions on Earth, while others have much more conjugate directions. This leads to the concept of magnification factors.

For a given flux of incoming UHECRs in some direction at the entrance of the Galaxy, the resulting flux actually observed on Earth will be larger if that direction has a large number of conjugate pixels on Earth than if it has a small number of conjugate pixels. As consequence, the flux of UHECRs from that (former) direction will be \emph{magnified} compared to the latter direction. On the other hand, for the very same reason, the corresponding UHECRs will be distributed over a broader region of the sky, when observed from Earth.

%The results of the back-propagation of charged particle cannot be exploited to build simulated sky maps until an inversion is done to relate the arrival directions of cosmic rays at the entrance of the Galaxy (from their particular extragalactic sources) to the actual directions in which they are observed on Earth. We may call a “Galactic pixel” a pixel in the sky map where a back-propagated particle exit the Galaxy, or where a forward-propagated particle enters the Galaxy. Likewise, an “Earth pixel” is a pixel in the sky map where a back-propagated particle starts its trajectory or where a forward-propagated particle is eventually observed.
%The inversion is done by simply keeping track, for each UHECR incoming direction (i.e., for each Galactic pixel), of the different Earth pixels in the direction where back-propagated particles were initially sent to exit the Galaxy in that pixel. The number of Earth pixels related to a given Galactic pixel is not known a priori. On average, 256 pixels on the fine grid are associated with the Galactic pixels on the coarse grid. This is also, of course, the number of Earth pixels that would be associated with each Galactic pixel if there were no deflection at all. However, some directions turn out to be more likely fed by back-propagated particles than others, because the GMF can focus back-propagated particles from a wide range of directions into a smaller solid angle, or conversely. 

This is illustrated in Figs.~\ref{FigMagFac1} and \ref{FigMagFac2}, which show the magnification factors for each Galactic pixel, defined as the number of Earth pixels associated with that pixel divided by 256. For instance, a magnification factor of 2 (that is $\sim 0.3$ in the log-scale displayed) indicates that a source in that direction will contribute twice as much flux to the UHECRs observed on Earth as if there were no deflection (or as the average of what could be expected if it were anywhere in the sky). The cases displayed show the magnification maps for UHECR rigidities in a range of particular interest for the present study (i.e. corresponding to energies between $\sim 2.5$ and 10~EeV), for the two above-mentioned GMF models and various values of the coherence length, $\lambda_{\mathrm{c}}$, of the turbulent component. 
   
Quantitatively, one can see that for some source locations the magnification factor can reach high values and that these values can change significantly between two neighboring regions in the sky, all the more for the lowest coherence lengths displayed here. The direction-to-direction contrast in magnification is especially large for rigidities in the range between $10^{17}$ and  $10^{19}$~V, where the cosmic-rays are not anymore completely isotropised by the GMF (at least for the GMF models we consider) but still suffer very significant deflections. This phenomenon corresponds to the caustic curves studied in detail in \citet{Harari1999}. One of its most interesting consequences is that some sources can be partially or completely hidden to us by the GMF, while others can be magnified. In addition, since the magnification of a source in a given direction depends on the rigidity, the magnification factor can be very different for different species at a given energy, or for a given species at different energies. In particular, a given source location can result in a strong magnification (or demagnification) of the heavy component in a specific energy range, while the proton component may be affected differently. One should thus expect noticeable modifications of the composition and spectrum of individual sources or of distinct regions of the sky, in a way that however depends critically on the actual structure of the GMF, and thus, in the case of the simulations, on the specific assumptions regarding the GMF. In addition to the magnification factors, the inverted (i.e. forward) relation between Galactic pixels and Earth pixels allows us to determine where the UHECRs entering the Galaxy in a given direction will be observed on Earth. These are key elements of the datasets and skymaps generation procedure presented in the next section. 

It is of course interesting, in principle, to examine which regions are specifically magnified or demagnified, especially when it comes to examine the contribution of specific sources to the observed UHECR flux, or when trying to interpret possible localised excesses in the sky, for instance in relation with specific source catalogs (which will be addressed in the next paper). Here, we simply note that the explored magnetic field models tend to demagnify a large part of the Southern Galactic hemisphere, for all coherence lengths of the turbulent GMF and the particle rigidities (in the range under consideration). This is the case for both the JF12+Planck and the Sun+Planck GMF models.

Interesting sources such as CenA, M87 and M104 appear to be located near the region where a transition between magnification and demagnification occurs (thus with magnifications relatively close to one, on average: see Fig.~\ref{FigMagFac2}). As can also be seen on the figures, which of these sources are actually magnified or demagnified does depend on the value of $\lambda_{\mathrm{c}}$ as well as on the rigidity.  This implies, on the one hand, that precise predictions are be difficult to make without a good knowledge of the intervening magnetic fields, even if the sources were known, and on the other hand that the apparent flux and composition of a given source can be modulated in different ways in different energy bins, thereby modifying the resulting spectrum and composition.

Conversely, a source such as M82 can be seen to always lie in a magnified region of the sky (whatever the coherence length and rigidity), even though the value of the magnification factor is of course not always the same. In parallel with its large magnification, and for that very reason, a source located in that direction would distribute its UHECRs over a large region of the sky, as seen from Earth, since many pixels of the observer's sky are then associated with roughly the same direction at the entrance of the Galaxy. In the case of M82, for instance, even though the source itself is located outside the field of view of Auger, it turns out to provide an appreciable contribution to our datasets simulated with the Auger sky coverage, whenever this source is present in the source catalog.

Finally, it is interesting to examine the distribution of magnification factors, independently of the specific regions of the sky with which they are associated. In Fig.~\ref{FigMagFac3}, we show the cumulative distribution function of the magnification factors, $\mathcal{M}$, for different rigidities and different GMF coherence lengths. In all cases, the fraction of the sky that is magnified represents roughly between 30\% and 40\% of the entire sky, and is thus smaller than the fraction of the sky that is demagnified. A larger spread in the magnification factors is obtained, as expected, for lower rigidities, and for lower values of $\lambda_{\mathrm{c}}$ (although the latter effect is hardly visible for rigdities $R\ga 10^{19}$~eV). Interestingly, one sees that about 20\% of the sky is demagnified by at least a factor of 10 at a rigidity $R = 10^{18.4}$~V (i.e. $\simeq 2.5$~EV), and a by at least a factor of 4 at $R = 10^{19}$~eV. This can have major effects, depending on the source distribution, since it essentially switches off or at least strongly reduces the contribution of a significant part of the sky (and thus of potential sources).

\section{Dataset generation and analysis}
\label{sec:data}
\subsection{Generation of the simulated datasets}
\label{sec:datasets}

To generate a simulated UHECR dataset, one first needs to select the parameters of the underlying astrophysical scenario, namely:

\begin{itemize}
    \item the source spectrum and composition, to be chosen among models A, B, C or D in Tab.~\ref{table:1};
    \item the distribution of sources, which first implies choosing a source density, and then either drawing randomly out of the mother catalog one particular realisation of the set of sources with that density, or using the corresponding volume-limited catalog built from the 2MRS catalog by applying a cut in intrinsic ($K_\mathrm{s}$-band) luminosity (see above and Tab.~\ref{table:2}). As an alternative, we may also choose a source distribution drawn from an underlying uniform distribution or from one the biased catalogs mentioned in Sect.~\ref{sec:sourceDistrib};
    \item the strength of the EGMF (see Sect.~\ref{sec:EGMF}).
    \item the GMF model, among those described in Sect.~\ref{sec:GMF}.
\end{itemize}

Once these astrophysical hypotheses are made, we calculate the spectrum and angular distribution of the particles {\it entering the galaxy}, using an updated (see above) version of the Monte-Carlo treatment introduced in \citet{GAP2008}: we calculate propagated spectra and angular distributions of the events from sources at various distances\footnote{More precisely, by distance we refer to the  proper comoving  distance hereafter.} between 1 and 4000~Mpc. More precisely, distance bins with a half width of $\sim 2\%$ of their central value are implemented in this range.
For each given distance bin, say at distance $D$, the spectrum, composition and angular distribution of the UHECRs contributed by a source is obtained by integrating, over the whole range of (past) emission times, the UHECR flux coming from that source as it appears at the present epoch at a distance $D$ from its position. Then the total UHECR spectrum, composition and angular distribution are simply obtained by summing the contributions of all sources in the considered realisation of the source distribution, each at its respective distance from our Galaxy.

This numerical scheme allows us to build the following probability distributions (corresponding to the UHECRs at the entrance of the Galaxy):
\begin{itemize}
\item the energy distribution of the extragalactic UHECR,
\item their mass distribution at a given energy,
\item the effective source distance distribution at a given energy for a given species,
\item the relative contribution of the sources in a given distance bin to the UHECRs observed with a given energy and mass,
\item the angular distribution of particles coming from a source at a given distance and observed with a given mass at a given energy. 
\end{itemize}

If the GMF could be ignored and the detector were assumed to have uniform exposure over the entire sky, we could then easily produce simulated datasets of any size, using a simple Monte-Carlo method to draw UHECR events one by one according to the above distributions, until we reach the intended statistics. Now, to take into account the GMF, we proceed in two steps. First, for each event drawn in the previous way, we apply an acceptance/rejection method to either keep or reject the event according to the magnification factor of its particular arrival direction at the entrance of the Galaxy (given the particle rigidity), after normalising all magnification maps to the highest magnification factor over the entire range of rigidities. Second, if the event is “accepted” (i.e. the random number drawn between 0 and 1 is lower than the corresponding normalized magnification), its final arrival direction on Earth is determined by drawing randomly among the “Earth pixels” associated with the incoming direction at the entrance of the Galaxy, at the rigidity under consideration.

Finally, to take into account the specific exposure map of the simulated detector, we again apply a simply acceptance/rejection method to each event drawn (and accepted) as indicated above. In this paper, we investigate three different exposure maps, corresponding to either Auger, TA, or a uniform full-sky exposure, which we consider as an approximation of a potential space-borne observatory such as aimed at by the JEM-EUSO program (see e.g. \citet{JEMEUSO2013, JEMEUSO2015} and more recently \citet{Bertaina2019, POEMMA2021}). The case of a uniform exposure is also interesting to study potential limitations or biases introduced by partial-sky coverage observatories.

As a last step of our dataset building procedure, we take into account the fact that the detector's energy resolution is not perfect, and draw randomly the final energy of each individual event assuming a gaussian reconstruction error with a value of $\sigma$ of $15\%$ for the Auger-like datasets, $20\%$ for the TA-like datasets, and $15\%$ for the full-sky datasets. The angular reconstruction accuracy is also not perfect. However, since we study anisotropies on angular scales much larger than the angular resolution of the considered observatories, we do not apply any modification of the final arrival direction obtained from the above procedure.

%\subsection{Statistics and characteristics of datasets}
 
%The statistics we will assume in this study depends upon the choice of the sky exposition made that is whether the dataset is build assuming the Auger, the TA or a full-sky coverage. The later could be, in particular relevant for a space-borne observatory as considered within the JEM-EUSO program, see \citet{JEMEUSO2013} and more recently \citet{Bertaina2019} and  is in any case interesting to study the potential limitations or biases introduced by the use of partial exposure observatories.
 
In each case, we investigate realistic datasets by adapting the event statistics to the observatory under consideration. In the case of Auger, we choose the statistics collected at the time when the dipole modulation with a significance larger than $5\sigma$ was reported \citet{AugerDip2017}, which corresponds to 32,187 events recorded above 8 EeV, and a total exposure of 76,800 $\rm km^2\,sr\,yr$ for UHECR showers with zenith angle lower $80^\circ$. In the case of TA-like datasets, we simulate UHECR skies with the statistics reported in the joint Auger/TA anisotropy analyses \citep{AugerTAan2020} and take into account the $\sim 10\%$ systematic energy scale difference between Auger and TA reported in \citet{AugerTAspe2019}, choosing arbitrarily to scale down the TA energies. As a result, we simulate datasets with statistics of 4,321 events with zenith angle lower than $55^\circ$ above 9~EeV. For the full-sky datasets, we consider in most cases datasets with statistics twice as large as that of Auger above 8~EeV.

Note that our datasets are built to match these statistics for the (simulated) “reconstructed energy” rather than the true energy, and that we also keep the events reconstructed with an energy down to 4~EeV, to study the anisotropy and its energy evolution in the same energy range as Auger \citep{AugerDipLong2018, AugerICRC2019}.
 
For each set of astrophysical hypotheses (see above), we  build 300 different datasets based on 300 different realizations of the assumed source catalog. This allows us to explore both the cosmic variance, i.e. to what extent the characteristics of the observed UHECR sky depend on the specific realisation of a given astrophysics scenario rather than on the scenario itself, and the statistical variance, i.e. to what extent the exact same scenario and source distribution would give rise to similar characteristics if we would collect another dataset with the same statistics. This is where the use of either the mother catalog or the volume-limited catalogs come into play. In the former case, each new realisation (and corresponding dataset) corresponds to a new set of sources, located at their own specific location in the universe around us. In the latter case, on the contrary, the sources up to the distance $D_{\max}$ used to define the volume-limited catalog under consideration (see Sect.~\ref{sec:sourceDistrib}) remain the same for each realisation, and the cosmic variance is limited to the random choice of the missing sources completing the catalog beyond $D_{\max}$, with negligible influence on the main characteristics of the datasets compared to the statistical variance (as shown below). This is true for the spectrum as well as the composition and anisotropy patterns.

\subsection{Principle of the analyses}
 
For each simulated dataset, we proceed with a first level of analyses, extracting the resulting all-particle spectrum and the energy evolution of the composition. This allows us to verify that the underlying astrophysical models are potential good candidates to account for the observed UHECRs, as far as composition and spectrum are concerned. We also extract from the datasets the average particle rigidity, the fractional contribution of the different source distances to the observed flux above a given energy, as well as the contribution of the brightest sources. These allow to better understand, in physical terms, the origin and characteristics of the resulting anisotropy patterns. Likewise, we can isolate the contribution of individual sources to the recorded UHECRs, and examine their effective spectrum and their images on the simulated sky map.

On a second level of analyses, we perform specific anisotropy studies, to investigate whether the underlying astrophysical models can produce datasets exhibiting similar features as the actual UHECR datasets collected by existing observatories. In the present paper we concentrate on the search for dipole and quadrupole modulations, as well as on the global angular power spectrum of the datasets. We thus reproduce on our simulated datasets the very same analyses that have been recently reported by Auger and/or TA: i) we follow the so-called Rayleigh analysis in right ascension and azimuth, as presented in \cite{AugerDip2017}, and ii) we compute the angular power spectrum (both for full and partial sky coverage), as presented in \cite{AugerMulti2017}. We refer the reader to these two papers for the principles and details of these analyses. In addition, as a consistency check of the study of the dipole modulation, we also consider the 3D reconstruction method presented in \cite{Aublin2005}, which turns out to give quantitative results very close to those obtained with the Rayleigh analysis.

Note that we first checked the accuracy and reliability of our analysis tools by testing them on idealized cases, such as finite simulated datasets drawn from a perfect dipole distribution with the same amplitude and direction as those reported by Auger above 8~EeV. For all the tests, we reconstructed the right value of the input parameters and direction (with the expected statistical dispersion, but no systematic bias), for both partial and full sky coverages, and for the Rayleigh as well as the angular power spectrum analyses.

\begin{figure*}
   \centering
   \includegraphics[width=8.5cm]{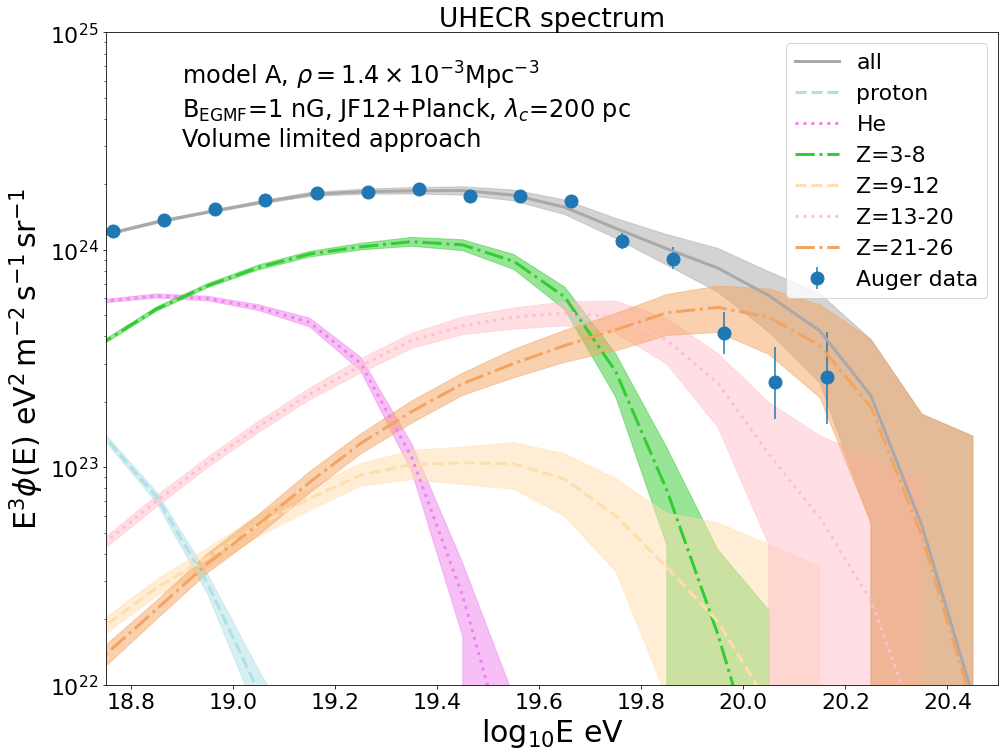}
   \includegraphics[width=8.5cm]{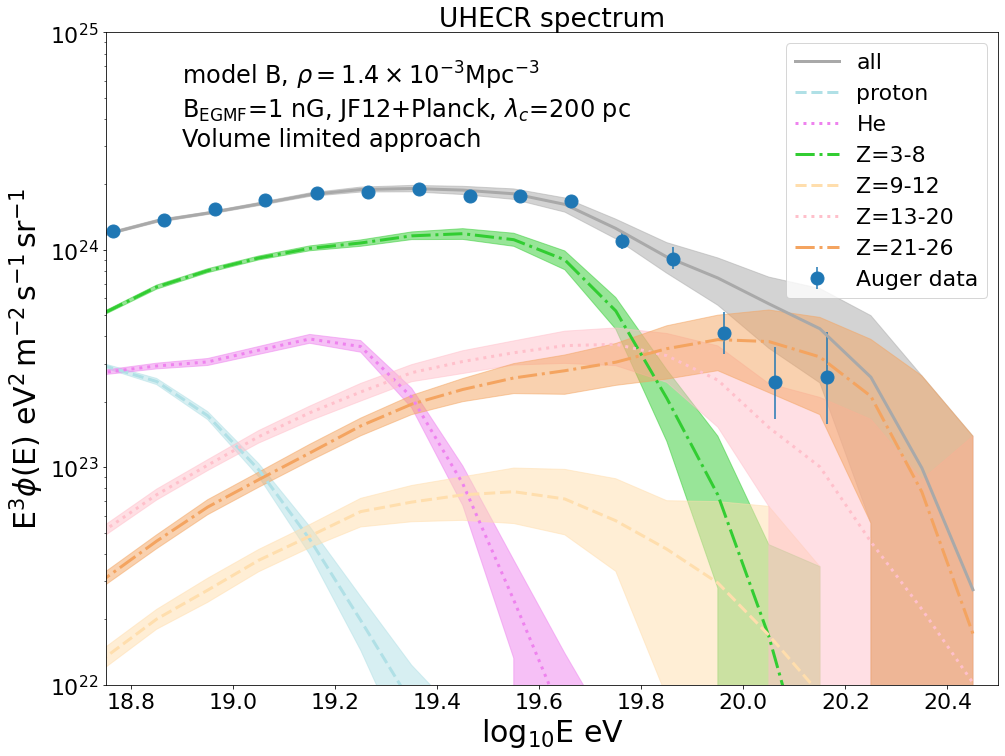}
   \includegraphics[width=8.5cm]{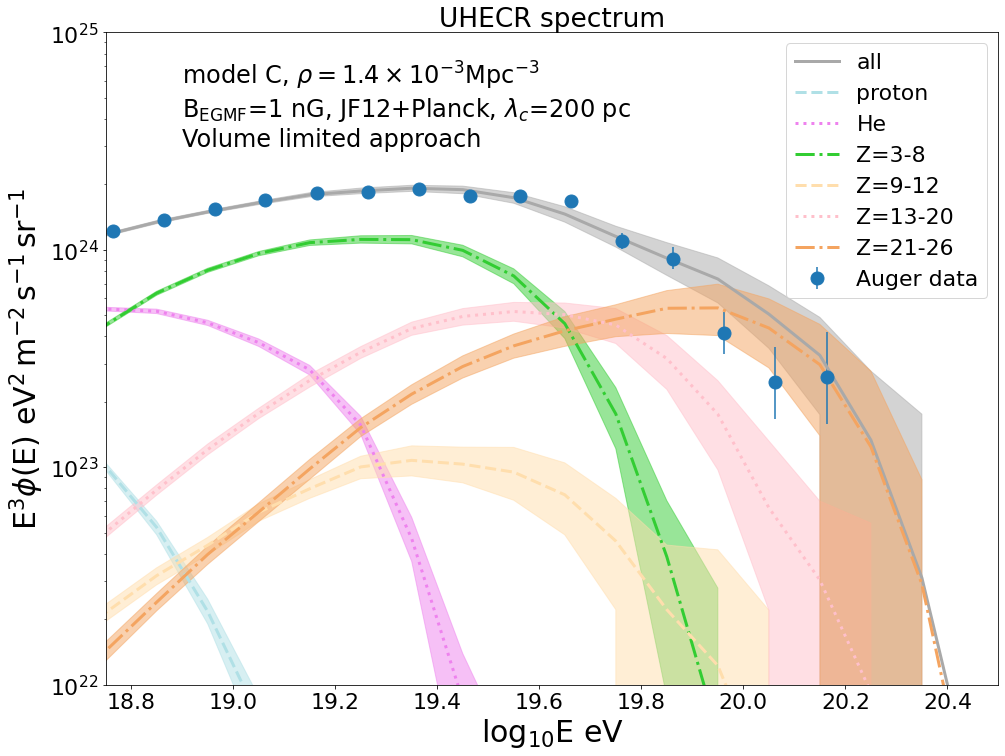}
   \includegraphics[width=8.5cm]{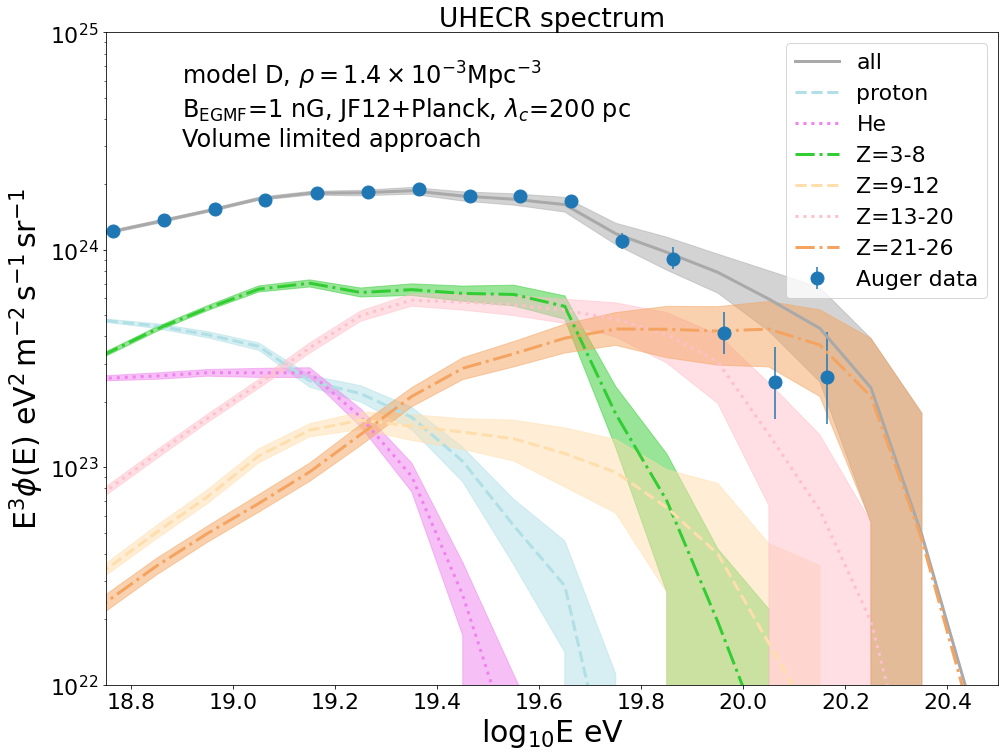}
      \caption{UHECR energy spectra obtained after propagation in the extragalactic and Galactic media, for 4 different models of source spectrum and composition, namely models A, B, C and D, from left to right and top to bottom. The assumed source distribution is that of the volume-limited catalog extracted from the 2MRS with a density $1.4\times10^{-3}\,\rm Mpc^{-3}$, which is complete up to 104~Mpc (cf. Table~\ref{table:1}) and completed beyond that distance as explained in the text. An EGMF of 1~nG and the "JF12+Planck" GMF model with $\lambda_{\mathrm{c}}=200$~pc are assumed (see text). The simulated sky coverage is that of the Pierre Auger Observatory. The lines show the mean expected fluxes, averaged over 300 realizations, and the shaded areas cover the range in which $90\%$ of the realizations are found (energy bin by energy bin). The different colors show different groups of nuclear species, as indicated on each plot.
              }
         \label{FigSpec}
   \end{figure*}

\begin{figure}
   \centering
   \includegraphics[width=8.5cm]{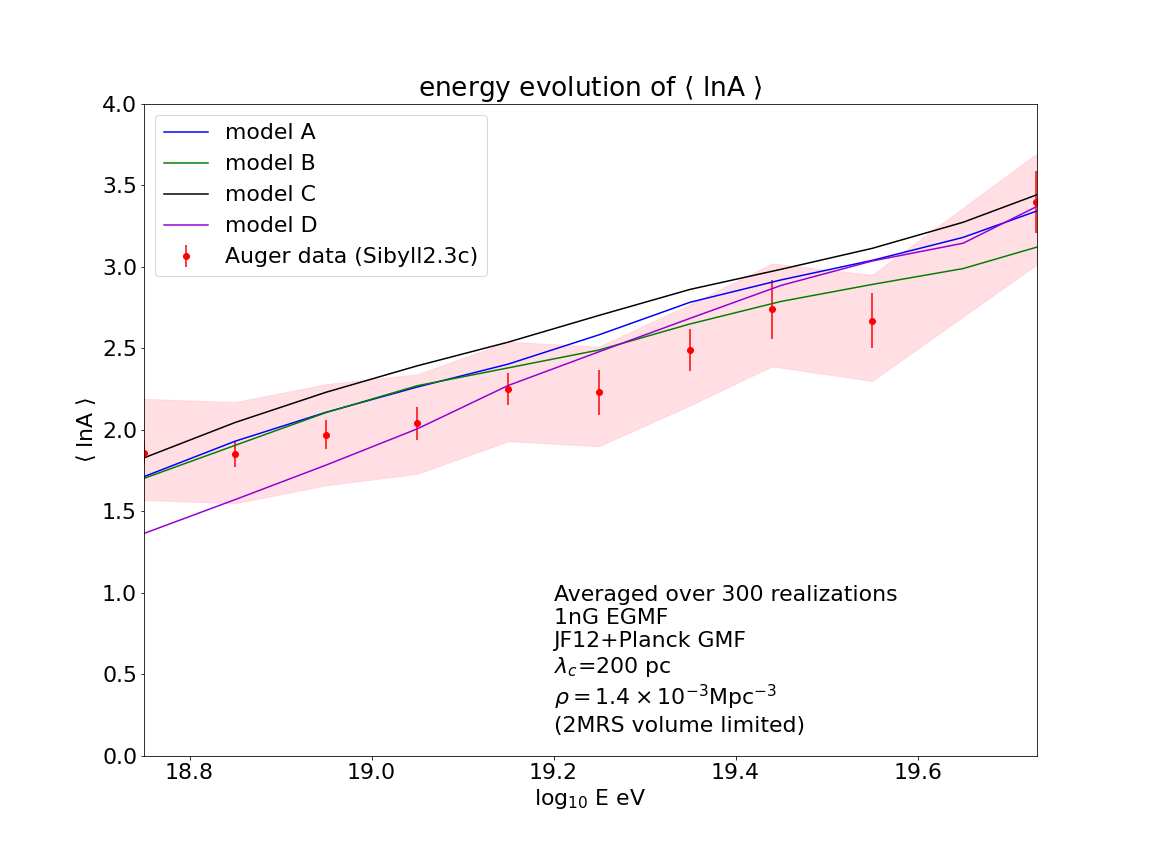}
   \includegraphics[width=8.5cm]{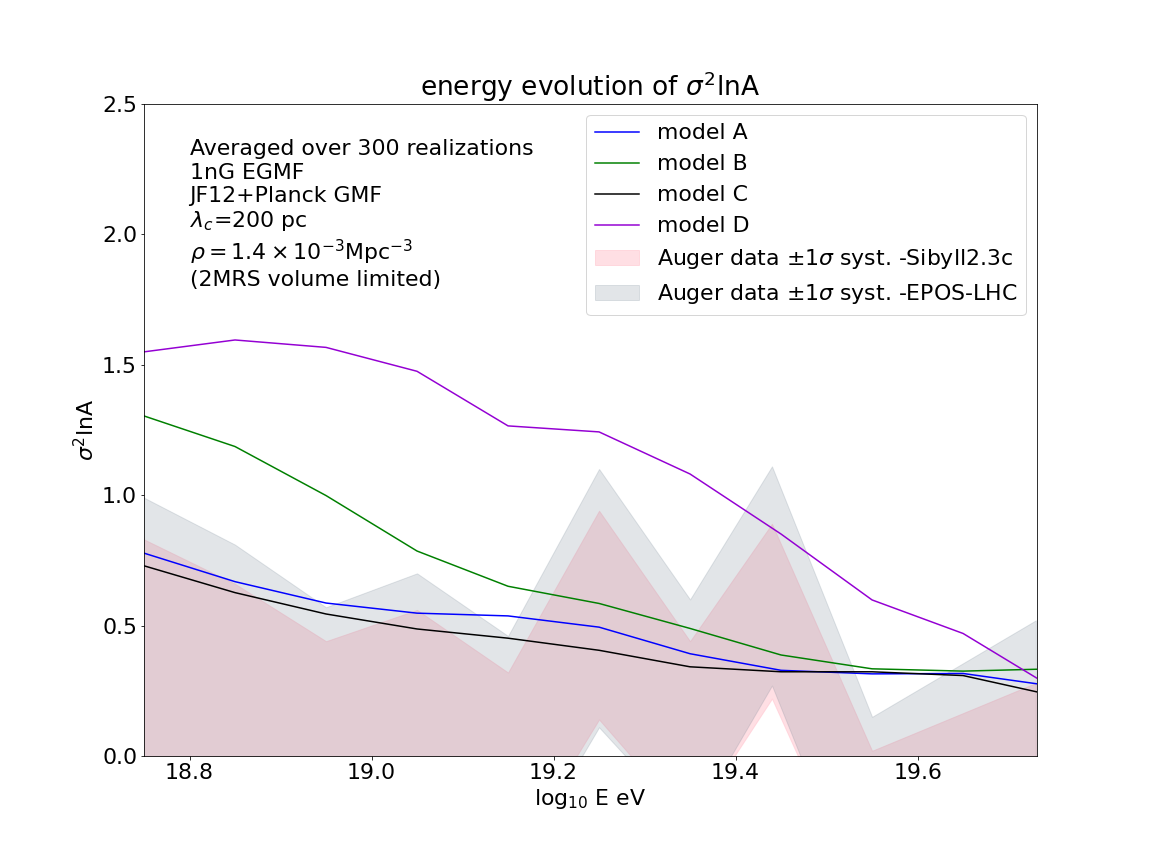}
      \caption{Top panel : Energy evolution  of $\langle \ln A \rangle$ (top panel) and  $\sigma^2 \ln A $,averaged over 300 realizations, for the four models displayed in Fig.~\ref{FigSpec}, compared to that estimated at the Pierre Auger Observatory assuming the Sibyll2.3c or EPOS-LHC hadronic models. 
              }
         \label{FigLnA}
   \end{figure}
   
\begin{figure}
   \centering
   \includegraphics[width=7.5cm]{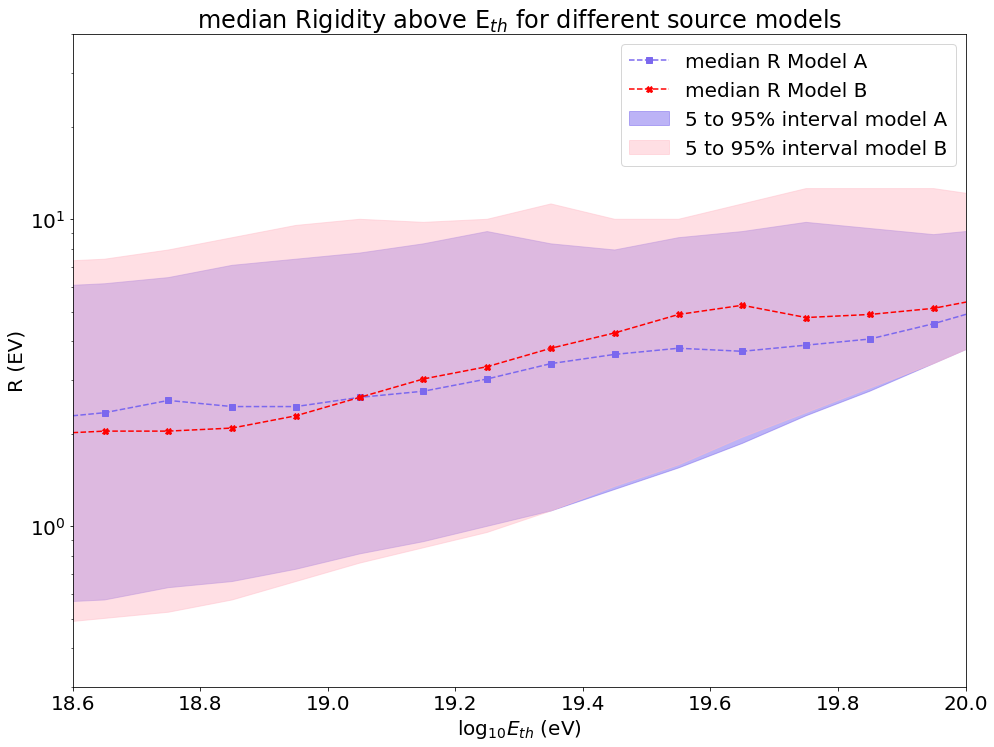}
   \includegraphics[width=7.5cm]{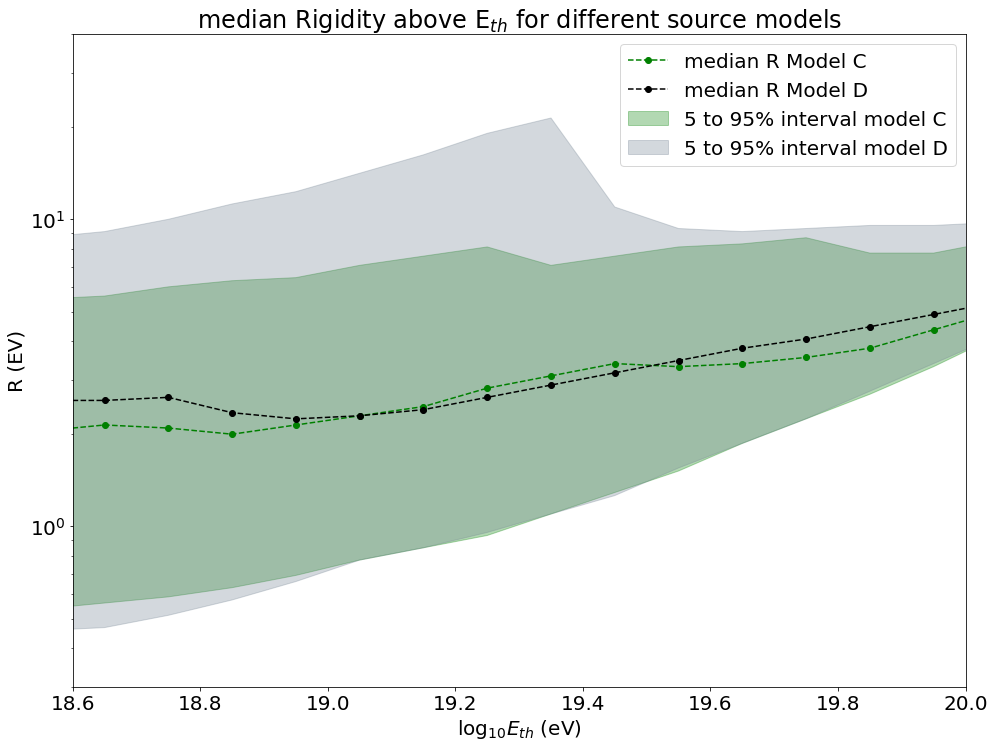}
      \caption{ Energy evolution of the rigidity for four models A and B (top panel), C and D (bottom-panel), averaged over 300 realizations, the lines show the evolution of the median rigidity and the shaded area the evolution of $90\%$ range in which particles with energy above $E_{th}$ are found. 
              }
         \label{FigRigid}
   \end{figure}

\section{General content of the simulated datasets}
\label{sec:content}
\subsection{Spectrum and composition of the simulated datasets}

As mentioned above, before turning to the discussion of anisotropies, we need to verify that the investigated astrophysical models, with their specific choice of (effective) source spectrum and composition, provide a fair account of the main observed spectral and composition features. To this end, we compared our simulated datasets with the Auger dataset, taken as a reference because of its larger statistics.

Figure~\ref{FigSpec} shows a direct comparison of the spectra obtained with many realisations of models A, B, C and D with the Auger surface detector spectrum, presented in \citet{AugerICRC2019}. In these examples, the source distribution is taken from the 2MRS-based volume limited catalog with density $1.4\times 10^{-3} \rm Mpc^{-3}$ (see Sect.~\ref{sec:sourceDistrib}) and the extragalactic magnetic field is assumed to be of 1~nG. The lines show the mean values obtained for the different nuclear components, calculated over 300 realizations, and the shaded area shows the dispersion which is almost exclusively due to the statistical variance of the datasets (see below).

The agreement between the different models and the Auger spectrum appears satisfactory. Yet, while models A, B, C and D lead on average to similar all-particle spectra, they differ by the relative contribution of the different nuclear components. The light component (H+He) of model B is weaker than that of models A and~C below $\sim10^{19}$~eV, but has a larger contribution of protons, extending up to higher energies. For this model the contribution of CNO nuclei to the energy range of interest for the present anisotropy studies is larger, and that of the heavier elements lower. As a result its composition around $\sim5\,10^{19}$~eV is lighter. As for model~D, it has the largest light nuclei abundance, dominated by protons. It has the most mixed composition of the four models and shows the most abrupt change in composition above the ankle. As mentioned above, model~C does not differ much from model~A, and is essentially a slightly heavier version of model A.

These different nuclear contents can be seen in a more synthetic way by examining the energy evolution of the average logarithm of the UHECR mass, $\langle \ln A \rangle$. It is shown for all four models on the top panel of Fig~.\ref{FigLnA}, where we also show for comparison the values deduced from the Auger $X_{\mathrm{max}}$ measurements \citep{AugerLNA2013}, as reported in \citet{AugerICRC2019}, using the Sibyll2.3c \citep{Sib2018} hadronic model. The models appear to reproduce the composition trend suggested by the Auger data, i.e. an increasing value of $\langle \ln A \rangle$ with increasing energy. Admittedly, barring the underlying statistical and systematic uncertainties, the agreement with the shape of the evolution estimated by Auger is not perfect. However, fine tuning the models to match exactly the data would make very little sense, as already discussed in Sect.~\ref{sec:sourceSpectrumAndComposition}, and we also have to keep in mind that the conversion from shower observables to actual values of $\langle \ln A \rangle$ suffers from large uncertainties associated with the hadronic interaction models, which all lead to predictions incompatible with observations, at least with respect to the muon content of the atmospheric showers (see \citet{AugerNewMuon2021} for the most recent account). In view of these uncertainties, we consider the explored models as giving a fair global account of the observed UHECRs, while not limiting the investigation to a single effective composition, somewhat arbitrarily defined. In this way, we can study a wider range of possible compositions, both compatible with the data and likely to cover a broader range of possibilities, representative of currently acceptable underlying astrophysical models.

From this perspective, the different models are interesting as exhibiting different features likely to influence the resulting anisotropies and their evolution with energy. More specifically, models A and~B can be seen to show very similar values of $\langle \ln A \rangle$ below $\sim 10^{19.2}$~eV, and then slightly diverge: model~A is characterised by a stronger evolution of the average mass, towards elements heavier than CNO around $\sim5\,10^{19}$~eV (as could be suggested by the last point of the Auger data, when taken at face value), while model~B shows a flatter evolution, probably more in qualitative agreement with trend suggested by the surface detector composition observables \citep{AugerCompoSD2017, AugerICRC2019}. On the other hand, model~D shows a transition from light to heavy that is significantly more pronounced than that suggested by the Auger data. From the point of view of the mass dispersion $\sigma^2\ln A$ (bottom panel of Fig.~\ref{FigLnA}), models A and C hug the 1$\sigma$ upper bound reconstructed by Auger when using the Sibyll2.3c or the EPOS-LHC \citep{Werner2006, Pierog2013} hadronic models, while model B slightly overshoots it, in particular below $10^{19}$~eV, because of the larger relative abundance of protons which contribute to the increase of $\sigma^2\ln A$. However, this model does not show any strong disagreement with the data from that point of view either, taking into account the above-mentioned uncertainties preventing a precise determination of the observed composition. Being more mixed, especially below $\sim 2\,10^{19}$~eV, model~D shows a more significant disagreement with the data from the point of view of mass dispersion as well, whether interpreted using the Sibyll2.3c or the EPOS-LHC hadronic models.
 Model~A explores a  somewhat different type of transition from light heavy, intermediate between that of model~B and model~D, with a mean composition as heavy as model~B at low energy (where model~D is significantly lighter), and as heavy as model~D at high energy (where model~B is significantly lighter). Besides the lower maximum rigidity (which results in a weaker proton component), the lower mass dispersion of model A (and C) is due to the harder source spectral index, which allows a less mixed composition at all energies even though the composition still passes from a light (here He) dominated to a heavy dominated composition over the energy range of interest for our study.

%{\color{red}[I would suggest to present things a bit differently, mentioning that model~A explores a different type of transition, intermediate between that of model~B and model~D, with a mean composition as heavy as model~B at low energy (where model~D is significantly lighter), and as heavy as model~D at high energy (where model~B is significantly lighter). Model C, on the other hand, is a somewhat heavier version of model A.]}

Figure~\ref{FigRigid} shows the evolution of the rigidity of the UHECRs above a given energy, as a function of that energy. More precisely the shaded are shows the rigidity range in which lie $90\%$ of the particles with an energy reconstructed above a threshold energy $E_{th}$. Note that this dispersion of the rigidity is related to above mentioned $\sigma^2\ln A$), while the lines show the evolution of the median rigidity. A crucial feature of all models, which is directly relevant to the evolution of the UHECR anisotropy with energy, is the very slow energy evolution of the rigidity distribution. This is due to the joint increase of the energy and the mass (thus the charge) of the particles. One also sees that a relatively narrow range of rigidities, say between $\sim 2$ and 6~EV, is common for UHECRs over the entire energy range.

\begin{figure}
   \centering
   \includegraphics[width=8.5cm]{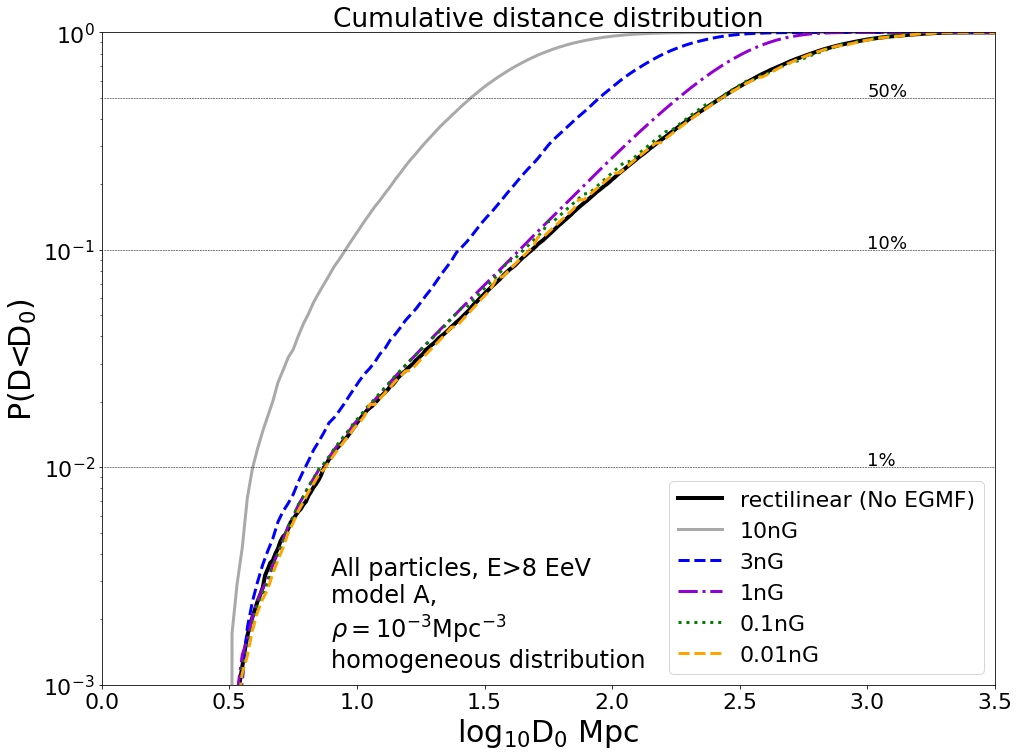}
   \includegraphics[width=8.5cm]{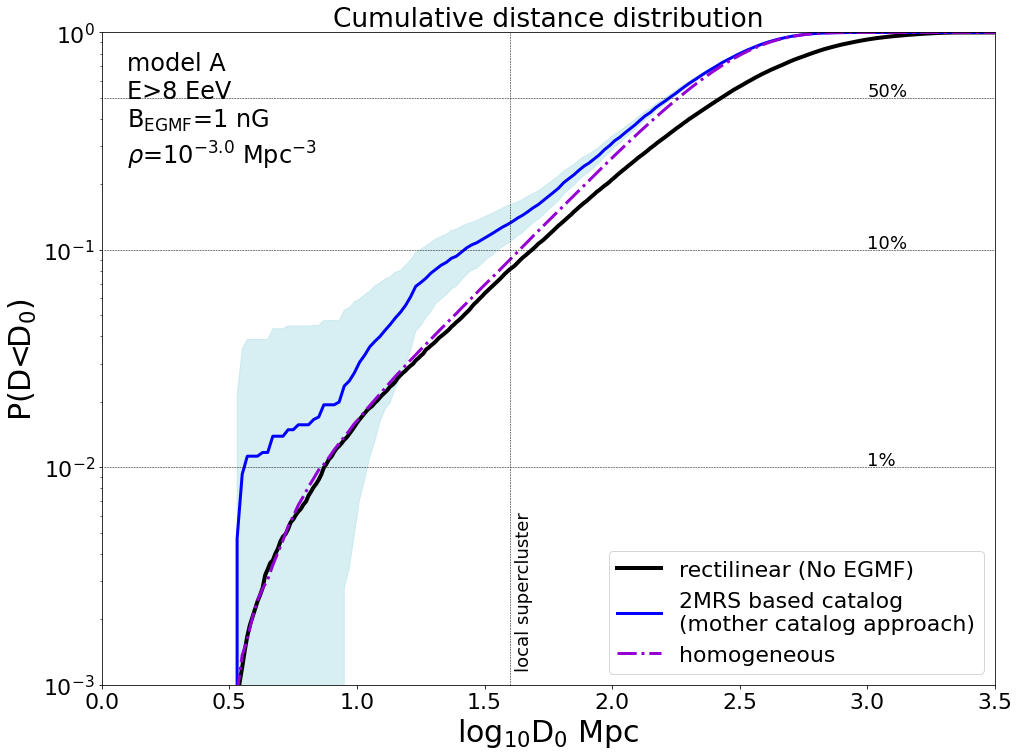}
      \caption{Evolution with the proper comoving 
      distance $D_0$ of the fractional contribution to the UHECR total flux (all species) above 8 EeV of the sources enclosed within that distance,  $P(D<D_0)$, assuming the source model A. Top panel : various hypotheses on the intensity of the EGMF between $10^{-2}$ and 10~nG are displayed. The source catalogs assumed are 300 realizations of an homogeneous and isotropic distribution of discrete sources with a $10^{-3}\,\rm Mpc^{-3}$ source density, the different lines shows the  values obtained after averaging over the 300 realizations. The thick black line shows the case of the rectilinear propagation ({\it i.e}, without extragalactic magnetic fields) assuming a continuous source distribution for the same source model A. In all cases the minimum source distance is set to 3 Mpc.
      Bottom panel : the 1~nG EGMF and rectilinear cases, shown in the top panel, are compared with scenarios assuming 2MRS-based catalogs in the "mother catalog approach" for a $10^{-3}\,\rm Mpc^{-3}$ source density and a 1~nG EGMF. For the latter, the average value (obtained over 300 realizations) is shown together with  the interval where $90\%$ of the simulations are found, represented by the shaded area. 
              }
         \label{FigHorizon}
   \end{figure}

\begin{figure}
   \centering
   \includegraphics[width=7.5cm]{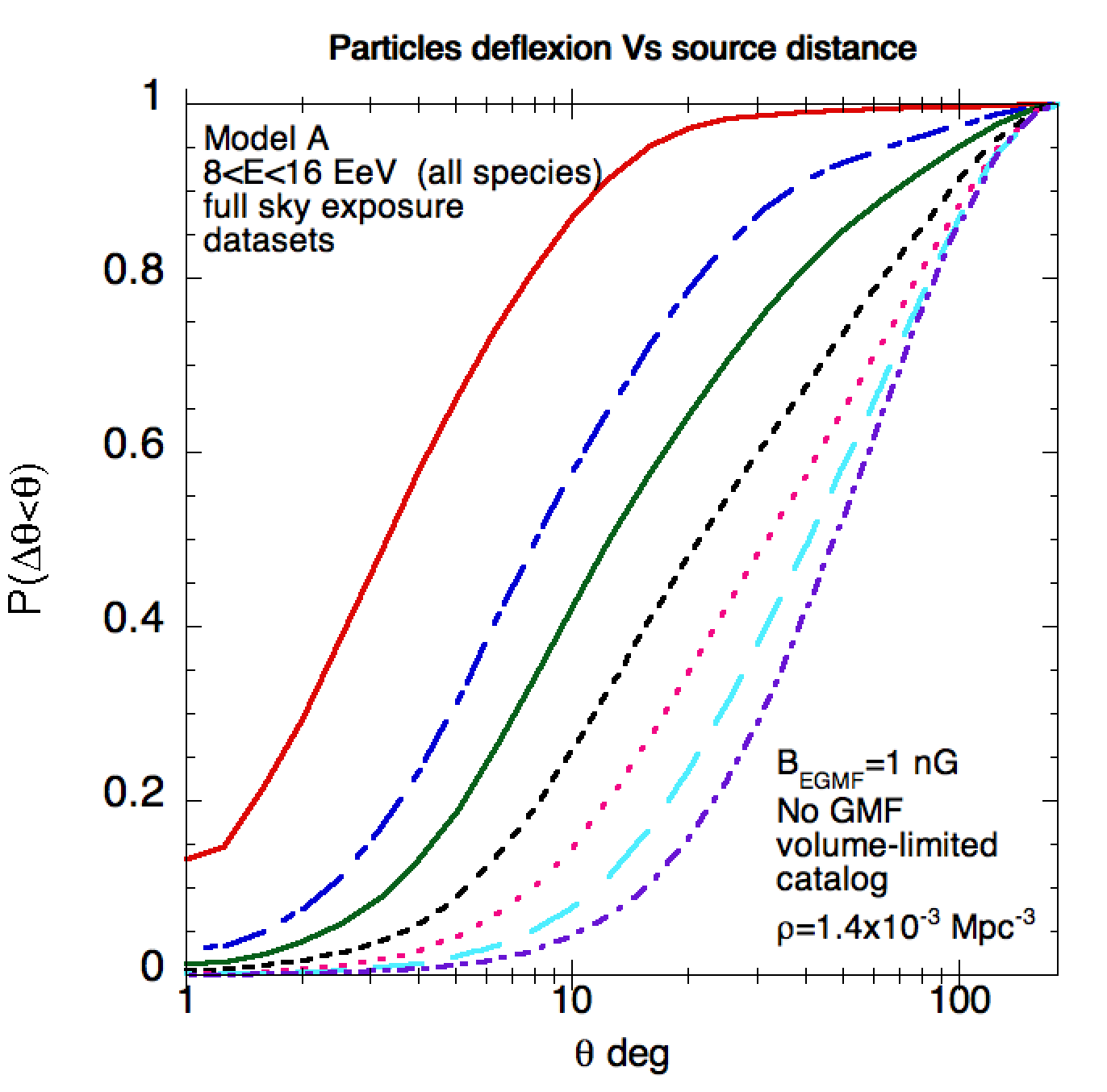}
   \includegraphics[width=7.5cm]{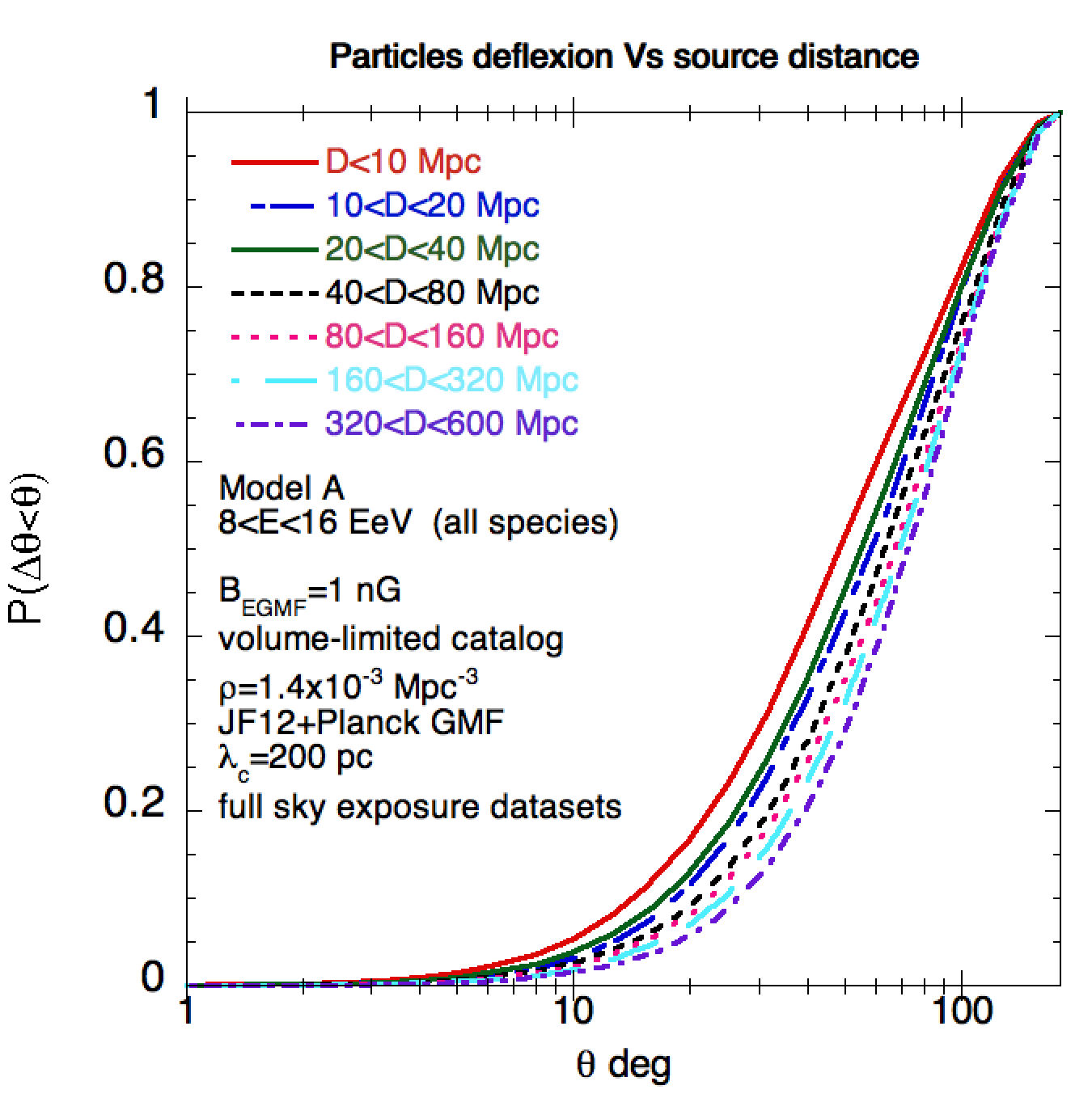}
      \caption{Cumulative distribution function of angular deflexions of UHECRs (including all species) with energy reconstructed on Earth between 8 and 16~EeV coming from sources in various distance bins (see legend). The cases displayed assume the source model A, 2MRS-based catalogs with source density $1.4\times10^{-3}\,\rm Mpc^{-3}$ in the volume-limited approach and a full sky exposure (with twice Auger statistics). The top panel shows the deflexion assuming a 1 nG and no GMF while on the bottom panel the same EGMF is assumed together with the "JF12+Planck" GMF model with $\lambda_{\mathrm{c}}=200$~pc. 
              }
         \label{FigDef}
   \end{figure}

\subsection{Magnetic horizons}

The influence of extragalactic and Galactic magnetic fields on the trajectory of UHECRs, even at the highest energies, is the main obstacle to direct source identification. It obviously modifies the arrival direction of particles, in a way that cannot be inverted without a very precise knowledge of the intervening fields, but it may also modify the other characteristics of the observed UHECRs, namely their spectrum and composition, by introducing delays in the transport from the sources to the Earth with respect to a simple, rectilinear propagation. In the presence of a sufficiently large EGMF, UHECRs of a given energy and mass injected in the intergalactic medium by a source at a distance $D_0$ well within the theoretical GZK horizon may be prevented from reaching the Earth if their effective propagation time (along the curved trajectory reshaped by the EGMF) becomes larger than their energy loss time. This so-called “magnetic horizon” phenomenon can greatly affect the cosmic-ray energy spectrum either at low energy \citep{Aloi2005, Lemoine2005} or even at the highest energies (see {\it e.g}, \citet{Deligny2004}), potentially modifying the GZK argument itself if the time-of-flight distribution of the detected cosmic-rays is indeed significantly different from that expected in the rectilinear case. The larger the magnetic field, the stronger the effect, but for a given magnetic field configuration, the importance of a resulting modification of the spectrum also depends on the source distribution and/or its granularity, as demonstrated in \citet{Aloi2004} (see also \citet{Parizot2004}).

To illustrate and examine this effect in the specific case of our different astrophysical scenarios, we have extracted from our simulated datasets the relative contribution of sources located at different distances, to the overall UHECR flux. In the top panel of Fig.~\ref{FigHorizon}, we show the cumulative distribution functions of the fractional contribution of the sources closer than a certain distance, $D_0$, as a function of that distance. The assumed set of discrete sources for this plot is drawn from a homogeneous (thus also isotropic) underlying distribution, with a source density of $10^{-3}\, \rm Mpc^{-3}$. The source model is model~A, and the different curves correspond to different values of the EGMF, from $10^{-2}$~nG to 10 nG. The case of a purely rectilinear propagation is also shown for comparison (thick solid line). All UHECRs arriving on Earth with an energy larger than 8~EeV, whatever their nuclear species, have been included.

While the cases with an EGMF of $10^{-2}$~nG or even $10^{-1}$~nG are essentially indistinguishable from the rectilinear case, the emergence of a magnetic horizon is clearly visible for larger values of the EGMF, with the largest effect in the 10 nG case, for which almost all the particles originate from sources closer than 100~Mpc. However, this does not necessarily mean that the propagated spectra, as observed on Earth, depend on the value of the EGMF. The resulting effect on the spectrum depends on the actual source distribution, and in particular no effect at all should be expected if the latter satisfies the conditions of application of the so-called "propagation theorem" \citep{Aloi2004}.

An important feature of the 2MRS catalog, already mentioned above and visible in Fig.~\ref{Fig2MRSDens}, is the higher density of galaxies in the nearby universe, compared to the average expectation in a homogeneous universe. This is true even when compensating for the distance selection effect, as clearly visible in our volume-limited catalogs. This local overdensity is directly reflected in the distance distribution of the sources contributing to the observed UHECR flux (if the UHECR source distribution is indeed assumed to roughly follow that of the galaxies). This is shown in Fig.~\ref{FigHorizon}, where we compare the cumulative distribution function of the source distances obtained with source catalogs drawn from the 2MRS-based “mother catalog” and with homogeneous source catalogs of the same density, here chosen as $\rho_{\mathrm{s}} = 10^{-3}\,\rm Mpc^{-3}$. As can be seen, the local overdensity of the 2MRS-based catalogs leads to a significant over-representation of nearby sources up to $\ga 100$~Mpc. Important variations can however be seen between different realisations of the source distribution (cosmic variance), as well as different realisations of the dataset for a given source distribution (statistical variance). The latter, however, is  smaller. The amplitude of these variations can be seen on the bottom panel of Fig.~\ref{FigHorizon}, where the shaded area shows the variance of this cumulative distance distribution for the 300 realisations at a density $\rho_{\mathrm{s}} = 10^{-3}\,\rm Mpc^{-3}$.

The local overdensities inherited from the 2MRS catalog also affect the expected spectrum on Earth, which is on overage slightly harder than what would be obtained from a homogeneous distribution. Note that the larger contribution of nearby sources also modifies the above mentioned time-of-flight distribution, as compared with the homogeneous case, for a given value of the magnetic field. As a result, even though source model A gives a satisfactory account of the observed Auger spectrum, its agreement with the data is poorer when applied to a set of sources drawn from an underlying homogeneous distribution. Therefore, when we will be considering below the case of a homogeneous distribution of sources, we will use a slightly modified version of model A (referred to as model~$\rm A^\prime$), in order to improve the agreement with the observed spectrum.

\subsection{Global deflections}

The main goal of UHECR anisotropy studies is to identify and quantify structures in the distribution of the UHECR arrival directions that could not only vouch for the fact that the observed dataset is not a collection of events sampled from an underlying isotropic distribution of cosmic rays (with a high enough confidence level), but also discriminate between different source models and general astrophysical scenarios. The latter task is of course much more useful, but also much more difficult. This is notably due to the large global deflections of the UHECRs, i.e. the large angle between the arrival direction of a given particle on Earth and the actual direction of its source.
%The key obstacle to deriving constraints on the nature and position of the sources from the observed arrival directions is the large global deflections of the UHECRs.

To give a quantitative idea of the magnitude of these angular deflections, we plotted on Fig.~\ref{FigDef} the cumulative distribution function of the deflections experienced by UHECRs observed on Earth between 8 and 16 EeV (all species together), in the case of source model~A and the 2MRS-based volume limited catalog with density $1.4\times 10^{-3}\, \rm Mpc^{-3}$, for different source distances. The curves were obtained with large statistics by adding up 300 realizations of our datasets with full-sky coverage (each with twice the Auger statistics, see Sect.~\ref{sec:datasets}).

The top panel shows the deflections obtained with a 1~nG EGMF, which we use as a baseline, in the absence of a Galactic magnetic field (i.e. turning off the Galactic magnifications and deflections), for various source distance intervals. Obviously, the deflections increase with the source distance: for sources closer than 10~Mpc, $\sim 50\%$ of the UHECRs have global deflections smaller than $3^\circ$, while $\sim50\%$ of the particles coming from source between 20 and 40 Mpc from the Earth are deflected by more $10^\circ$. The deflections are of course smaller for smaller values of the EGMF.

However, even if the deflections in the EGMF remain moderate, the inclusion of the GMF strongly modifies the overall picture. In the bottom panel of Fig.~\ref{FigDef}, we show the same cumulative distributions as above when turning on the "JF12+Planck" GMF model, with a coherence length $\lambda_{\mathrm{c}}=200$~pc. As can be seen, the GMF causes much larger deflections than the EGMF, which remains true even for an EGMF as large as 3~nG, or even as large as 10~nG in the case of the nearest sources. In addition, these deflections do not only consist in a blurring of the source image on the sky, but result in a systematic shift of their average apparent position on sky, mostly due to the contribution of the regular component of the GMF to the global deflections. In that case, the fraction of particles with deflections smaller than $10^\circ$ is of the order of $5\%$, whatever the source distance, i.e. even for the nearest sources and even assuming an essentially negligible $10^{-2}$~nG EGMF. This small fraction is mostly made of a subset of the light nuclei still present in this energy range.

With this situation in mind, we now turn to the analysis of the arrival direction of our simulated datasets.

\begin{figure}[ht!]
   \centering
   \includegraphics[width=8.7cm]{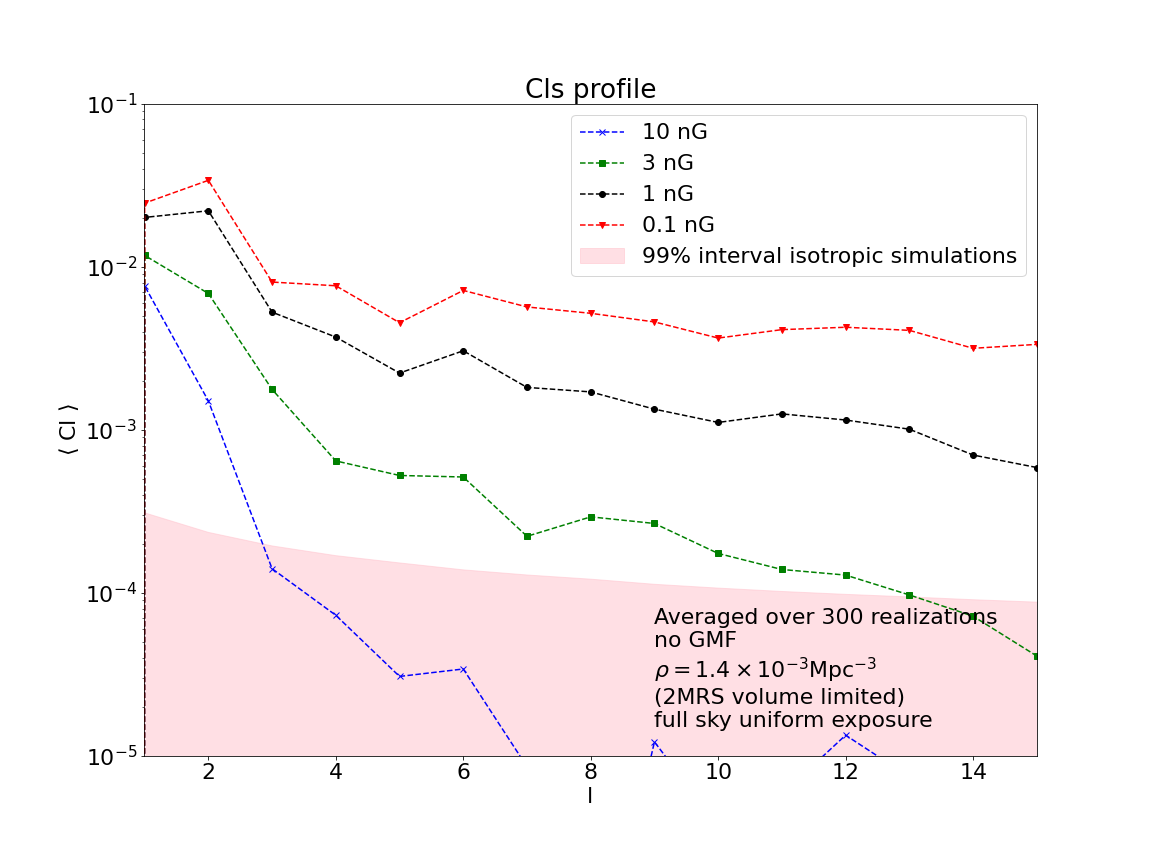}\\
   \vspace{-12pt}
   \includegraphics[width=8.7cm]{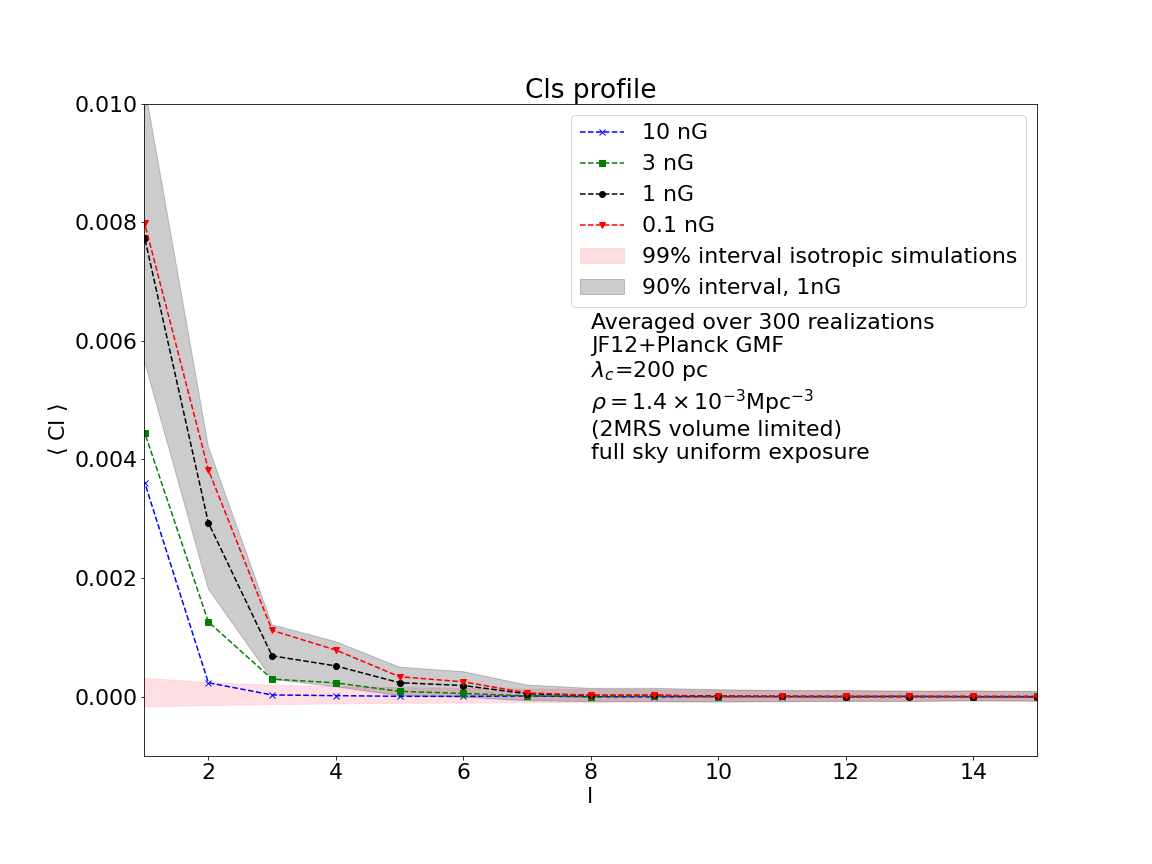}\\
   \vspace{-12pt}
   \includegraphics[width=8.7cm]{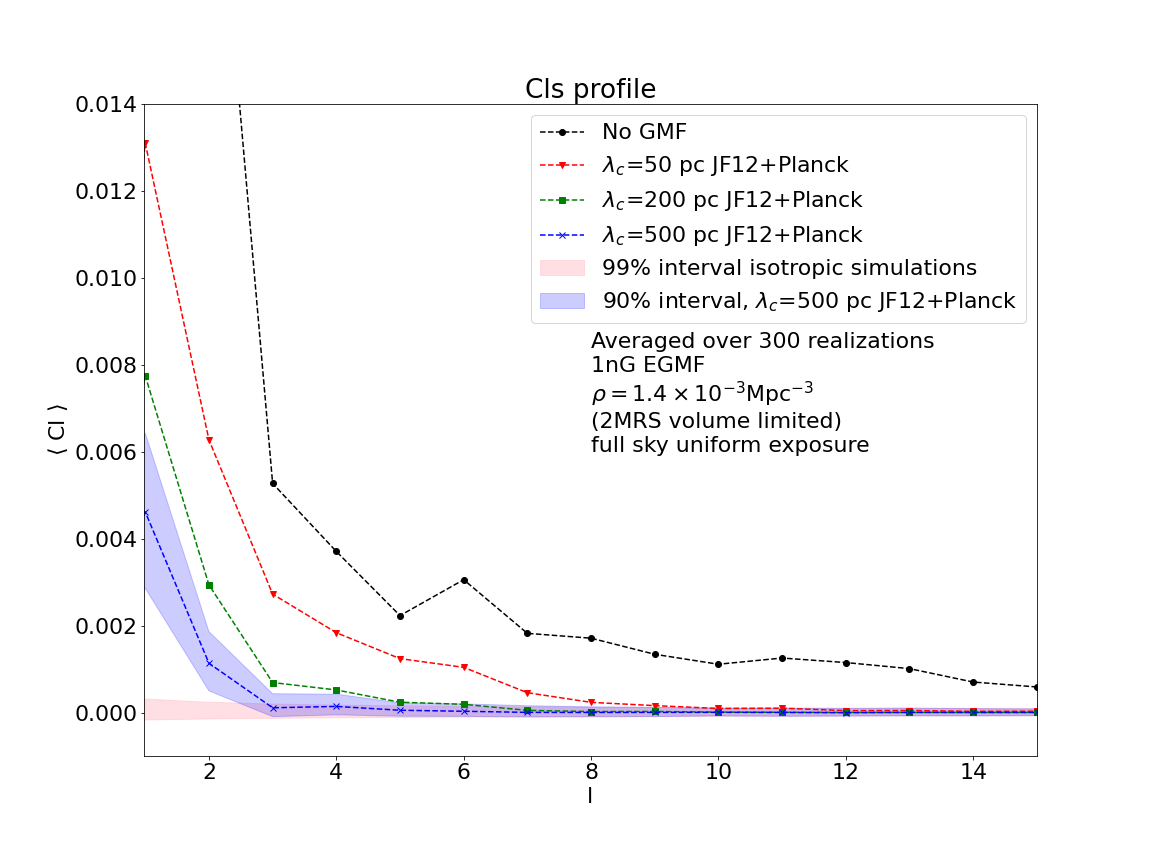}
      \caption{Angular power spectrum of our simulations for the source model A, 2MRS-based catalogs with source density $1.4\times10^{-3}\,\rm Mpc^{-3}$ in the volume-limited approach and various hypothesis on the galactic and extragalactic magnetic fields. The lines show the average values obtained over 300 realizations, the (red) shaded area show the range in which $99\%$ of isotropic UHECR simulations are found with the same statistics as our datasets . The top panel shows the predictions for various values of the EGMF intensity and no GMF assuming a full-sky exposure. The middle panel shows the same cases but now assuming the "JF12+Planck" GMF model with $\lambda_{\mathrm{c}}=200$~pc. The bottom panel shows the result in the case of a 1~nG EGMF for various assumptions regarding the GMF coherence length (assuming full-sky coverage).  
              }
         \label{FigCl}
   \end{figure}

\begin{figure}[ht!]
   \centering
    \includegraphics[width=8.7cm]{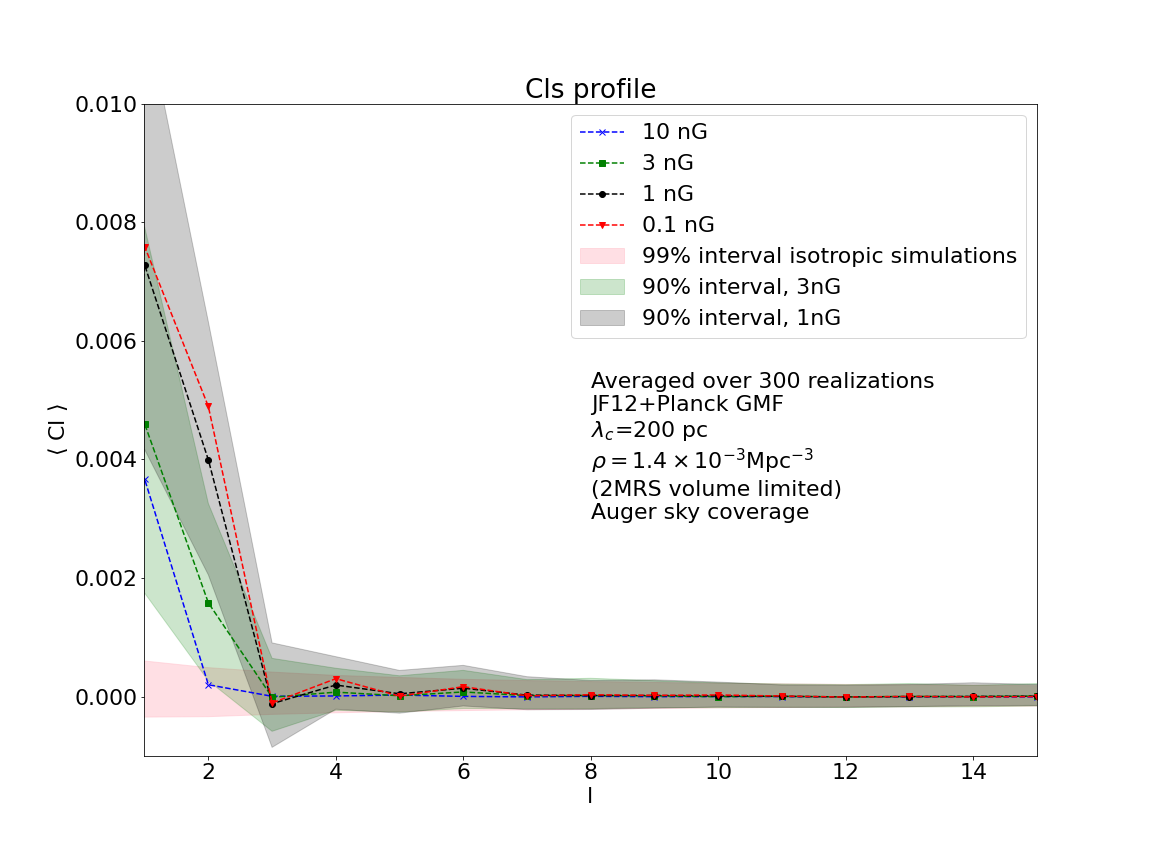}
   \includegraphics[width=8.7cm]{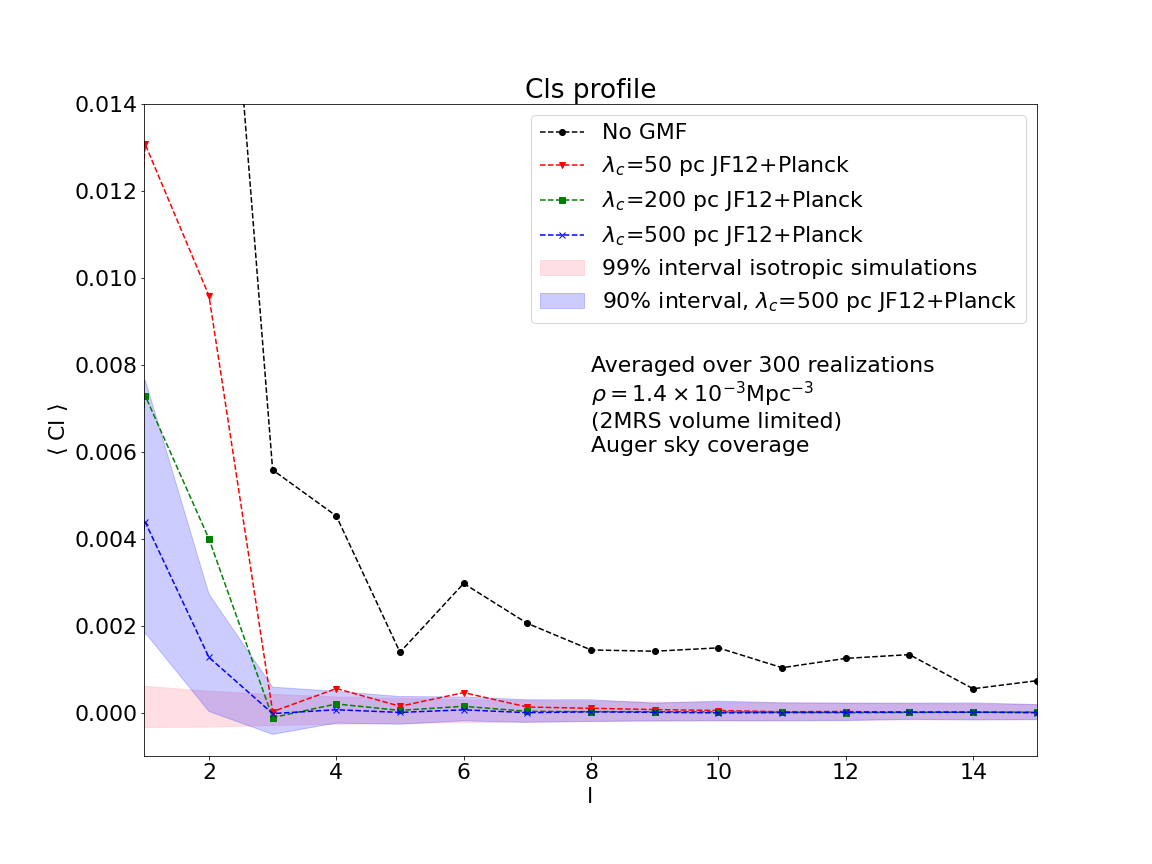}
      \caption{Same as Fig.~\ref{FigCl}, central and bottom panel, but with the Auger coverage instead of a full-sky coverage.
              }
         \label{FigCl_Auger}
   \end{figure}

\section{Anisotropy results}
\label{sec:results}
\subsection{Angular power spectrum}

For each of our datasets, we have computed the angular power spectrum, given in terms of the $C_\ell$ values for $\ell = 1$ to 65, of all UHECR events reconstructed with an energy larger than 8 EeV. In Fig.\ref{FigCl}, we show the result obtained for our baseline volume-limited catalog (with density $\rho_{\mathrm{s}}=1.4\times 10^{-3}\,\rm Mpc^{-3}$, see Sect.~\ref{sec:sourceDistrib}), assuming the source model~A and different hypotheses regarding the EGMF intensity and the GMF coherence length. The different lines show the power spectra obtained after averaging over 300 realizations of full-sky datasets (the $C_\ell$ values are shown only up to  $\ell = 15$ for clarity). %(for the two top panels and the bottom left panel) and of datasets with the Auger sky coverage (bottom right panel).
The plots also show as a light red shaded area the range of expectations for 99\% of our simulated isotropic distributions of events with the same statistics, so that the significance of the $C_\ell$ values can be estimated. 

On the top panel, we show the angular power spectrum of the UHECRs just before entering the Galaxy, for different values of the EGMF (i.e. we turn off the GMF). As can be seen, for EGMF intensities smaller than a few nG, significant anisotropies can be expected at all angular scales, down to the smallest (corresponding to the highest values of $\ell$). As expected, increasing the EGMF makes the global anisotropy levels lower, particularly at small angular scales.

However, the addition of the GMF has a dramatic effect on the high-$\ell$ (small scale) modes, as clearly shown on the middle panel of Fig.~\ref{FigCl}, where we show the angular power spectrum obtained with the JF12+Planck GMF model and a coherence length of the turbulent component $\lambda_{\mathrm{c}} = 200$~pc. Even for sub-nG values of the EGMF, all the high-$\ell$ modes are suppressed, and only the dipole and quadrupole modes are expected to show significant excesses compared to isotropic expectations, at least for some realisations, for all EGMF cases we considered. To illustrate the statistical variance, we also showed on the plot, as a black shaded area, the interval in which 90$\%$ of the datasets are found in the case of a 1~nG EGMF. However, the level of anisotropy of the datasets also depends rather strongly on the assumed value of the GMF coherence length. The bottom panel of Fig.~\ref{FigCl} shows the $C_\ell$ values for $\lambda_{\mathrm{c}} = 50$~pc, 200~pc, and 500~pc (in the case of a 1~nG EGMF). As could be expected, larger coherence lengths lead to smaller anisotropy level at all angular scales. This is because the behaviour of a charged particle in a turbulent field is very sensitive to the ratio of its Larmor radius to the coherence length of the turbulence, the overall deflections being larger for lower values of this ratio. In the cases displayed on the plot, a coherence length as large as 500~pc appears to be needed to bring back the reconstructed quadrupole values marginally within the range of the isotropic expectations for at least a small fraction of the datasets. Note that the blue shaded area shows the range where 90\% of the realisations lie (all except the 5\% most isotropic and the 5\% least isotropic).

This result is important in view of the comparison with the Auger data. Indeed, while Auger reported a significant dipole anisotropy (which we examine below), its dataset does not show any significant power in the quadrupole mode (given its current statistics). To better compare our $C_\ell$ predictions with the Auger data, we did the same angular power spectrum analysis as above for our Auger-like datasets, i.e. assuming the actual Auger sky coverage and the same statistics. The results are shown in Fig.~\ref{FigCl_Auger} for different values of the EGMF (top panel) and of the GMF coherence length (bottom panel). The reconstructed angular power spectra show larger dispersion because of the smaller statistics (by a factor of two compared to Fig.~\ref{FigCl}). The quadrupole mode is more difficult to distinguish from isotropic expectations in the 500 pc case and, to a lesser extent, in the 200~pc case (see below for a more detailed discussion). However, it is important to keep in mind the degeneracy between the EGMF value and the GMF coherence length, which prevents any reliable conclusion on either of these parameters. With an EGMF of 1~nG, none of the 300 realisations have a low enough power in the quadrupole mode to be compatible with the data\cite{AugerMulti2017}, and a coherence length of 500~pc or larger appears to be required, while a 3~nG EGMF allows some realisations to have quadrupoles statistically compatible with isotropic simulations even with a 200~pc GMF coherence length. It should also be noted that the cosmic variance (due to the specific galaxies assumed as actual UHECR sources) is in fact larger than the statistical variance (see below).

Nevertheless, a conjunction of a coherence length $\lesssim 100$~pc and a weak EGMF (say $<1$~nG) appears to be difficult to reconcile with the observed weakness of the quadrupole moment of the UHECR angular distribution, at least for the explored scenarios. This remains true even in the case of a larger source density, such as that of the mother catalog itself.

\begin{figure*}[ht!]
   \centering
   \includegraphics[width=7.5cm]{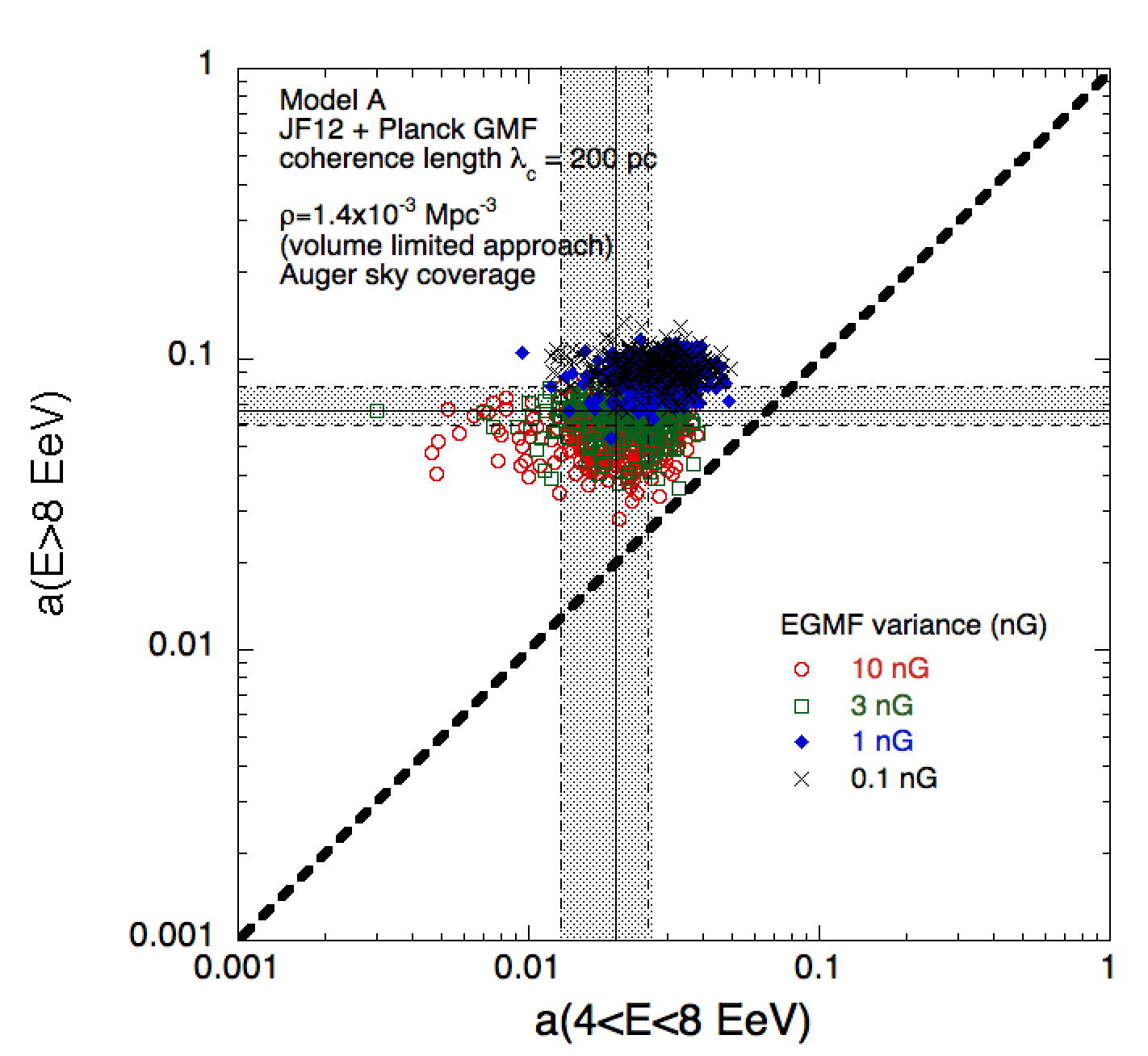}
   \includegraphics[width=7.5cm]{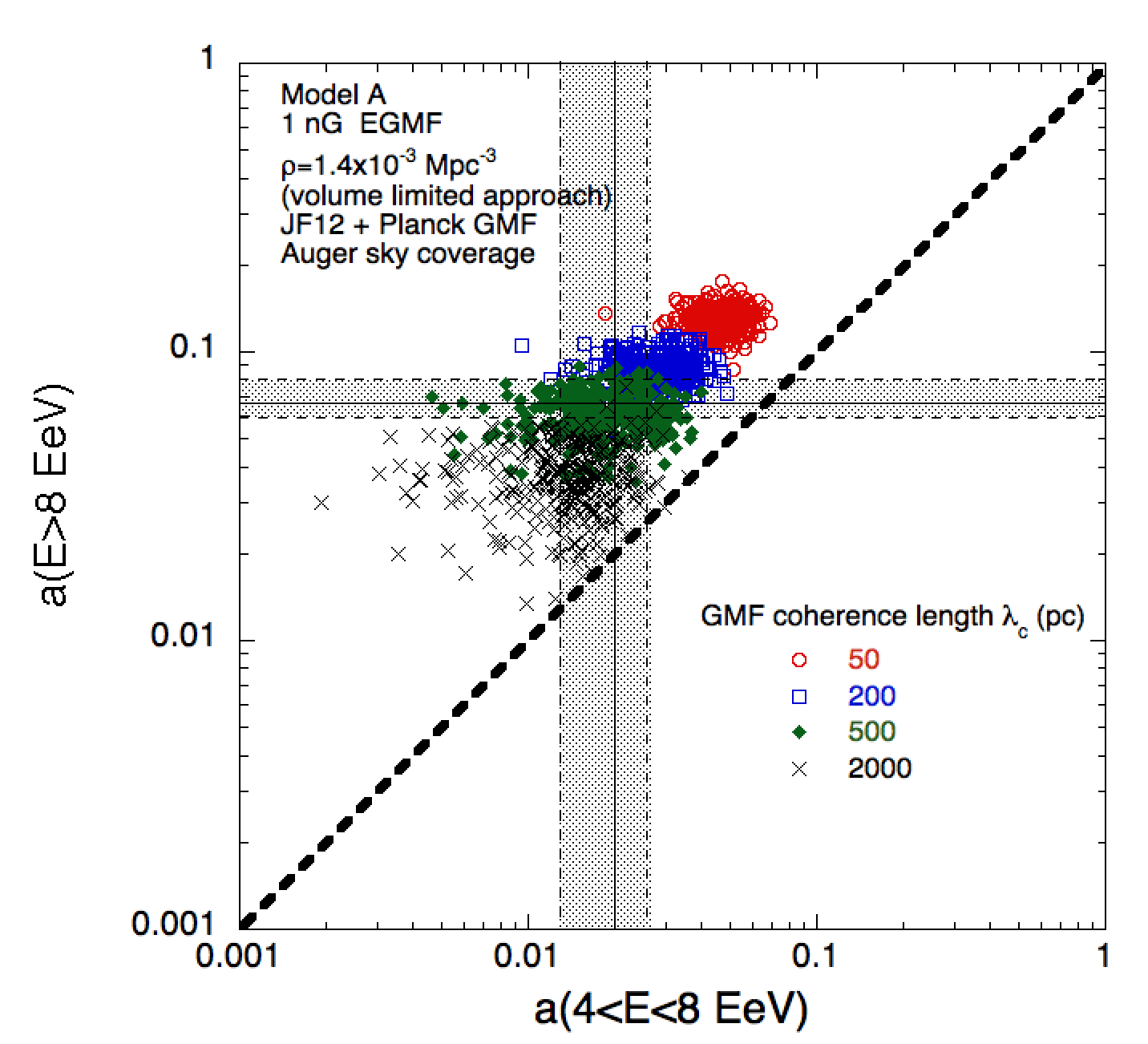}
   \includegraphics[width=7.5cm]{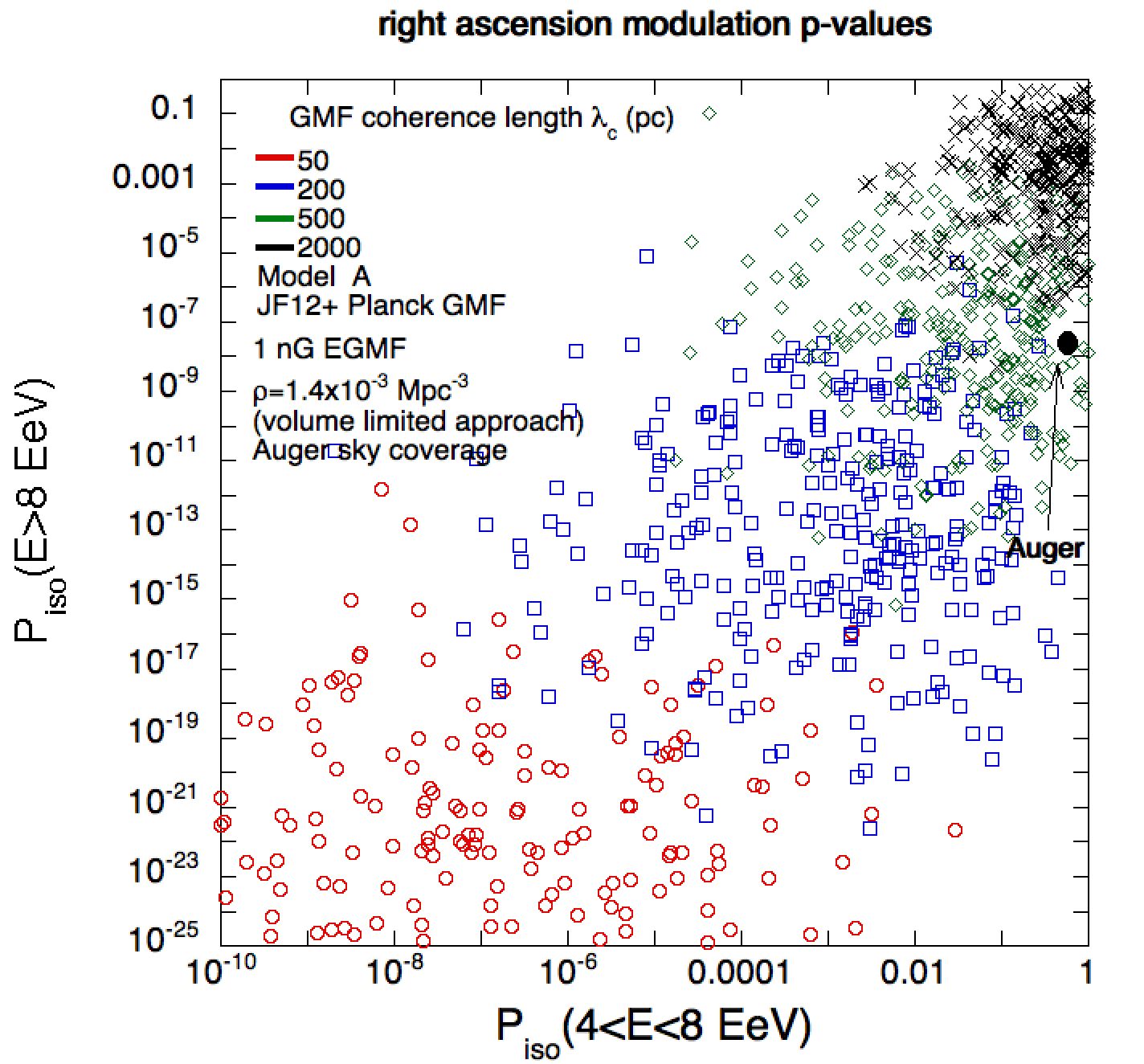}
    \includegraphics[width=7.5cm]{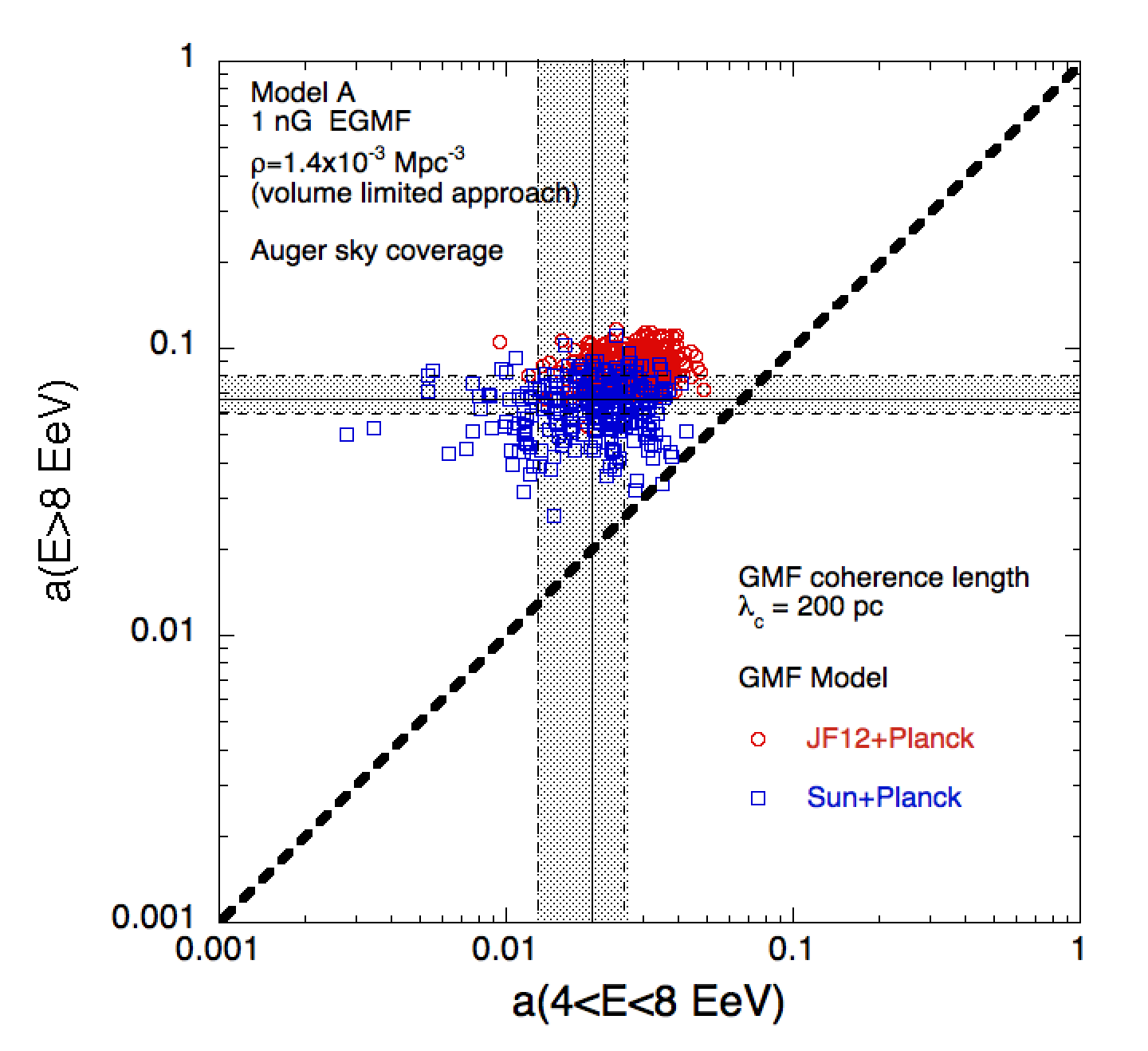}
      \caption{Top-left : scatter plot of the dipole amplitude reconstructed for 300 datasets computed for various astrophysical scenarios, based on the baseline volume-limited catalog, source model A and the JF12+Planck GMF with $\lambda_{\mathrm{c}}=200$~pc in the energy bins $4 \le E < 8$~EeV and $E\geq 8$~EeV. The four scenarios displayed differ by the assumed variance of the EGMF, 10, 3, 1 or 0.1 nG and the Auger $\pm 1\sigma$ confidence intervals are shown with shaded areas. Top-right : Same as the top-left panel, but the scenarios now differ by the assumed value of $\lambda_{\mathrm{c}}$, 50, 200, 500 and 2000~pc. In all case a 1~nG EGMF is assumed. Bottom-left : Scatter plot of the estimated p-values corresponding to the cases displayed on the top-right panel, the p-values found by Auger are shown by a full black marker. Bottom-right :  scatter plot of the dipole amplitude reconstructed for 300 datasets for the same astrophysical scenario as on the top-right panel, the two cases shown differ by the assumed GMF model, JF12+Planck or Sun+Planck (in both cases $\lambda_{\mathrm{c}}=200$~pc.
              }
         \label{FigDipScat1}
   \end{figure*}
   
\begin{figure}[ht!]
   \centering
   \includegraphics[width=7.5cm]{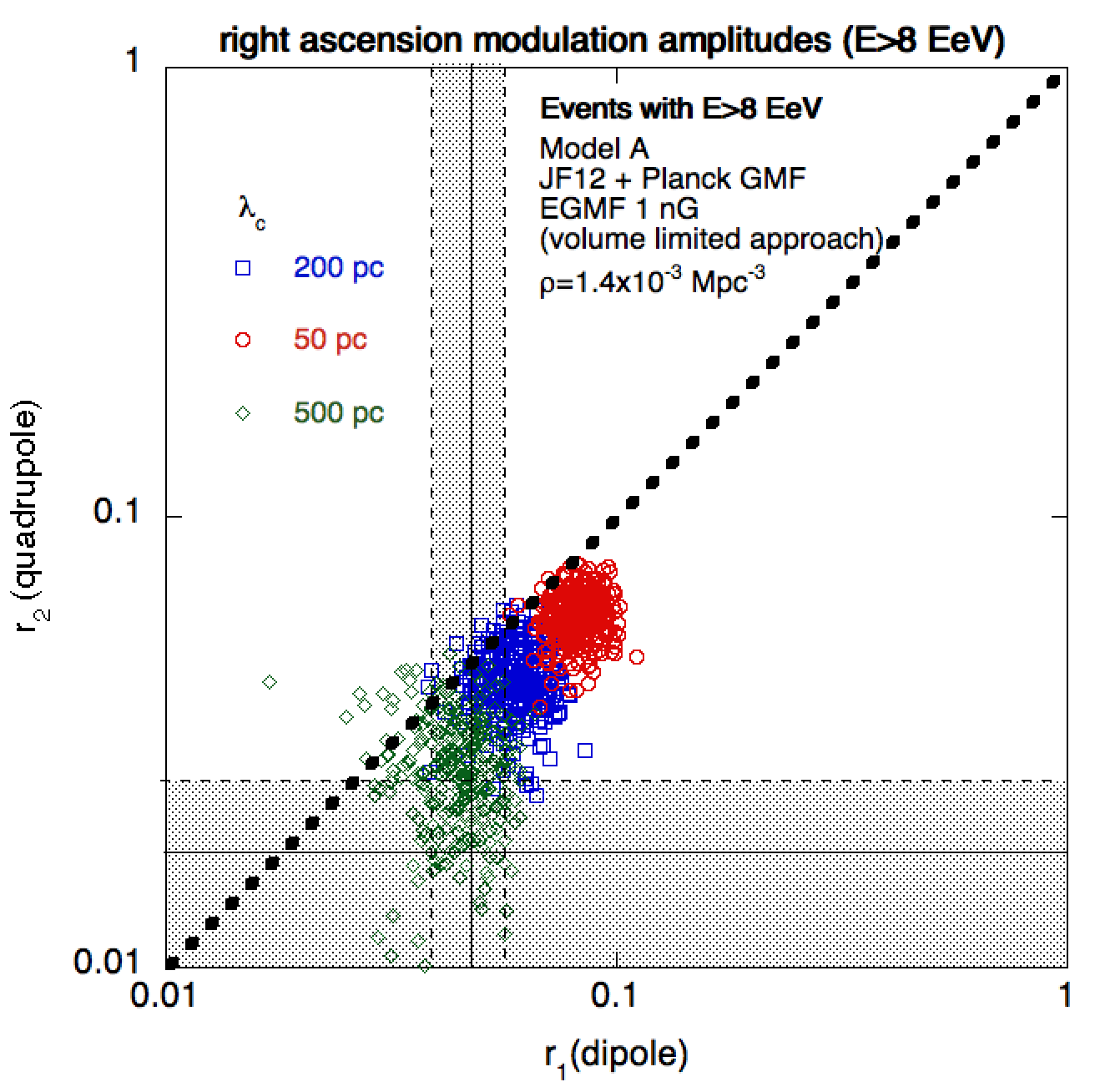}
   \includegraphics[width=7.5cm]{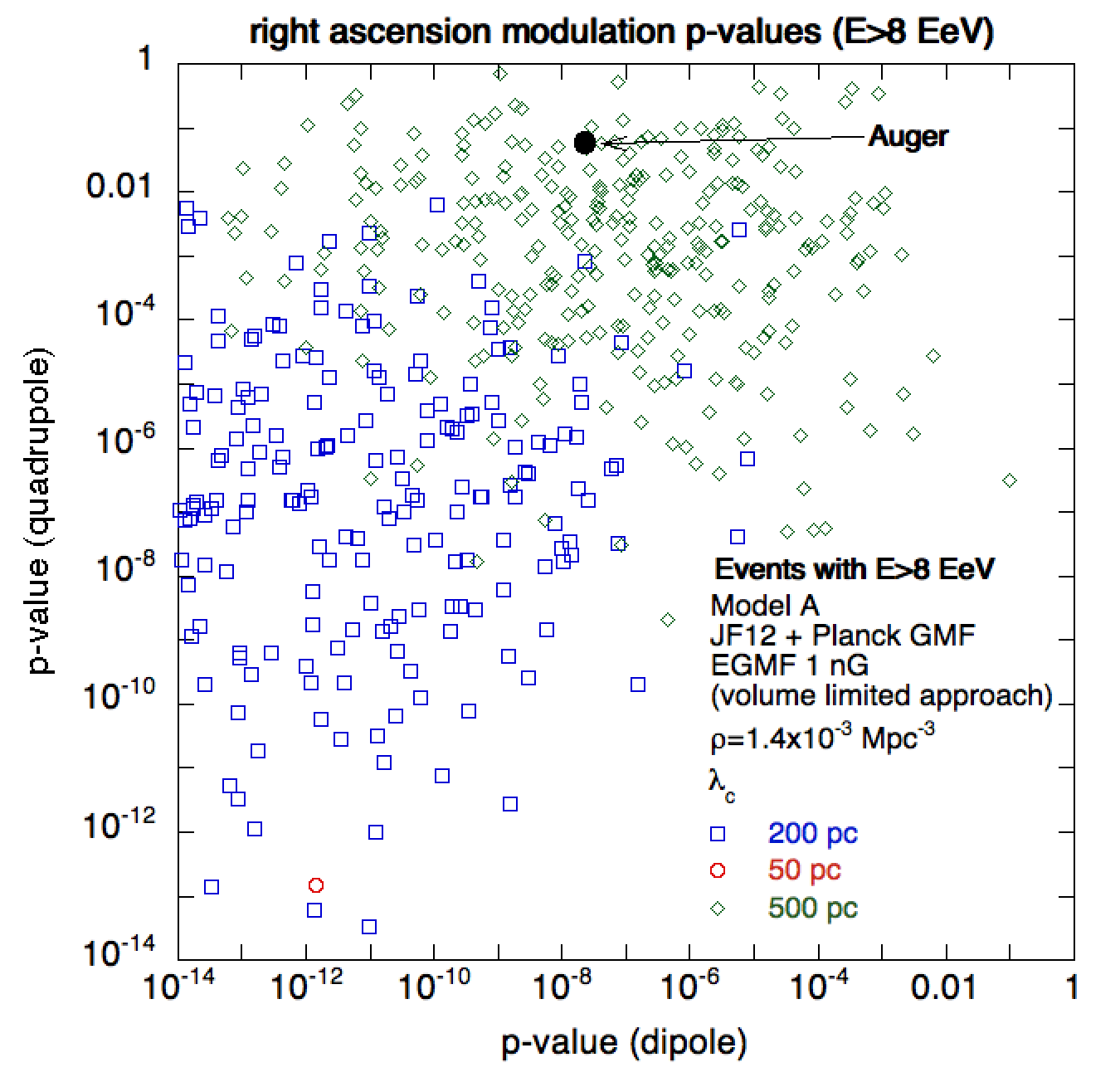}
      \caption{Top : scatter plot of the right ascension dipolar and quadrupolar modulation amplitude reconstructed for 300 datasets computed for the same astrophysical scenarios as on the top-right panel of Fig.~\ref{FigDipScat1}, the Auger $\pm 1\sigma$ confidence intervals are shown with shaded areas.  Bottom : scatter plot of the corresponding p-values, the p-values found by Auger are shown by a full black marker.  
              }
         \label{FigQuadScat1}
   \end{figure}   
   
\subsection{Rayleigh analysis : impact of physical and astrophysical parameters}
\label{sec:Rayleigh}
We now turn to the analysis of our datasets with the Rayleigh method, as discussed in \citet{AugerDip2017, AugerDipLong2018}. In the 2017 paper, the Auger collaboration applied this analysis to determine an effective dipole amplitude in two different energy bins, namely between 4 and 8~EeV and above 8~EeV. We follow the very same procedure for our simulated datasets, concentrating mostly on datasets produced with the Auger sky coverage (unless otherwise indicated), and examining more specifically the impact of the various physical and astrophysical parameters on the predicted dipole amplitude. To this end, we fix the source distribution  by using the  volume-limited catalogs (cf. Sect.~\ref{sec:sourceDistrib}). (NB : Our extragalactic source models  account for the whole UHECR above 5~EeV, as a result a residual (probably) heavy Galactic component is expected to still be present in the first above-mentioned between 4 and 5~EeV. This missing residual component represents between $\sim$3 and 7\% (depending on the source model) of the total number of the Auger events in the $4\leq E<8$~EeV. We do not correct for it as it should not have a strong effect on the expected anisotropy level in that energy bin.)

\subsubsection{Influence of the magnetic field}

The main results regarding the influence of the magnetic field on the reconstructed dipole amplitudes are summarised in Fig.~\ref{FigDipScat1}, which consists of scatter plots in which each point corresponds to the analysis of one particular simulated dataset (i.e. one of the 300 realisations of a given astrophysical scenario), and each colour corresponds to a specific choice of the magnetic field parameters.

The top-left panel shows the amplitude of the dipoles reconstructed in the two energy bins considered in \citet{AugerDip2017}, for 300 realisations of our baseline volume-limited catalog, assuming source model A. The amplitude in the higher energy bin ($E \ge 8$~EeV) is shown on the y-axis and that in the lower energy bin (4~EeV $\le E < 8$~EeV) is shown on the x-axis. The different colours correspond to different values of the assumed EGMF, as indicated (the 0.01~nG case is omitted here, being essentially indistinguishable from the 0.1~nG case). The assumed GMF is the JF12+Planck model with $\lambda_{\mathrm{c}} = 200$~pc and the $\pm 1\sigma$ interval reconstructed by Auger is shown by the shaded area.

As can be seen, the dipole amplitudes are on average decreasing with increasing EGMF values, which is consistent with the results of the previous subsection (see Fig.~\ref{FigCl_Auger}). However, the statistical dispersion is quite large with this dataset size (and dipole amplitude), and it appears essentially impossible to separate the different EGMF scenarios on the basis of a single realisation. (NB: we have checked that the obtained dispersion is similar in the test case of perfect dipoles of similar amplitudes simulated with the same statistics.) %This dispersion which critically depends on both the dataset statistics and the true value of the amplitude is quite large and consistent  with what we obtained with test cases of perfect dipoles with similar amplitudes.

Similarly, the top-right panel of Fig.~\ref{FigDipScat1} shows the influence of the GMF coherence length on the reconstructed amplitudes. A 1~nG EGMF and the JF12+Planck GMF model are assumed. The different colours correspond to four different values of $\lambda_{\mathrm{c}}$, from 50 to 2000 pc. The expected trend, namely an increase of the amplitude with decreasing coherence length, is clearly visible (also consistent with the results of the previous section). Based on this plot, a coherence length around 500~pc (green points) appears to be favoured by the Auger data, for this particular choice of the source model and source distribution. However, the statistical variance is such that coherence lengths as large as 2000~pc (black points) or as low as 200~pc (blue points) cannot be easily excluded. We also wish to stress that the attempts to derive a quantitative constraint on the magnetic field from this observation can only be valid in the limits of a particular choice of the underlying source scenario. This comment applies to all the quantitative statements of this section. 

%While for the particular case of this volume-limited catalog, for this source model case, and assuming a 1~nG EGMF, the 500 pc case seems to give a better account of the observed data, the realization to realization separation with the 200 pc and 2000 pc remains however difficult and it is important to stress that any attempt to make a quantitative statement from this observation would be only valid for this particular scenario where all the physical parameters but $\lambda_{\mathrm{c}}$ were fixed. This comment basically to all the quantitative statements of this section. 

The bottom-left panel of Fig.~\ref{FigDipScat1} shows a scatter plot of the corresponding p-values for the right ascension modulation amplitude, that is the probability for an intrinsically isotropic distribution of cosmic-ray events with the same dataset statistics to give an amplitude as high or higher. As already seen for the amplitudes, the 500~pc case is the most compatible with the raw p-values values reported by Auger,  i.e. a very significant dipole in the energy bin above 8~EeV and a p-value close to 1 in the lower energy bin. 

%Finally a comparison of the obtained amplitude values for the JF12+Planck and Sun+Planck GMF assuming for both $\lambda_{\mathrm{c}}$=200 pc is shown on the bottom-right panel. A systematic shift to slightly lower values is visible for the Sun+Planck model, probably owing to the slower evolution of the toroidal halo field with the galactic height predicted by this model. 

Finally, the difference between the JF12+Planck and the Sun+Planck GMF models is shown in on the bottom-right panel of Fig.~\ref{FigDipScat1} in the case of a GMF coherence length $\lambda_{\mathrm{c}}$=200 pc (still for model A, a source density of $1.4\times10^{-3}\,\rm Mpc^{-3}$, and an EGMF of 1~nG). A systematic shift to slightly lower values is visible for the Sun+Planck model, probably owing to the slower evolution of the toroidal halo field with the Galactic height implied by this model. 

Keeping in mind the Auger results on the quadrupole, we also examined the influence of the magnetic field on the amplitude of the second harmonic of the right ascension modulation, by applying the Rayleigh analysis on the same realizations as on the top-right panel of Fig.~\ref{FigDipScat1}. The top panel of Fig.~\ref{FigQuadScat1} shows a scatter plot of the first two harmonics ($r_1$ and $r_2$), together with the $\pm 1\sigma$ intervals obtained by Auger. 

Again, for this particular volume-limited catalog, the datasets obtained in the cases when $\lambda_{\mathrm{c}}$ is 50~pc or 200~pc have too large values compared to the observed one, while the case of a 500~pc coherence length appears compatible with the data. Around 10\% of the realisations show a value of $r_2$ lower than that reconstructed by Auger, while $\sim40\%$ are within the $\pm 1\sigma$ interval. Likewise, the corresponding p-values displayed on the bottom panel of Fig.~\ref{FigQuadScat1} show $\sim10\%$ realisations with a p-value larger than that of Auger, $\sim30\%$ above $10^{-2}$ and a median p-value of the order of $10^{-3}$.

This particular case study illustrates the fact that the current absence of a significant signal for a quadrupolar modulation in the arrival directions of UHECR events does not particularly challenge the models discussed throughout the present study. We will come back to this point later on when discussing the impact of the cosmic variance.

\subsubsection{Influence of the source density}

\begin{figure}[ht!]
   \centering
   \includegraphics[width=7.5cm]{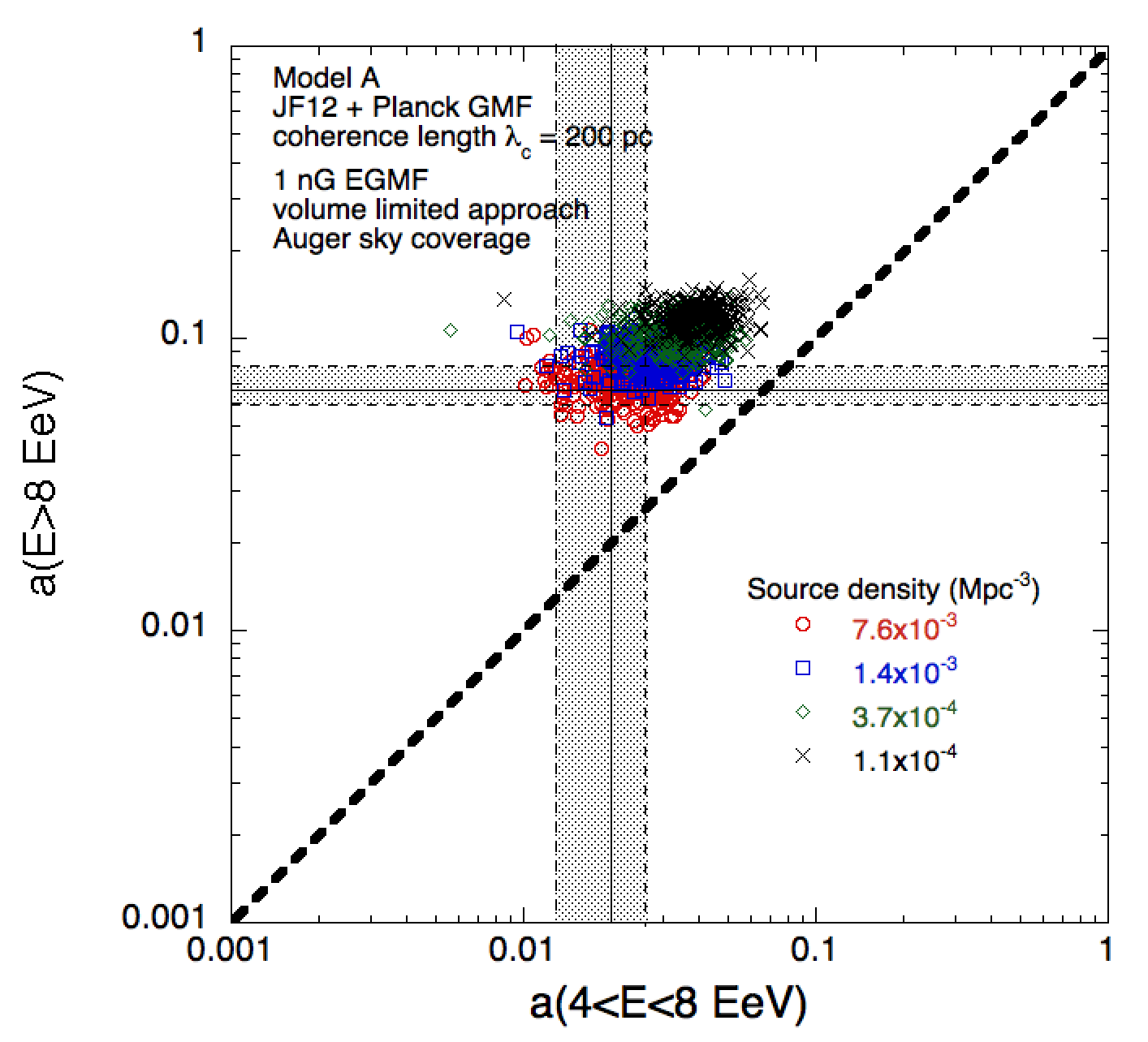}
      \caption{Scatter plot of the dipole amplitude reconstructed for 300 datasets computed for various astrophysical scenarios, based on source model A, a 1~nG EGMF, the JF12+Planck GMF with $\lambda_{\mathrm{c}}=200$~pc, in the energy bins $4\le E < 8$~EeV and $E\geq 8$~EeV. The cases of the four volume-limited catalogs listed in Tab.~\ref{table:1}, which differ by their densities, are considered. 
              }
         \label{FigDipScat2}
   \end{figure}

One of the key parameters influencing the expected level of anisotropy for a given astrophysical scenario is the source density. Obviously, in a lower source density scenario, the all-sky UHECR flux is contributed by a smaller number of more powerful sources, resulting on average in more pronounced anisotropies (all other things being equal). In addition, the cosmic variance is expected to be larger in the case of a low source density (as suggested by Fig.~\ref{Fig2MRSDens}), if a candidate source distribution is to be obtained by sub-sampling the same given parent distribution. This point will be addressed below. Here, we consider the four volume-limited catalogs listed in Tab.~\ref{table:1}. We remind the reader that these catalogs are not unrelated with one another, since they are obtained from the same mother catalog by applying increasingly stringent cuts on the $K_\mathrm{s}$-band luminosity. As it turns out, given the local  spatial and luminosity distributions of the sources, increasing the threshold luminosity does result in larger local overdensities for the three lower density catalogs (compared to that of the mother catalog shown in Fig.~\ref{Fig2MRSDens}). 

The scatter plot of the dipole amplitudes reconstructed in the two Auger energy bins is shown in Fig.~\ref{FigDipScat2}, assuming again the source model A, a 1 nG EGMF and the JF12+Planck GMF model with $\lambda_{\mathrm{c}}$=200 pc. Unsurprisingly one sees a shift of the distributions toward higher values as the density decreases. This same trend is observed for the quadrupolar modulation. From this plot, one could be inclined to conclude that a source density as low as $10^{-4}\,\rm Mpc^{-3}$ is disfavoured by the data, while datasets obtained directly from the mother catalog (with a source density of $7.6\times10^{-3}\,\rm Mpc^{-3}$) provide a good account of the data. However, such a conclusion is dependent on the (currently unknown) value of the turbulent field coherence length, and as it appears, the distribution of the amplitudes obtained here from the mother catalog is essentially indistinguishable from that obtained above with the baseline volume-limited catalog ($1.4\times10^{-3}\,\rm Mpc^{-3}$), assuming a coherence length of $\lambda_{\mathrm{c}}=500$~pc. This is yet another illustration of the high degree of degeneracy of the problem at hand, at least as far as the amplitude of the dipole is concerned.

%and consider the four volume-limited catalogs listed in Tab.~\ref{table:1}. Again these catalogs are deduced from each other (starting with the mother catalog) by applying increasingly  stringent cuts on the $K_\mathrm{s}$-band luminosity, the resulting overdensities of the three lower density volume-limited catalogs obtained are larger than that of the mother catalog shown in Fig.~\ref{Fig2MRSDens}. The scatter plot of the dipole amplitude reconstructed in the energy bins between 4 and 8 EeV and above 8 EeV is shown in Fig.~\ref{FigDipScat2}, assuming source model A, a 1 nG EGMF and the JF12+Planck model with $\lambda_{\mathrm{c}}$=200 pc. Unsurprisingly one sees a slight systematic shift of the distributions toward higher values as the density decreases. This trend is also true for the quadrupolar modulation. Note that, distributions obtained assuming the mother catalog ($7.6\times10^{-3}\,\rm Mpc^{-3}$) together with the above mentioned hypothesis on the magnetic fields are basically indistinguishable from what was obtained with the baseline volume-limited catalog ($1.4\times10^{-3}\,\rm Mpc^{-3}$) assuming $\lambda_{\mathrm{c}}=500$~pc. This is another illustration of high degree of degeneracy of the problem at hand at least as far as the amplitude of the dipole concerned. 

\subsubsection{Influence of the source composition model}
\label{sec:compo}
   \begin{figure}[ht!]
   \centering
   \includegraphics[width=8.5cm]{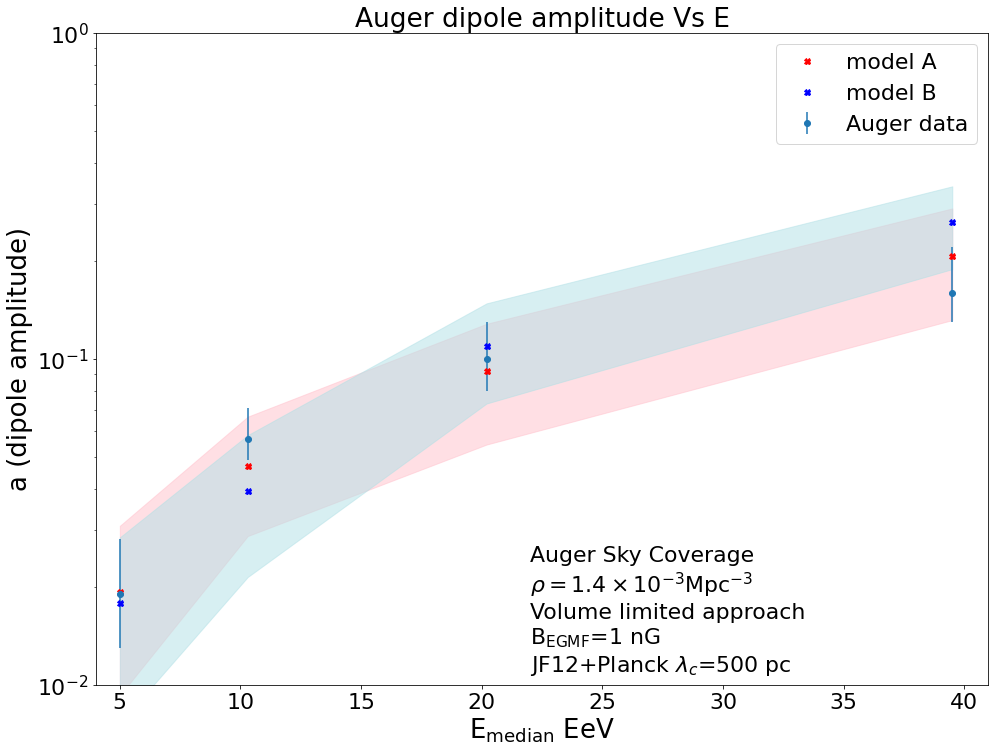}
   \includegraphics[width=8.5cm]{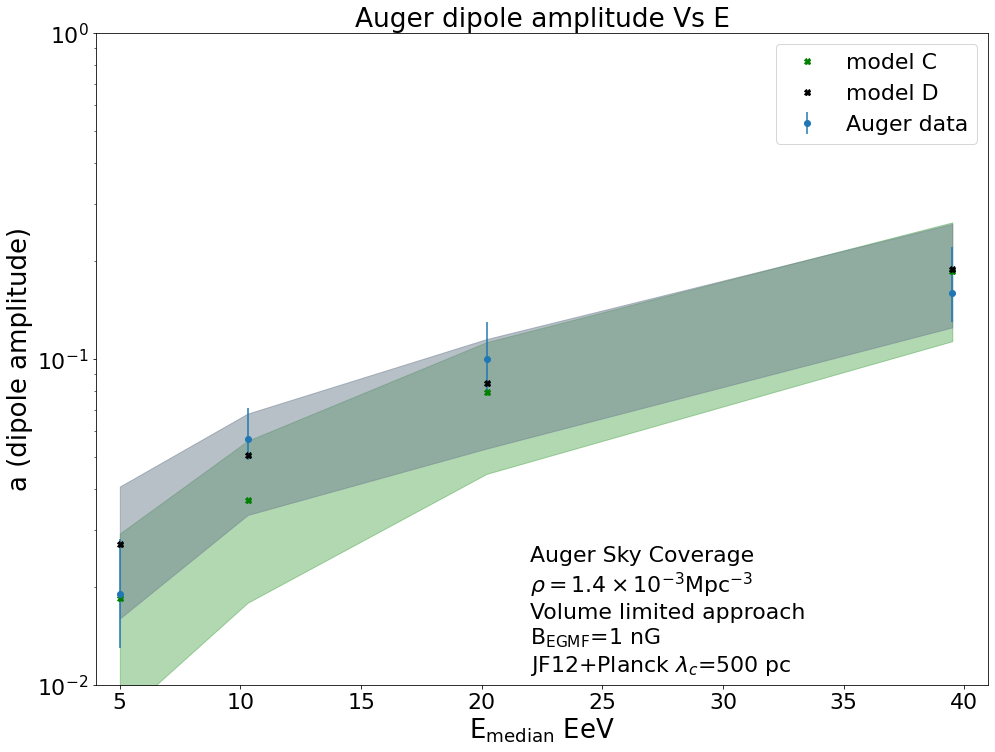}
   \includegraphics[width=8.5cm]{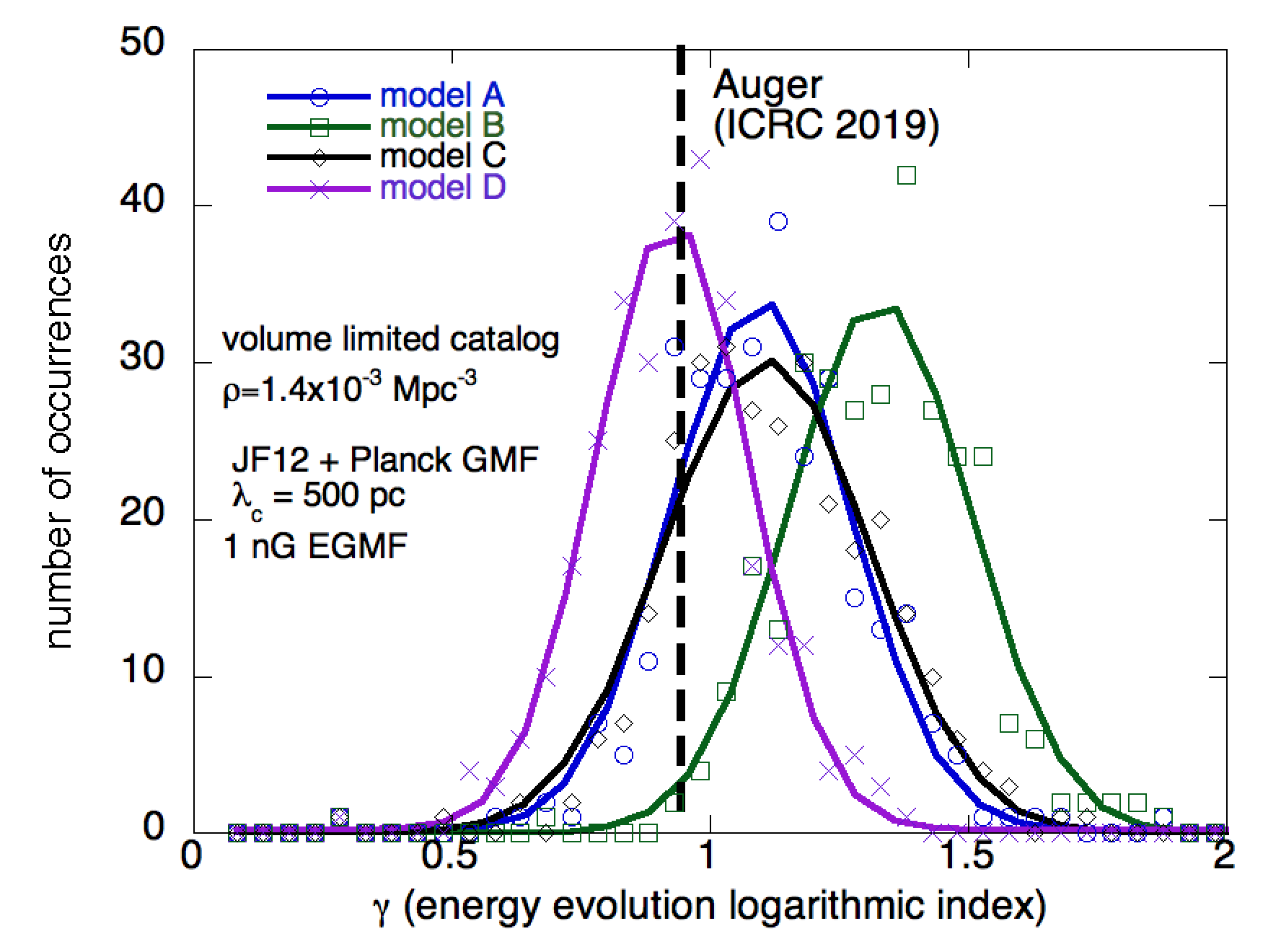}
      \caption{Top and central panels : energy evolution of the amplitude of the dipole for the four source models A, B, C and C, compared to that estimated by Auger in  four energy bins : $4\leq E<8$~EeV, $8\leq E<16$~EeV, $16\leq E<32$~EeV, and $E\geq 32$~EeV (the values are shown at the median energy of the bins). The markers show the mean values average over 300 datasets while the shaded areas shows the intervals which contain 90\% of the datasets. The scenarios considered are based on the baseline volume-limited catalog, a 1~nG EGMF and the JF12+Planck GMF with $\lambda_{\mathrm{c}}=500$~pc. Bottom panel :  histograms of the logarithmic index $\gamma$ of  the energy evolution of the dipole amplitude, extracted from a power law fit of this  evolution for each of the 300 datasets (see text). The solid lines show gaussian fits of the histograms.
              }
         \label{FigDipVsE_compo}
\end{figure}

Another key parameter influencing the amplitude of the reconstructed dipole is the composition of the UHECR. To better distinguish between different models, we now also consider the four energy bins discussed in the update of the Auger Rayleigh analysis \citep{AugerDipLong2018}, namely $4\leq E<8$~EeV, $8\leq E<16$~EeV, $16\leq E<32$~EeV, and $E\geq 32$~EeV. The reference experimental values will be taken from \citet{AugerICRC2019}. As we demonstrate below, the evolution of the dipole amplitude as a function of energy may be considered an observable in its own, and can in principle provide additional constraints on the different parameters, even though, as emphasized above, the degeneracy between the multiple influencing model parameters prevents one from drawing any conclusion regarding one of them in particular in the current state of knowledge. It is nevertheless useful to study the main consequences of specific assumptions regarding composition, and compare the general trends with the data.

The energy evolution of the dipole amplitude expected for models A, B, C and D is shown in Fig.~\ref{FigDipVsE_compo}, in the case of our baseline volume-limited catalog, a 1~nG EGMF and the JF12+Planck GMF with $\lambda_{\mathrm{c}}=500$~pc (which was shown above to give a good agreement with the data for model A). Note that the amplitude for each energy bin is represented at its median energy, more exactly the median energy found for each bin in Auger data. The models are shown on two separate panels for the sake of clarity. As can be seen from the shaded areas representing 90\% of the individual realisations with the Auger statistics and sky coverage, the four models are difficult to distinguish on the basis of a single realisation with the current statistics, even when all the other parameters in the underlying astrophysical scenario are specified, as it is the case here. However, the models do have distinct average evolution.

Model B, which corresponds to the flattest evolution of $\langle \ln A \rangle$ as a function of energy (see Fig.~\ref{FigLnA} and Sect.~\ref{sec:content}), and the lightest composition at the highest energies, shows the steepest average evolution of the amplitude and tends to overshoot the highest energy data point, which however has a large error bar. At the opposite, model D, which has the strongest evolution of the composition in the energy range of interest with the largest proton component in the ankle region, shows the flattest evolution of the dipole amplitude with energy. It is interesting to note that, for the chosen value of $\lambda_{\mathrm{c}}$, this model appears to be compatible with the observed amplitude of the dipole in the lowest energy bin, despite the large proton abundance, which is mainly responsible for the large mass dispersion predicted by this model, at odds with current data (see Fig.~\ref{FigLnA}, bottom panel). For the higher energy bins, the energy evolution is also well reproduced by this model.

As expected, the energy evolution predicted for models A and~C are very similar, which reflects the similar evolution of their average rigidity (Figs.~\ref{FigLnA} and~\ref{FigRigid}). Model~A reproduces well the observed amplitude, while model C is systematically shifted to lower amplitudes, which is in line with its heavier composition. Of course, this does not exclude model~C, but it shows that a lower value of $\lambda_{\mathrm{c}}$ would be required to best match the data with this assumption on the sources configuration. For the particular set of astrophysical parameters use in these examples, model A and~D reproduce quite well the Auger data : with a $\chi^2$ comparison with the observed values of the amplitude we find that 75 to 80\% of the realizations have a corresponding p-value larger than 0.05.

To emphasise the potential interest of the energy evolution of the dipole amplitude as a complementary observable, we have plotted in Fig.~\ref{FigDipVsE_compo} the histograms of the logarithmic index, $\gamma$, of the energy evolution of the dipole amplitude, assumed to be a power law. In other words, for each realisation of the datasets, we extract the value of $\gamma$ from a fit of the energy evolution of the amplitude by the function $a(E)=A\times E^\gamma$, where $A$ and $\gamma$ are free parameters. The solid lines in Fig.~\ref{FigDipVsE_compo} are gaussian fits of the resulting histograms (shown by light symbols), for the four composition models. Models A, C and~D appear to be compatible with the observed trend, while the evolution predicted for model B is clearly too steep (with this set of the other parameters). However, for most datasets the quadrupole modulation predicted with model~D is stronger than observed. Yet, the cosmic variance tends to alleviate this apparent tension with the data. This is also true to a lower extent for the too step evolution of the dipole amplitude in the case of model B. (NB : more details about the impact of the UHECR composition on the expected energy evolution of the amplitude are given in Sect.~\ref{sec:contribspecies}.)

\subsubsection{Influence of possible biases in the UHECR source distributions}
\label{sec:bias}
Given our current ignorance of the origin of UHECRs, a generic assumption is usually that the distribution of sources tends to reflect the distribution of matter in the universe, and thus of galaxies, as a reasonable proxy. This is the case for the source catalogs discussed above. However, it is clear that specific astrophysical scenarios may be associated with biases with respect to the general galaxy distribution. For instance, in scenarios involving regions of intense star formation (which may be considered attractive given the currently observed UHECR composition), the most massive galaxy clusters could be less favourable environments for UHECR sources, because most of the gas present in these galaxies is usually processed earlier in the cosmic history, during the cluster formation process (see {\it e.g} \citet{Guglielmo2015}). On the contrary, these regions may be regarded as preferred source locations for acceleration scenarios involving cluster accretion shocks (see {\it e.g} \citet{Inoue2007}). We thus studied the influence of such biases on the anisotropy features of our simulated datasets. To this end, we used the information provided by the Kourkchi-Tully group catalog \citep{Kourk2017} about which of the 2MRS galaxies belong to galaxy groups or associations within $\rm 3500\,km/s$, and modified the galaxy catalogs in one of three different ways: (i) excluding galaxies from the Virgo association, (ii) excluding galaxies from the Virgo and Centaurus associations (which are the two largest concentrations of galaxies within 50 Mpc), and (iii) excluding galaxies from all the groups and associations with luminous mass larger than $10^{14} \rm M_\sun$ (according to the group catalog). These cuts remove respectively $\sim 7$\%, 14\% and 25\% of the total number of galaxies with luminosities larger than $10^{10} L_{\sun}$ in the $K_\mathrm{s}$-band within 45 Mpc. Although these modifications are admittedly crude and somewhat arbitrary, they allow us to investigate the role and weight of the nearby galaxy concentrations on the expected amplitude of the anisotropies, as well as on the direction of the direction of an underlying dipole moment in the UHECRs arrival directions (see below).

%In the following, we only discuss in detail the possibility that the presence of UHECR sources are disfavored in large nearby galaxy concentrations.
 
\begin{figure}[ht!]
   \centering
   \includegraphics[width=8.2cm]{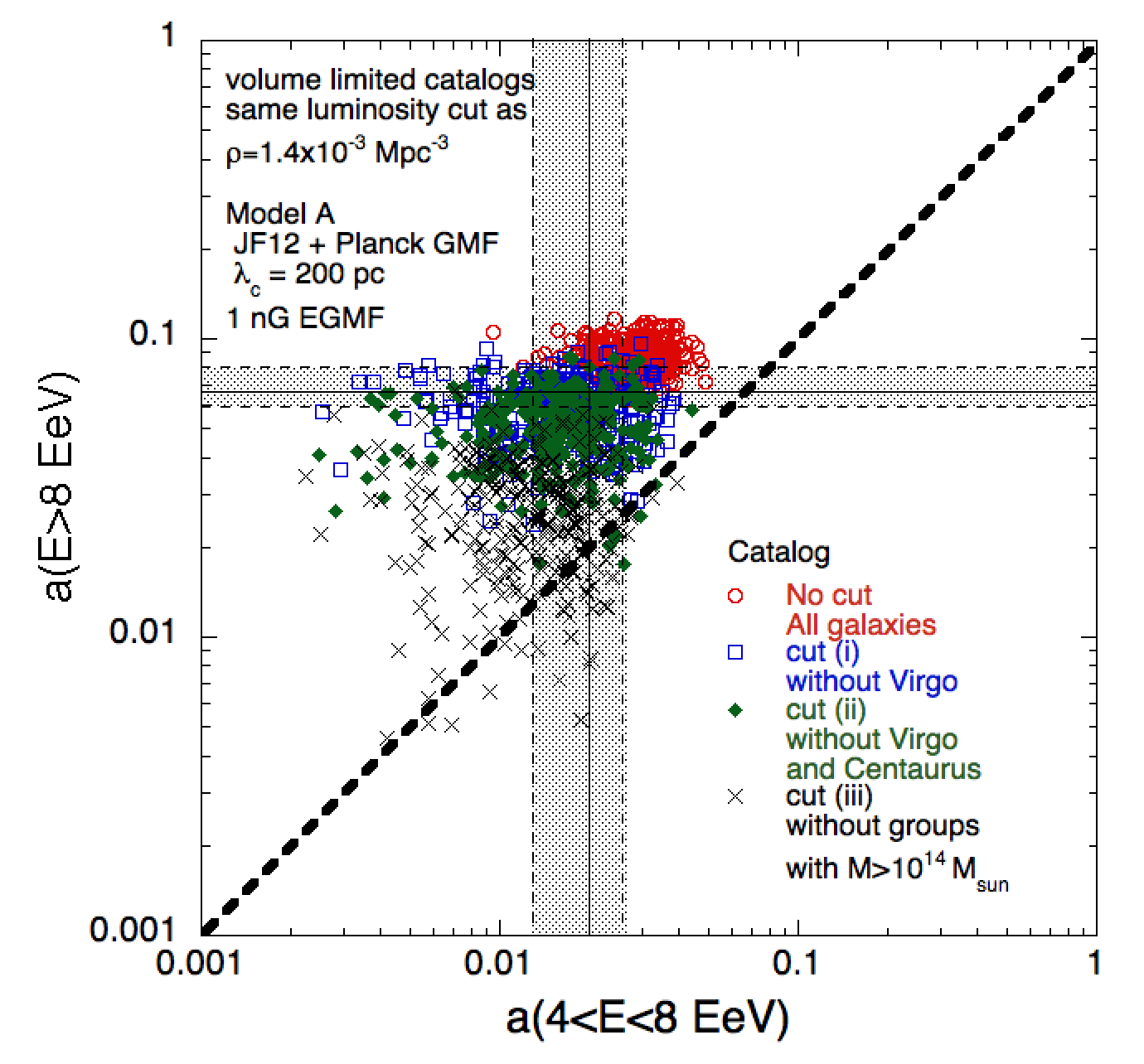}
   \includegraphics[width=7.5cm]{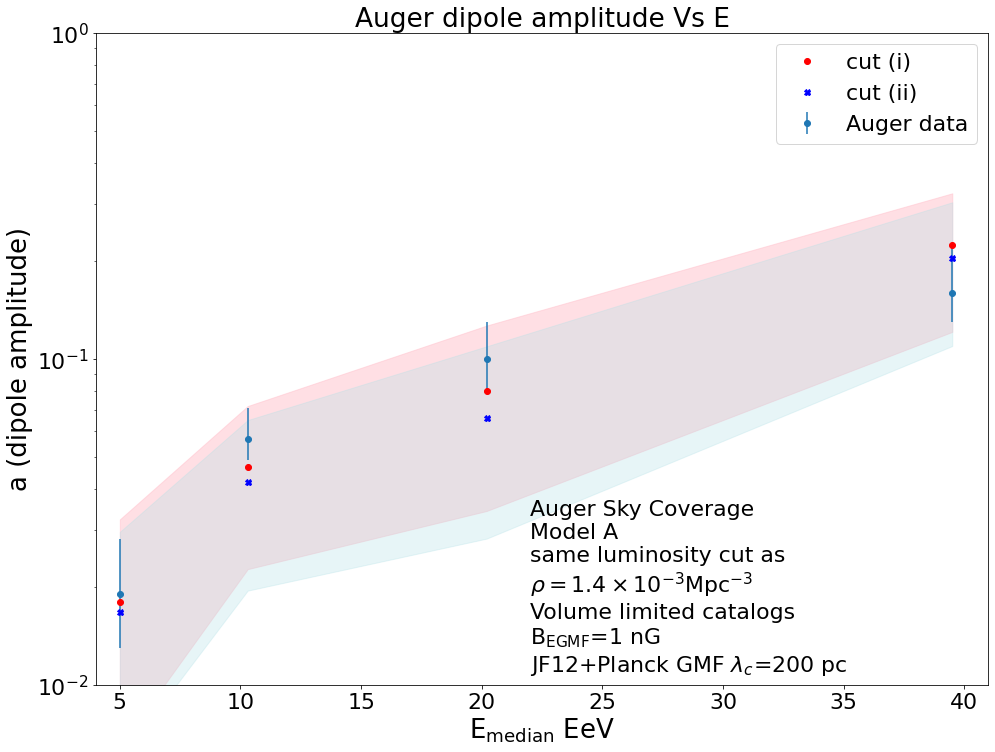}
   \includegraphics[width=8cm]{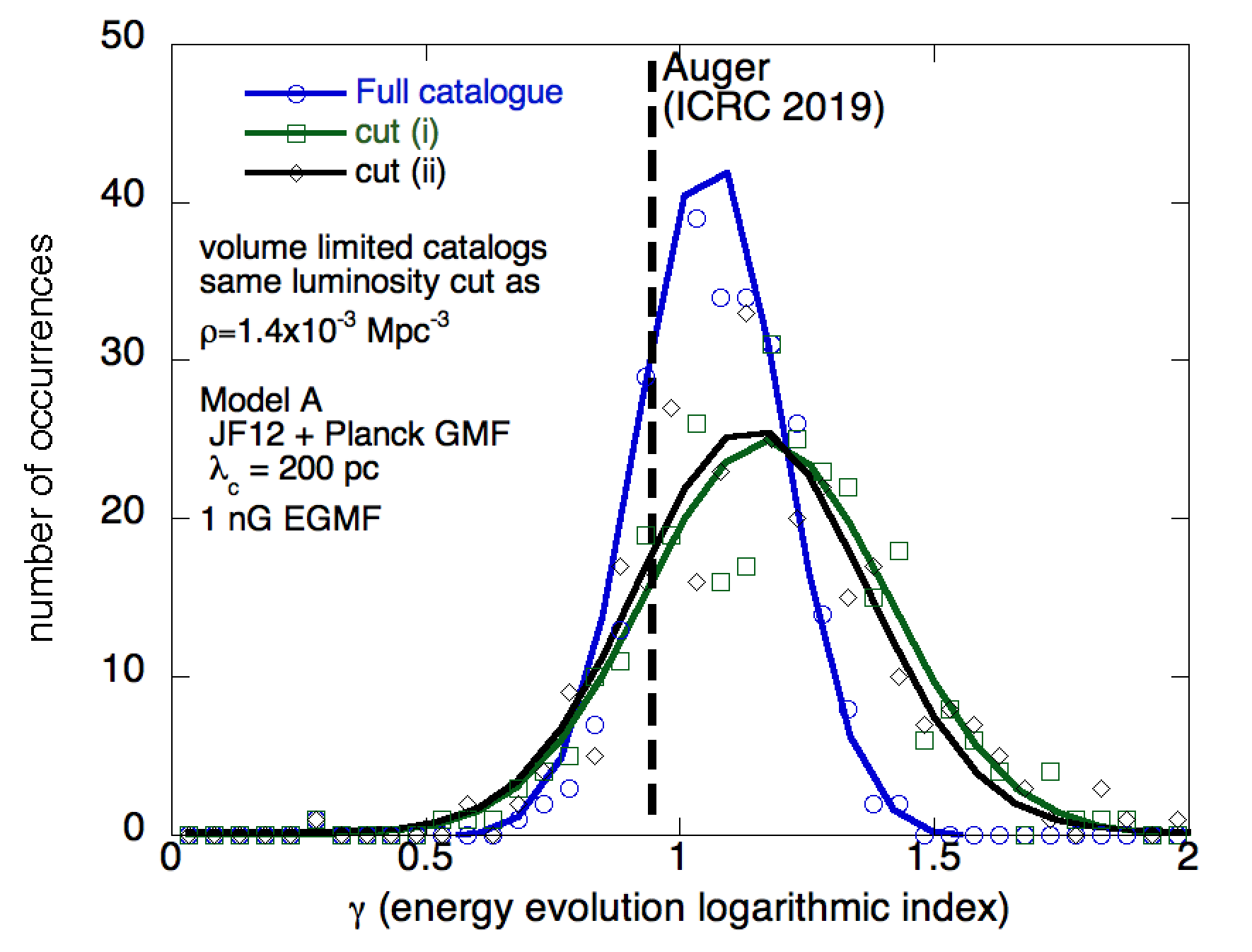}
      \caption{Top panel :  Scatter plot of the dipole amplitude reconstructed for 300 datasets computed for  scenarios based on source model A, a 1~nG EGMF, the JF12+Planck GMF with $\lambda_{\mathrm{c}}=200$~pc and assuming the various biases to the original 2MRS discussed in the text, in the energy bins $4\le E < 8$~EeV and $E\geq 8$~EeV. Central panel : Corresponding energy evolution of the dipole in the same energy bins as in Fig.~\ref{FigDipVsE_compo}, only cuts (i) and (ii) are shown for clarity. Bottom panel :  histograms of the logarithmic index $\gamma$ of  the energy evolution of the dipole amplitude for cuts (i) and (ii) compared to the case of the baseline volume-limited catalog. The solid lines show gaussian fits of the histograms.
              }
         \label{FigDipVsE_bias}
\end{figure}

%Some results are shown in Fig.~\ref{FigDipVsE_bias}, where we used

\begin{figure*}[ht!]
   \centering
   \includegraphics[width=8.5cm]{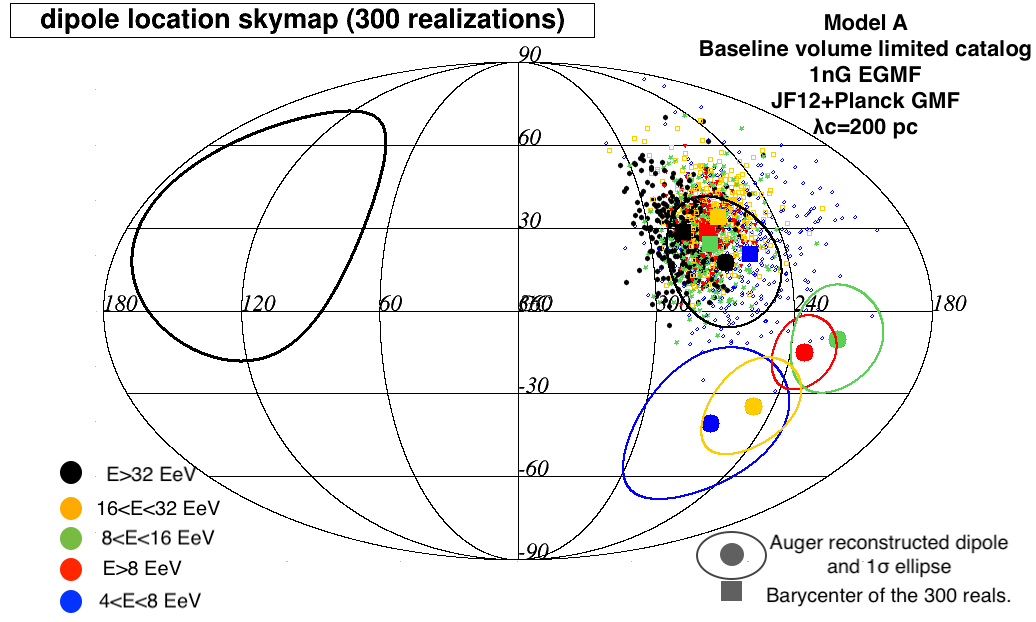}
   \includegraphics[width=8.5cm]{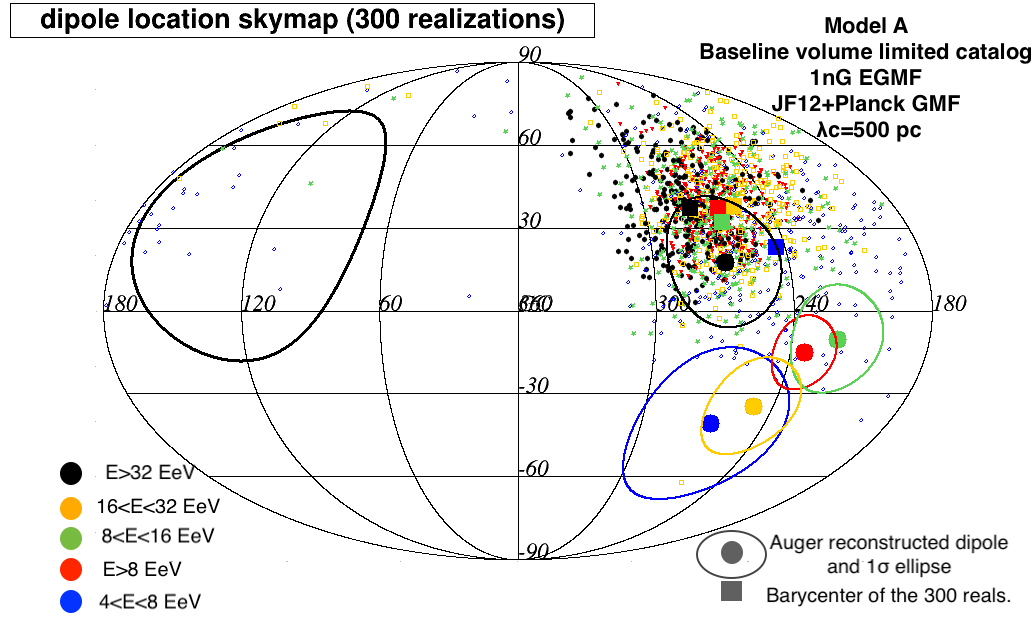}
   \includegraphics[width=8.5cm]{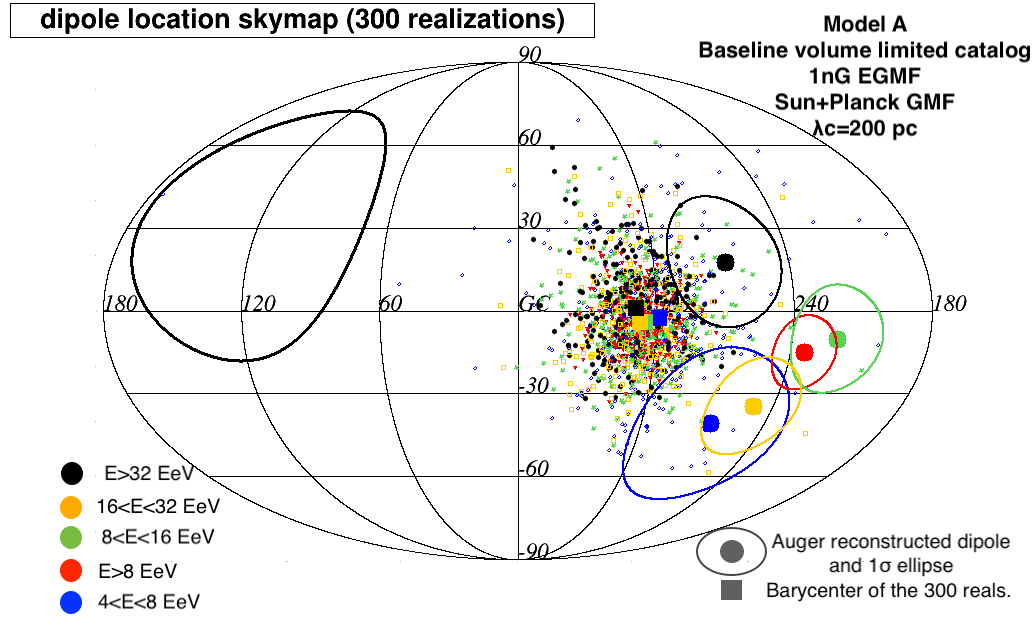}
    \includegraphics[width=8.5cm]{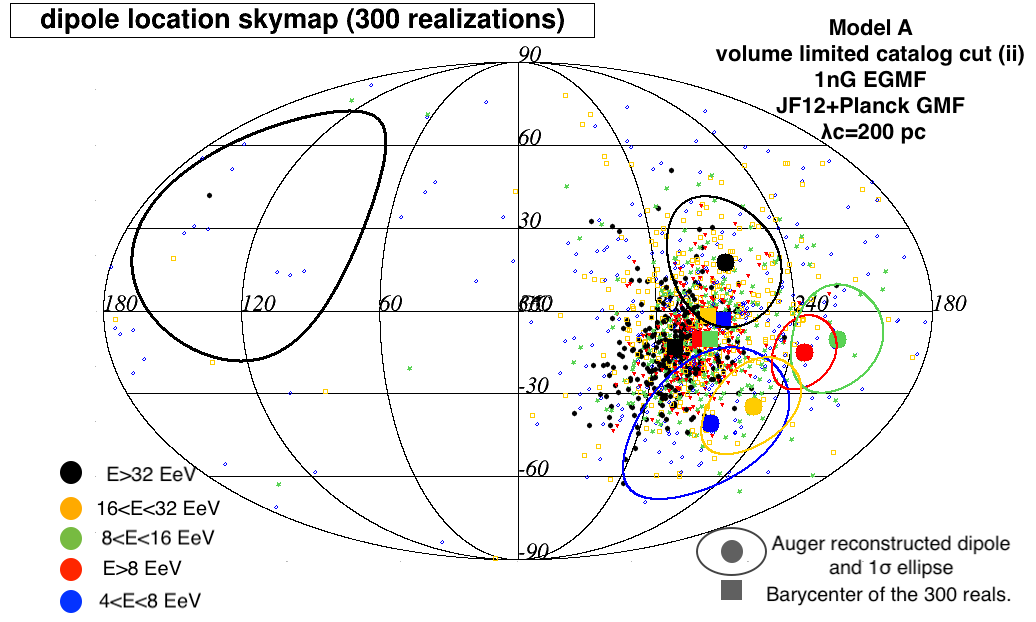}
      \caption{Skymaps in galactic coordinates showing the reconstructed directions of the dipoles  for various astrophysical scenarios for each of the 300 datasets in five different energy bins (see the color code legend and text). The barycenters of the 300 datasets are shown by large squared markers. The predictions can be compared to the directions reconstructed by Auger shown by full circles and their $1\sigma$ error ellipses. The four panels show different scenarios for the GMF model and coherence length based on source model A and 1~nG EGMF (see legend and text).
              }
         \label{FigLocation1}
   \end{figure*}

We applied these biases to build three new volume-limited catalogs, using the same initial luminosity cut as for our baseline catalog (see above). The propagated UHECR spectra obtained for cuts (i), (ii) and (iii) for a given source model are slightly softer at high energy than with the original catalog, as could be expected, with the largest difference for the largest cuts. However, the agreement with observed spectrum remains good enough so that we do not need to modify our source models, and can thus study the influence of the cuts independently of other modifications.

A scatter plot of the dipole amplitudes in the two energy bins, $4<E\leq8$~eV and $E>8$~EeV, is shown on the top panel of Fig.~\ref{FigDipVsE_bias}, in the case of source model~A, a 1~nG EGMF and the JF12+Planck GMF model with $\lambda_{\rm c}$=200~pc, for the three different cuts, as well as in the unbiased case, for comparison. As expected the amplitude of the obtained dipole decreases as the cuts become increasingly stringent. In the case of cut (iii) the result is very close to what is obtained when isotropising all the galaxies of the unbiased volume-limited catalog. The decrease of the amplitude for cuts (i) and (ii) is more moderate, but they would still require (among other possibilities of parameter adjustments) a lower value of $\lambda_{\mathrm{c}}$ to improve the agreement with the observed data, compared with the unbiased case.

Concerning the energy evolution of the amplitude, the results obtained when dividing the datasets into four energy bins is shown on the central panel of Fig.~\ref{FigDipVsE_bias} for cuts (i) and (ii). Besides the systematic shift downwards of the average values (all other parameters being fixed), the shape of the evolution is similar to that of the unbiased case, and compatible with the data as can be seen on the bottom panel showing the histogram of the evolution logarithmic index for 300~realisations. The similarity of the results obtained with cuts (i) and (ii) are the consequence of the larger expected weight of the Virgo association in the predicted anisotropy patterns. Indeed while the Virgo and Centaurus associations host a similar number of galaxies with luminosities larger than $10^{10} \rm L_\sun$ in 2MRS, the Centaurus association is located roughly two times further away from Earth.

\subsection{Direction of the reconstructed dipole}
\label{sec:direction}

The amplitude of the reconstructed dipole moment in the distribution of the UHECR arrival direction, as well as its evolution with energy, are not the only observations to account for. Successful astrophysical models of the current UHECR data must also be able to understand the direction of the dipole moment. Now, as our simulations show, this is much more difficult to reproduce.  While the observed dipole amplitude and energy evolution could be obtained for many different combinations of the source and transport parameters, the dipole directions of the simulated datasets appear at odds with the Auger findings for essentially all the investigated models, with the assumed models of the Galactic magnetic fields.

%Together with the above-discussed amplitude, the reconstructed dipole location and its energy evolution is a crucial piece of information provided by the Rayleigh analysis. While the amplitude of the dipole, even for a fixed distribution of sources, turned out to be possible to reproduce correctly for many combinations of physical parameters, it is worth studying whether it is also the case for the location, or if this observable will turn out be more constraining for our astrophysical scenarios.

The dipole direction reconstructed in the four energy bins for the 300 realizations of a given scenario are shown in the different panels of Fig.~\ref{FigLocation1}. They are compared with those estimated by Auger (shown by the filled circles at the center of the ellipses representing the 1~$\sigma$ uncertainty). Each point of a given colour shows the direction of the reconstructed dipole moment in the corresponding energy bin for one realisation (see the colour code on the plot), and the filled square marker shows the barycenter of the directions obtained for the 300 directions. The top-left panel corresponds to the baseline volume-limited catalog, model A, and a 1~nG EGMF together with the JF12+Planck model with $\lambda_{\rm c}=200$~pc for the GMF. The top-right panel shows the same scenario but with $\lambda_{\mathrm{c}}=500$~pc. As can be seen the barycenters of the reconstructed dipole directions are essentially the same, but with a larger dispersion in the case of a larger coherence length. This is in line with the smaller amplitudes already commented in the previous sections (recall that the 500~pc case is in better agreement with the data: cf. Fig.~\ref{FigDipScat1}).

On the bottom-left panel, the GMF model has been replaced by the Sun+Planck model, with $\lambda_{\mathrm{c}}=200$~pc. As for the bottom-right panel, it corresponds to the case of the biased volume-limited catalog implementing the above-mentioned cut (ii), for model A, 1nG EGMF, and the JF12+Planck model with $\lambda_{\rm c}=200$~pc.
 
It is striking that none of the models shows a satisfactory agreement with the data, especially for the energy bin with $E \ge 8$~EeV, for which the Auger measurement if the most significant. Even in the case of a 500~pc coherence length of the GMF, where the dispersion in the reconstructed dipole direction is larger, the model predictions appear incompatible with the observed data, except for the $E>32$~EeV energy bin.

Interestingly, the dipole directions reconstructed with the Sun+Planck GMF model (bottom-left panel) are significantly different, being shifted towards the Galactic plane. We obtained a somewhat similar behaviour with the JF12 model, when switching off the X-shaped halo field. This illustrates how much a change in the GMF model can impact the expected dipole direction. However, in both cases, the latter remains far from the observed directions. Unfortunately, in the current state of knowledge, it would be careless to conclude either that the usually favoured GMF models need major revisions, or that the actual UHECR sources simply do not fit in the range of astrophysical scenarios investigated here for the origin of UHECRs. Nevertheless, it is important to realise that any of the above combinations appears to be rather strongly disfavoured by the data.  NB: a more complete discussion of this issue also requires a more systematic investigation of the cosmic variance, which will be addressed in Sect.~\ref{sec:cosmicVariance} below. 

Finally, in the case of the biased catalog shown in the bottom-right panel, the directions of the dipole moment are also systematically shifted towards the galactic plane, closer to the observed ones, which illustrates how the contribution (or absence of contribution) of the  Virgo region impacts the reconstructed dipole direction when all the other parameters are fixed. As for the reconstructed amplitude, the cuts (i) and (ii) yield very similar results in terms of dipole direction, again mostly due to the larger contribution of the Virgo association compared to that of the Centaurus association. It is also the case for cut (iii) when a lower value of $\lambda_{\rm c}$ is chosen to better match the observed amplitude (which would be best achieved with a coherence length between 20 and 50 pc for the JF12+Planck GMF). Note that the use of the biased catalog has much less impact on the dipole direction in the case of Sun+Planck GMF model, which can be understood as a consequence of the fact that the Virgo region is strongly demagnified for this particular GMF model in the rigidity range of interest. Finally, we found that the results shown in Fig.~\ref{FigLocation1} are not modified when we use either model B, C, or D instead of model A. Quite generally, we found very little influence of the details of the source composition model on the resulting dipole direction, especially at the current level of statistics of the UHECR dataset.

\begin{figure}[t!]
   \centering
   \includegraphics[width=\hsize]{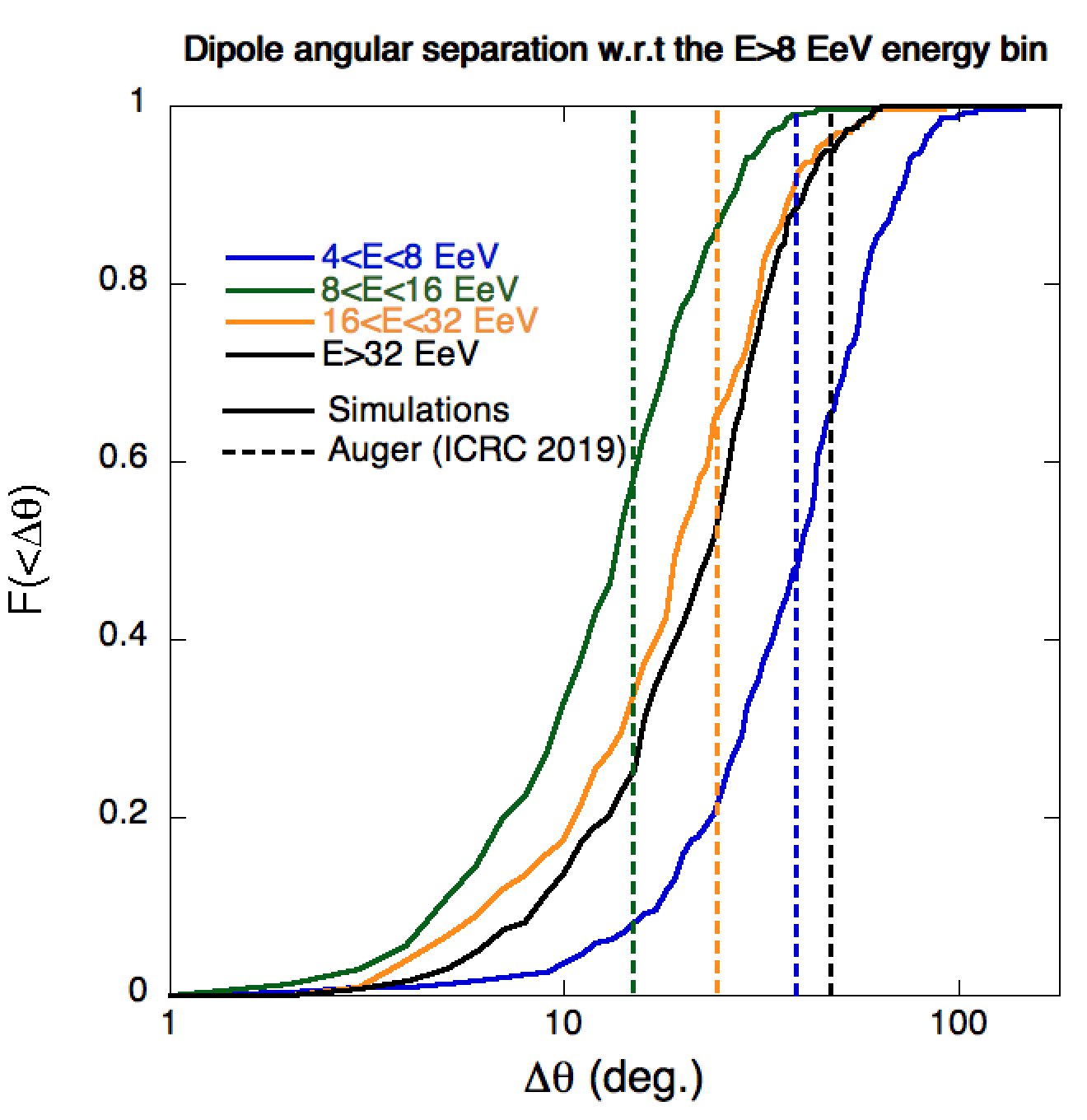}
      \caption{Cumulative distributions built over 300 datasets of the angular distance between the dipole direction reconstructed in a given energy bin and that reconstructed in the $E\geq 8$~EeV energy bin. The actual angular distances implied by Auger measurements are shown with dashed lines.
              }
         \label{FigAngDistDip}
\end{figure}

Concerning the energy evolution of the dipole moment direction, one notices that in all cases, the barycenters of the 300 realizations for each energy bins are relatively close to each other, say within 15 degrees. This gives the impression of a very small displacement of the dipole between the different energy bins, contrary to what can be observed on the Auger data. However, the differences in the dipole moment direction of each given realisation are much larger, with the current statistics. In this respect, it is interesting to compare the simulated data sets with the Auger measurements on a single realisation basis. This is done in Fig.~\ref{FigAngDistDip}, which shows the cumulative distribution function of the angular distance between the reconstructed dipole directions in a given energy bin and in the $E>8$~EeV energy bin, for all 300 realisations individually. The measured value for the Auger dataset is shown with a vertical dashed line, for each of the energy bins (indicated by different colors). One sees that the angular distance found by Auger are well within the distributions obtained with 300 realization, except for the $E>32$~EeV energy bin where the observed value is comparable with only the $\sim 5 \%$ largest ones in our simulations.

However, our simulations show that, as the statistics increases, smaller angular distances are expected between the reconstructed dipole moment directions in different energy bins, at least for the investigated scenarios, where all the cosmic-rays above the ankle have the same origin. Quantitatively, we found that for a statistic twice as large as that reported in the Auger dipole discovery paper, distances as large as those currently found for different energy bins become quite unlikely: the latter would be compatible with only $\sim 1$\% of the realisations (without additional hypotheses). 

With increasing statistics, this property may actually be turned into a criterion for the viability of single component scenarios, in contrast to scenarios where different types of sources, with different spatial distributions, provide a dominant contribution to the UHECR flux in different energy ranges above the ankle.

%At this stage, the disagreement of our predictions concerning the dipole location using our baseline volume-limited catalog and the observations is quite striking, even though it can be argued that the predictions for the two different GMF models show considerable differences. A full discussion however requires the discussion of the impact of cosmic variance and is thus postponed to later sections.  

\begin{figure}[t!]
   \centering
   \includegraphics[width=\hsize]{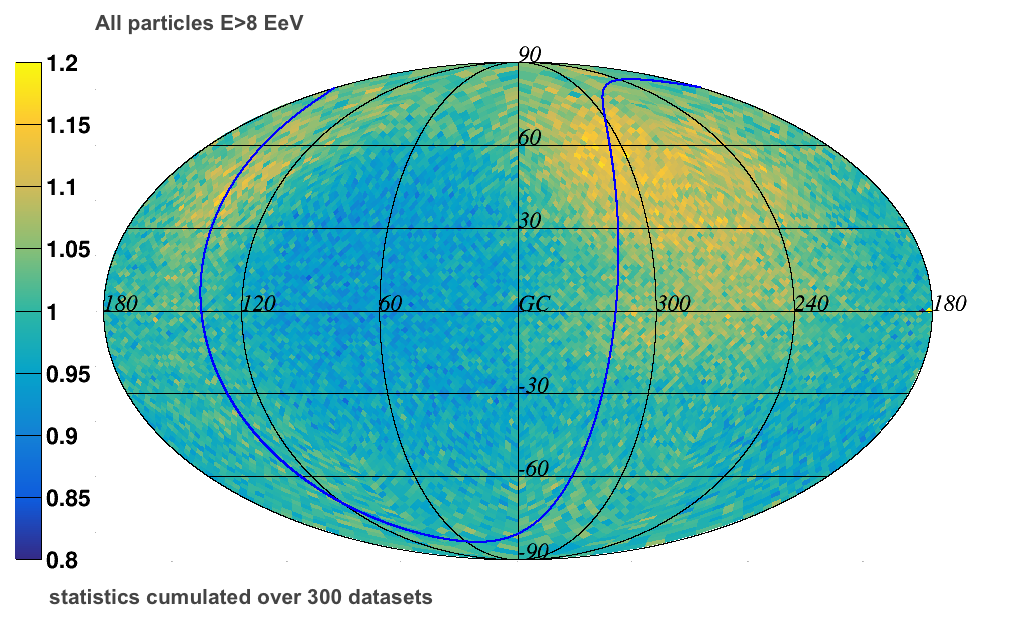}
      \caption{Density UHECR skymap obtained by summing 300 datasets (uniform full-sky exposure) for the astrophysical scenario assuming the baseline volume-limited catalog, source model A, a 1~nG EGMF, the JF12+Planck model with $\lambda_{\mathrm{c}}=500$~pc. The color scale is linear.
              }
         \label{FigSkymap1}
\end{figure}

\subsection{Expectations for Telescope Array}

So far, TA has not reported the measurement of a dipole modulation the northern hemisphere sky, such as that observed by Auger, but has accumulated a smaller dataset. It is thus interesting to examine whether the current absence of detection by TA is providing additional information or whether it can be attributed to the sole lower statistics, and to estimate the perspectives for a similar measurement when TA will have have collected larger datasets.

To address this question, we have simulated TA-like datasets (i.e. with the TA sky coverage) for two different statistics: {\it(i)} 4321 events above 9 EeV (\citet{AugerTAan2020} after rescalling down the TA energy scale by $\sim10\%$) and  {\it(ii)} 32187 event above 8~EeV (equivalent to the statistics that we used for Auger), in the case of an astrophysical scenario based on our baseline volume-limited catalog, model A, a 1~nG EGMF and the JF12+Planck EGMF with $\lambda_{\mathrm{c}}=500$~pc, for which the energy evolution of the dipole amplitude was shown to be in good agreement with the Auger data, when applying the Auger sky coverage. 

A quick look at the simulated high-statistics UHECR skymap obtained by adding up our 300 different simulations with full sky coverage, as displayed in Fig.~\ref{FigSkymap1}, reveals a potentially interesting difference between the Auger and TA skies, in the framework of the models under consideration. Indeed, the region around the direction of the M81 galaxy (barring M82, which is not present in the assumed volume-limited catalog, since it does not pass the corresponding $K_\mathrm{s}$-band luminosity cut) is associated with a significantly higher flux, both due to the very small distance to this source ($\sim 3.5$~Mpc) and to the significant magnification by the GMF (for both the JF12+Planck or the Sun+Planck models) in the rigidity range of interest. Yet, this region lies outside the Auger sky coverage. One might thus expect a significant anisotropy signal to manifest in the TA sky, already with smaller datasets. However, the study of the expected dipole amplitudes shows that there is no apparent conflict with the current data (assuming that there is indeed no significant dipole signal in the TA data, as the absence of a positive detection report may suggest).

%It is interesting to note that some differences may be naturally expected between the Auger and TA skies. This can be anticipated by studying the density skymap obtained by summing UHECR events above 8 EeV from 300 datasets assuming a uniform  full-sky  exposure  for  the  above cited astrophysical scenario displayed in Fig.~\ref{FigSkymap1}. Indeed, the region of M81\footnote{M82 does not pass the $K_\mathrm{s}$-band luminosity cut of the baseline volume-limited catalog.}, which is quite prominent on this skymap due to the fact that this source is very nearby and moreover magnified by the GMF (for both the JF12+Planck or the Sun+Planck GMF models) in the rigidity range of interest, lies outside of the Auger sky coverage. 

\begin{figure*}[ht!]
   \centering
   \includegraphics[width=8.5cm]{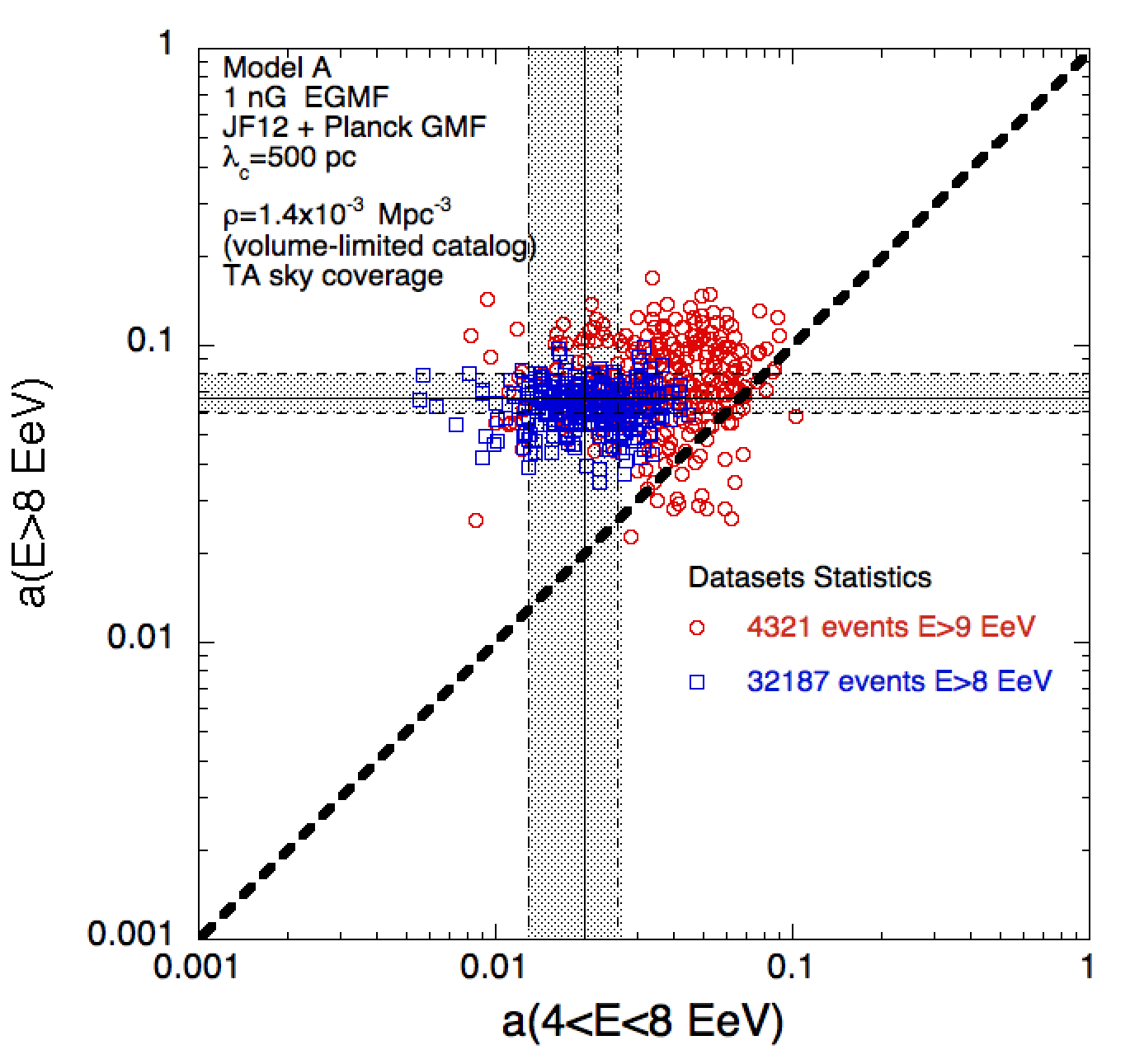}
   \includegraphics[width=8.5cm]{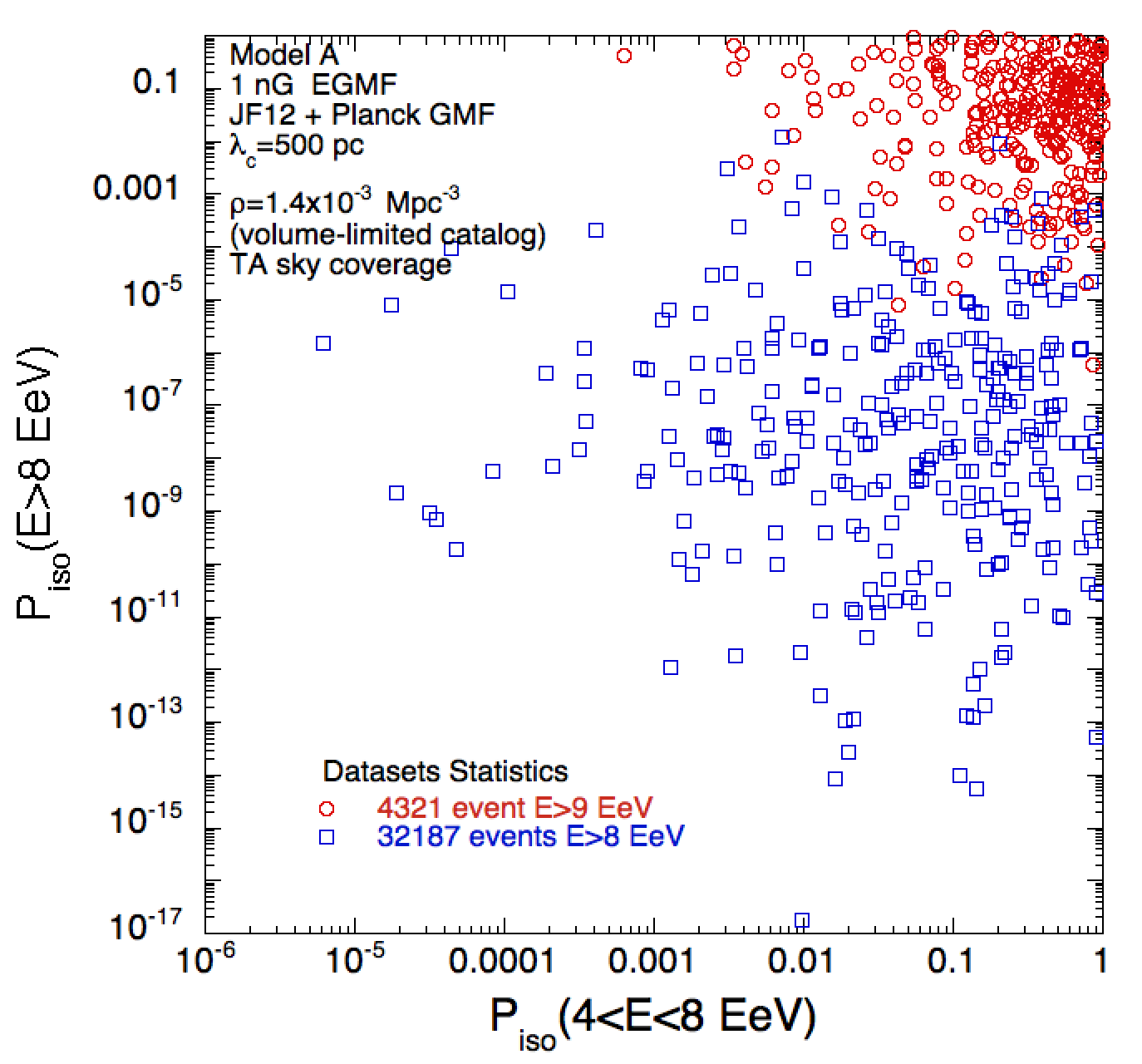}
   \includegraphics[width=8.5cm]{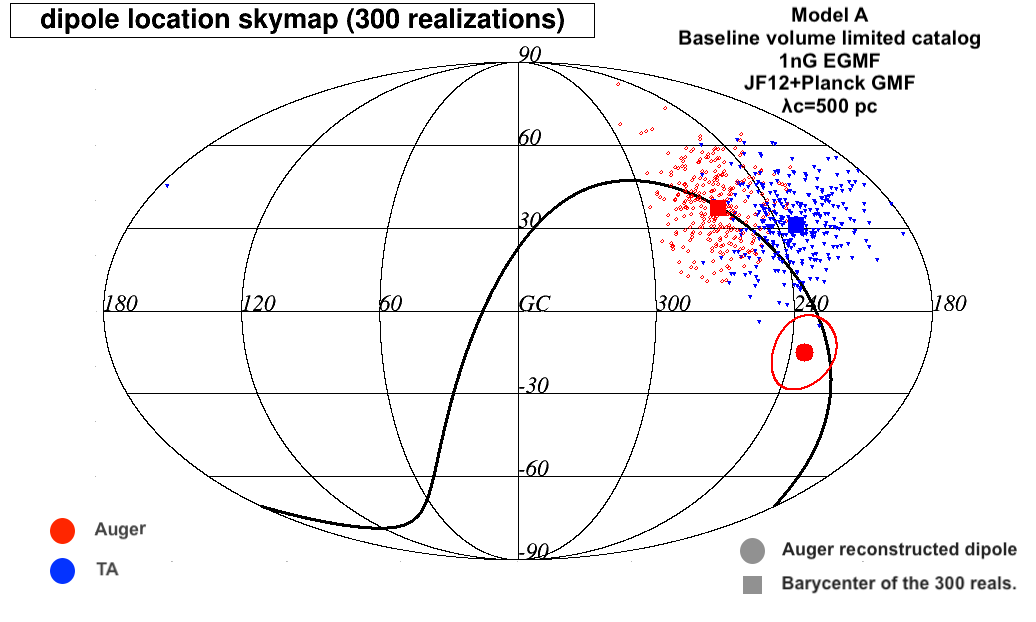}
    \includegraphics[width=8.5cm]{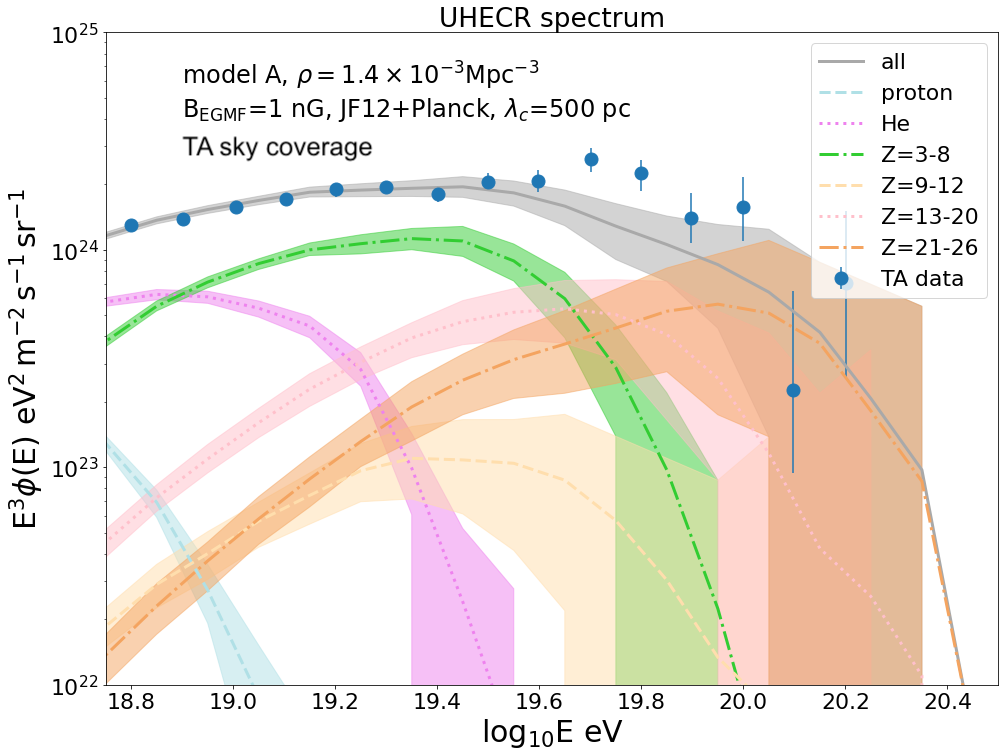}
    
      \caption{Top-left :  scatter plot of the dipole amplitude reconstructed for 300 datasets computed for an astrophysical scenario based on the baseline volume-limited catalog, source model A, a 1~nG EGMF and the JF12+Planck GMF with $\lambda_{\mathrm{c}}=500$~pc in the energy bins $4 \le E < 8$~EeV and $E\geq 8$~EeV assuming the TA sky-coverage. Two different statistics are considered : 4321 events above 9~EeV and 32187 events above 8 EeV. The $1\sigma$ confidence intervals estimated by Auger are shown by the shaded area. Top-right : Scatter plot of the corresponding p-values.  Bottom-left : Skymap in galactic coordinates showing the reconstructed directions of the dipoles reconstructed for each of the 300 datasets assuming Auger (red) and TA (blue) sky coverage. The statistics of the datasets is for both cases 32187 events above 8 EeV. The portion of the sky out of TA field of view lies below the thick black line. Bottom-right : UHECR energy spectrum obtained after propagation in the extragalactic and Galactic media The lines show the mean expected fluxes averaged over 300 realizations of the TA dataset (4321 events above 9 EeV) and the shaded areas cover the range in which 90\% of the realizations are found. The different colors show different groups of nuclear species. The obtained average spectrum is compared to the one measured by TA \citep{TASpec2020} after rescaling its energy scale down by 10\%.    
              }
         \label{FigDipTA}
   \end{figure*}

This can be seen on Fig.~\ref{FigDipTA}, where the top-left panel shows the scatter plot of the reconstructed dipole amplitudes in the two energy bins $4\leq E < 8$~EeV and $E\geq 8$~EeV, for the two above-mentioned statistics (one symbol for each of the 300 TA-like datasets), while the top-right panel shows the associated p-values. The plots clearly show that the datasets built with the current TA statistics (red circles) are dominated by statistical fluctuations, and the associated p-values do not show any significant departure from isotropy. In other words, for our models the (presumed) lack of a significant dipole modulation signal in the TA UHECR dataset at that level of statistics does not introduce any tension with the signal (in particular the dipole amplitude) observed by Auger. On the other hand, the results obtained assuming a statistics similar to that of Auger (blue points) show not only amplitudes very close to those obtained earlier for Auger sky coverage, but also very similar associated p-values.

Regarding the reconstructed dipole direction, the bottom-left panel of Fig.~\ref{FigDipTA} compares the dipole directions obtained after apply either the Auger sky coverage (in red) ot the TA sky coverage (in blue), to the above datasets (with Auger-like statistics). One sees a slight overall shift of the distribution of the reconstructed dipole directions in the case of the TA coverage, in the direction of the concentration of events influenced by M81/82. This is in line with the above comment regarding this particular source, not directly visible in the Auger sky.

Finally, we found that in spite of the different coverage and of the specific contribution of the M81/82 region, the measurements of the UHECR spetrum by Auger and TA should also be similar for the considered scenario. This can be seen on the bottom-right panel of Fig.~\ref{FigDipTA}, where the UHECR spectrum obtained after averaging over 300 TA datasets (with the TA statistics) is shown. Apart from the larger dispersion, due to the lower statistics, the spectrum is very similar to that previously obtained for the Auger sky coverage. In particular, the spectra that we obtain do not allow to match the well known excess present in the TA spectrum above $\sim 50$~EeV.  \footnote{Note that this excess seems to remain even though an energy dependent systematic error between the Auger and TA spectra (which is not accounted for in Fig.~\ref{FigDipTA}), as was pointed out by recent joint analyses \citep{AugerTAspe2019}}. It is worth noting that this apparent discrepancy with the data could be somewhat alleviated by significantly doping the assumed luminosity of M81 and/or M82 in our catalogs. In that case, the consequences on the expected anisotropy level (in particular the predictions for the dipole modulation) should become noticeable in the TA dataset already with a moderate increase of statistics, while the contribution of the M81/82 region to the Auger dataset would increase but remain very diffuse, thus not strongly modifying the predictions already discussed above. Note that these considerations are also related to the anisotropy expectations at higher energies, which will be discussed in a companion paper.

%\subsection{Partial contributions to the anisotropy signal}

%To that purpose, we take advantage of the possibility offered by the use of volume-limited catalogs, that is the fact of working with a fixed source distribution, to group the statistics of the 300 different datasets to build larger statistics "super datasets" whenever needed.   

%\subsubsection{Contribution of various source distances}
\subsection{Contribution of various source distances}
\label{sec:distance}
We have seen above that, barring the actual direction of the dipole moment, the observed dipole amplitude and its energy evolution are relatively easy to reproduce with an astrophysical scenario where the UHECR sources are globally distributed like the extragalactic matter, with a source composition allowing to reproduce the observed trend towards heavier nuclei at the highest energies. In this framework, it is interesting to investigate how this anisotropy signal builds up, i.e. what is the contribution of sources from different distances or of different groups of elements. While this cannot be done on the actual data, it is easy to isolate different contributions to the total UHECR flux in the simulated datasets. This is the subject of our discussions in the present section and the following one.

\begin{figure*}[ht!]
   \centering
   \includegraphics[width=8.5cm]{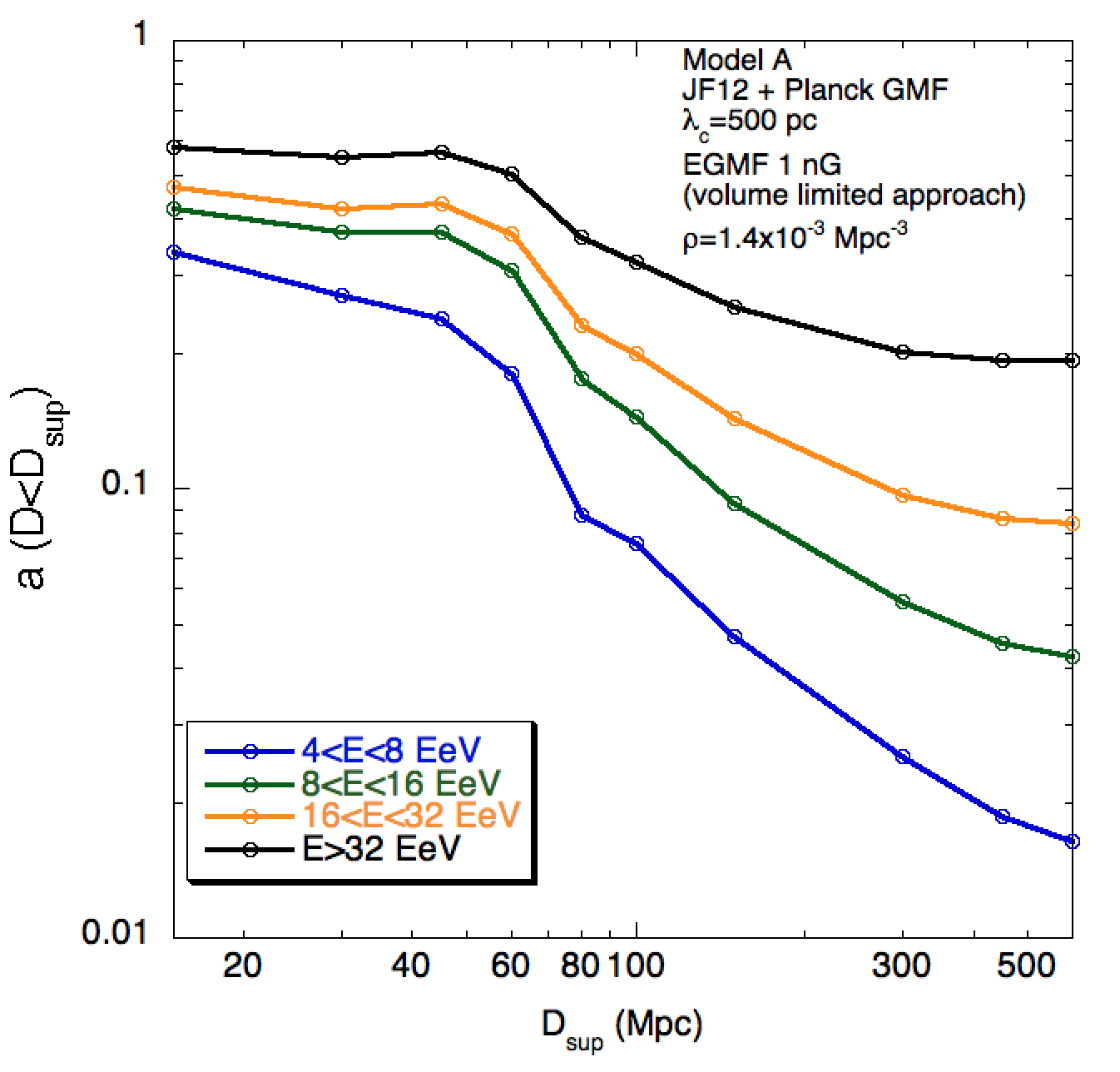}
   \includegraphics[width=8.5cm]{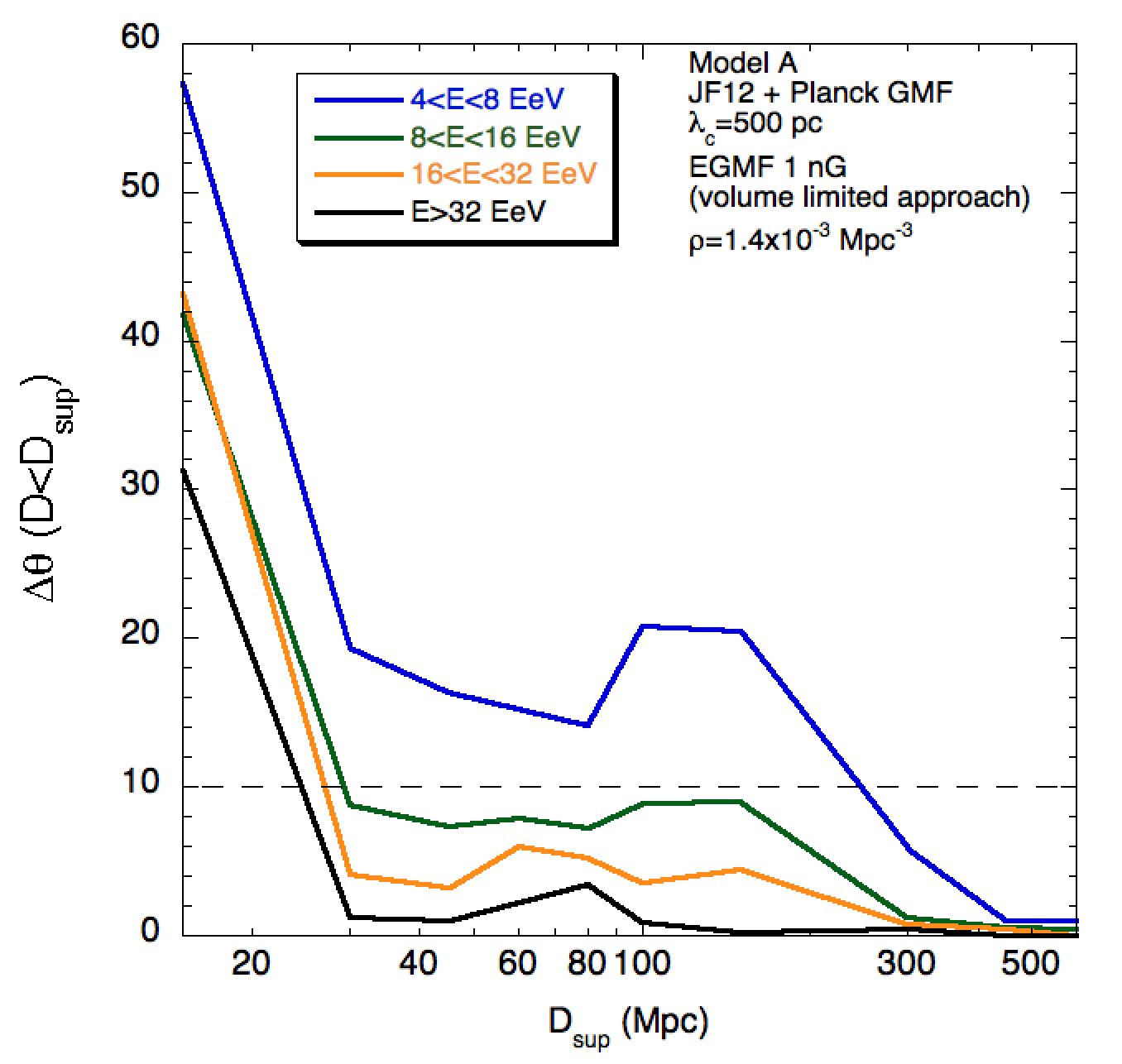}
   \includegraphics[width=8.5cm]{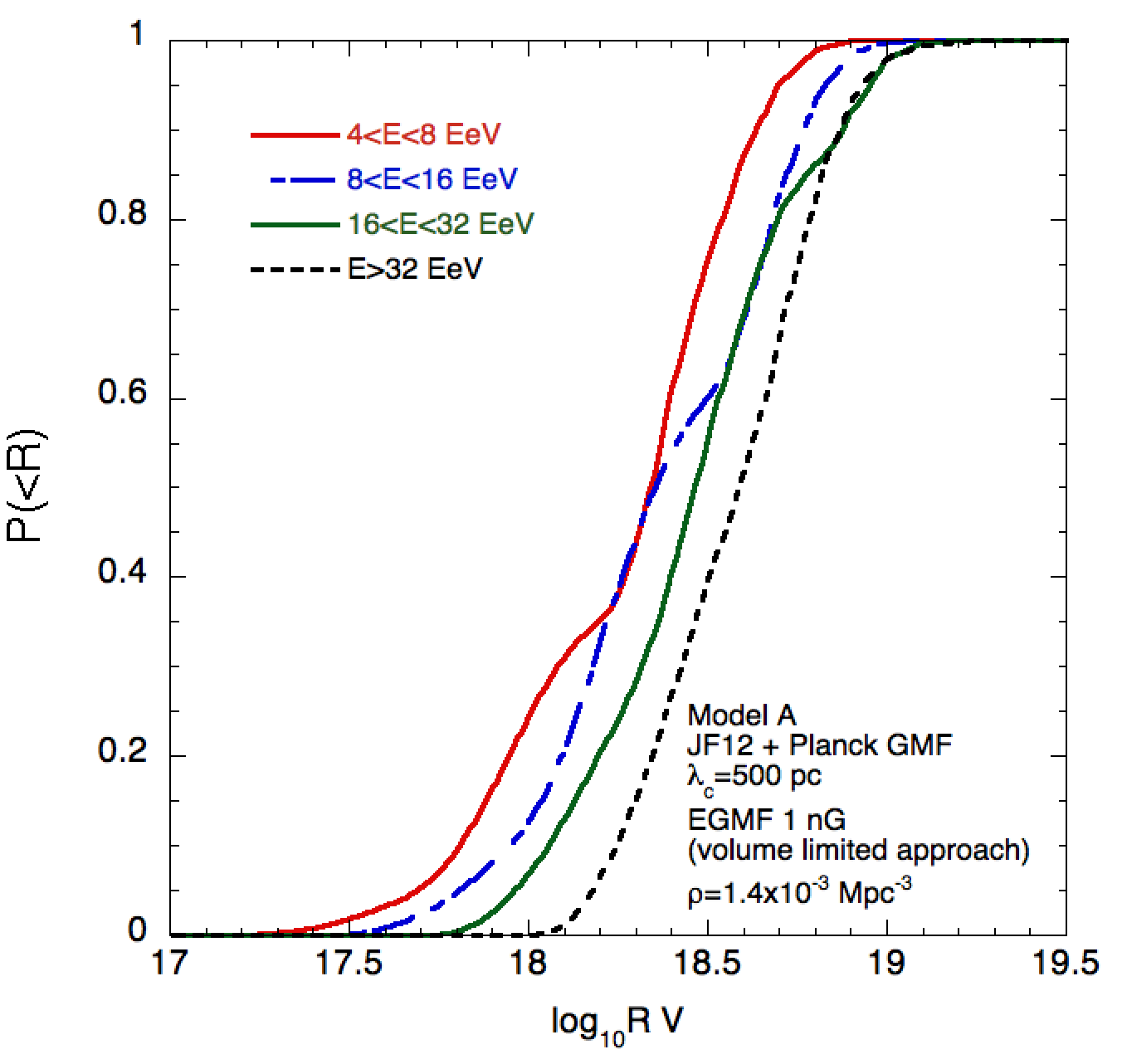}
    \includegraphics[width=8.5cm]{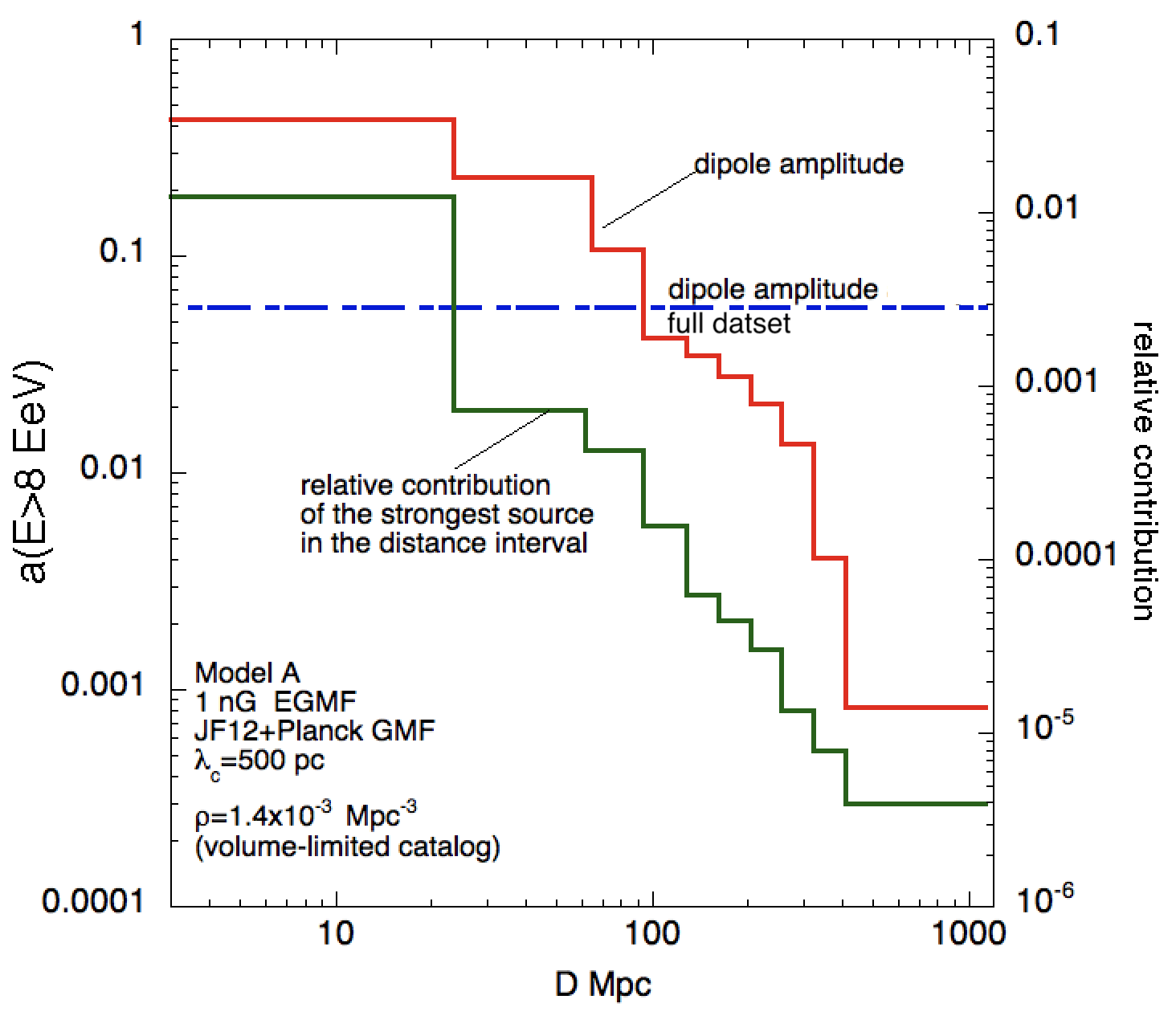}
      \caption{Top-left : Evolution of the dipole ampole amplitude with the distance $D_{\rm sup}$ (see text) for different energy bins. Top-right : Evolution of the angular distance between the dipole direction reconstructed considering only cosmic-rays originating from sources within a distance $D_{\rm sup}$ and that reconstructed for the full dataset as a function of $D_{\rm sup}$, for different energy bins. Bottom-left : Cumulative distribution of the UHECR rigidities in different energy bins. Bottom-right : amplitude of the dipole reconstructed in ten distance intervals each accounting for $\sim 10\%$ of the total flux in the $E\geq 8$~EeV energy bin. These amplitudes can be compared to that obtained for the full dataset (shown by a dashed-dotted line). The fractional contribution of the strongest source to the total flux in each distance interval is also shown. In all the panels, the astrphysical scenario considered is the baseline volume-limited catalog, source model A, a 1~nG EGMF, the JF12+Planck model with $\lambda_{\mathrm{c}}=500$~pc and the statistics is cumulated over 300 Auger datasets.     
              }
         \label{FigDipDist}
   \end{figure*}

To better understand the contribution of different distance bins to the expected anisotropies for each astrophysical scenario, we group the 300 realisations of the corresponding simulated datasets into a single, high statistics dataset, and compute the dipole amplitude in the same four energy bins as above obtained when selecting only the UHECRs reaching the Earth from sources closer than a given distance $D_{\rm sup}$. The result is shown in the top-left panel of Fig.~\ref{FigDipDist}, as a function of $D_{\rm sup}$, in the case of the baseline volume-limited catalog, with a 1 nG EGMF, the JF12+Planck GMF and $\lambda_{\rm c} = 500$~pc. By construction, the dipole amplitudes reach their full dataset values as $D_{\rm sup}$ tends to infinity. In practice, what takes the place of “infinity” is the GZK horizon distance (influenced by the EGMF) in the considered energy bin, which is smaller and smaller as the energy increases. This is why the dipole amplitude keeps decreasing up to larger distances for the lower energy bins, which results in the observed divergence of the different curves as $D_{\rm sup}$ increases. This is a generic expectation for extragalactic source scenarios, as already pointed out by several studies (see {\it e.g} \citet{Harari2015, No2017, Witt2018, Matteo2018}). Indeed, the anisotropy signal is mostly due to the contribution of the very anisotropic nearby universe, whose relative contribution to the UHECR flux increases as the energy increases (i.e. as the horizon shrinks).

As far as lower values of $D_{\rm sup}$ are concerned, i.e. when selecting only the most nearby sources well within the GZK horizon at all energies, the difference in the dipole amplitude in the different energy bins is mostly due to the different UHECR rigidities. However, the impact on the anisotropy level is less strong than that cause by the horizon effect at large values of $D_{\rm sup}$, because the evolution of the rigidity distribution with energy is relatively moderate, as a consequence of the gradual increase of the mean cosmic ray charge. This is shown in the bottom-left panel of Fig.~\ref{FigDipDist}, where one sees a relatively small shift towards the right (as energy increases) of the rigidity cumulative distribution functions. A clear demonstration of the transition from a rigidity effect to a horizon effect in the difference between the dipole amplitudes in different energy bins is that if one selects only the UHECRs with the same rigidity, say $3<R<4$~EV, in each of the four energy bins (i.e. UHECRs of larger charge at higher energy), one finds that the four curves are very close to one another for low values of $D_{\rm sup}$, and then progressively diverge as $D_{\rm sup}$ increases.

Note that the relatively flat evolution of the amplitude up to $\sim 60$~Mpc is due to the spatial configuration of the structures in the local universe and is in general not observed when sources are drawn randomly from an homogenous and isotropic distribution.  On the contrary, the evolution for larger values of $D_{\rm sup}$ remains similar in these cases.

We performed a similar study for the reconstructed dipole direction, rather than its amplitude. The result is displayed on the top-right panel of Fig.~\ref{FigDipDist}, where we show the angular distance between the dipole direction reconstructed from the sub-dataset of events coming from sources closer than $D_{\rm sup}$, and the dipole direction obtained with the entire dataset. Obviously, that angular distance goes to zero as $D_{\rm sup}$ goes to "infinity" (i.e. the horizon scale). Here again the effect of the energy evolution of the UHECR horizon and the predominance of the local universe in the building up of the dipole anisotropy is visible. For the energy bins above 8~EeV, the full-dataset dipole direction does not change by more than 10 degrees or even less whether one takes into account or not the sources further away than 30~Mpc. For the lower energy bin, however, the horizon scale is so large that distant sources influence the direction of the overall dipole significantly (but with a lower amplitude eventually). The plot also shows that the dipole that would result if one selected only the sources closer than 15~Mpc is relatively far from that of the full dataset, which indicates that the most nearby sources do not have a strong impact on the overall dipole location, at least for the scenarios considered here.

%A dipole location with 10 degrees of the all sky one is basically obtained for values of $D_{\rm sup}$ below 50~Mpc for the energy bins above 8~EeV. We however note that the dipole obtained from the contribution of the sources in the very local universe,  below 15 Mpc remains relatively far from the all-sky dipole location, indicating that this very contribution does not have a strong impact on the actual dipole location. Unsurprisingly, for the lower energy bin, the all-sky dipole location is less correlated that obtained at low $D_{\rm sup}$, although it remains close to it.   

Finally, to further illustrate the evolution of the dipole signal with the source distances, we divided the high statistics dataset into ten distance intervals contributing an equal flux to the energy bin above 8~EeV and calculated the dipole amplitude for each interval. The result is shown in the bottom-right panel of Fig.~\ref{FigDipDist}. A large predominance of the three first intervals, encompassing $\sim 30$\% of the total flux below 100 Mpc, can be seen. The effect of the intervals at larger distance is mostly to the weaken the anisotropy signal. This is because the number of sources increases with the distance while their relative contribution decreases and the angular distance between enighbouring sources becomes rapidly comparable to or smaller than the angular size of their UHECR image. 

Note that while the quantitative values that appear in Fig.~\ref{FigDipDist} are representative only of the particular astrophysical scenario displayed, the qualitative discussion about the predominance of the local universe (well within 100~Mpc) in the building up of the dipole anisotropy signal holds even for scenarios with a lower value of the EGMF, a lower source density, and a lower value of $\lambda$. For instance, we have investigated the influence of randomly redistributing all the sources beyond 50~Mpc (drawing their position from an underlying isotropic distribution), and compared the resulting dipole amplitudes and directions with those obtained from the original catalogs. In most cases, we did not find any noticeable difference in the results for the $E>8$~EeV energy bin (with the Auger statistics).

\begin{figure}[ht!]
   \centering
   \includegraphics[width=7.5cm]{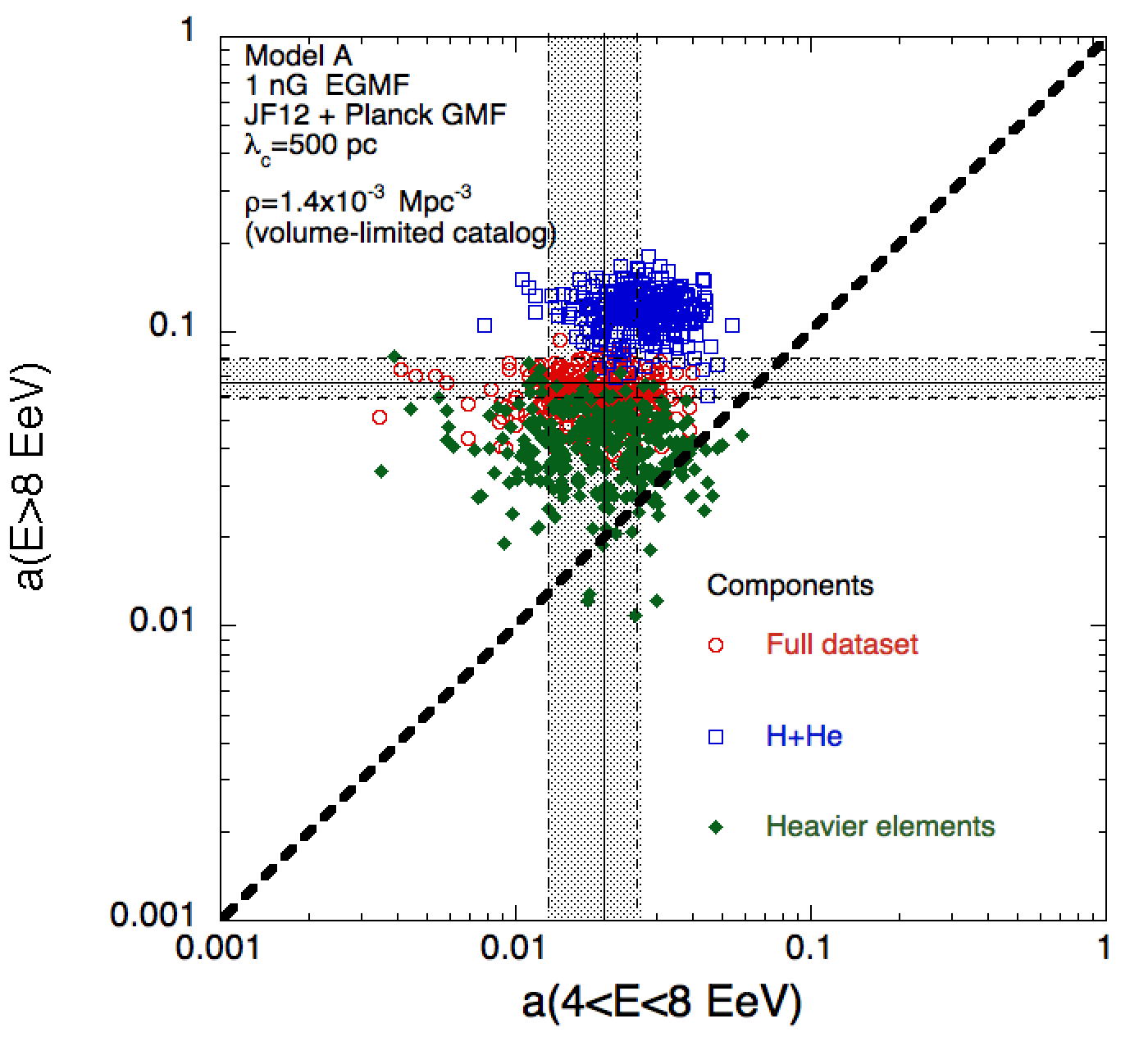}
   \includegraphics[width=7.5cm]{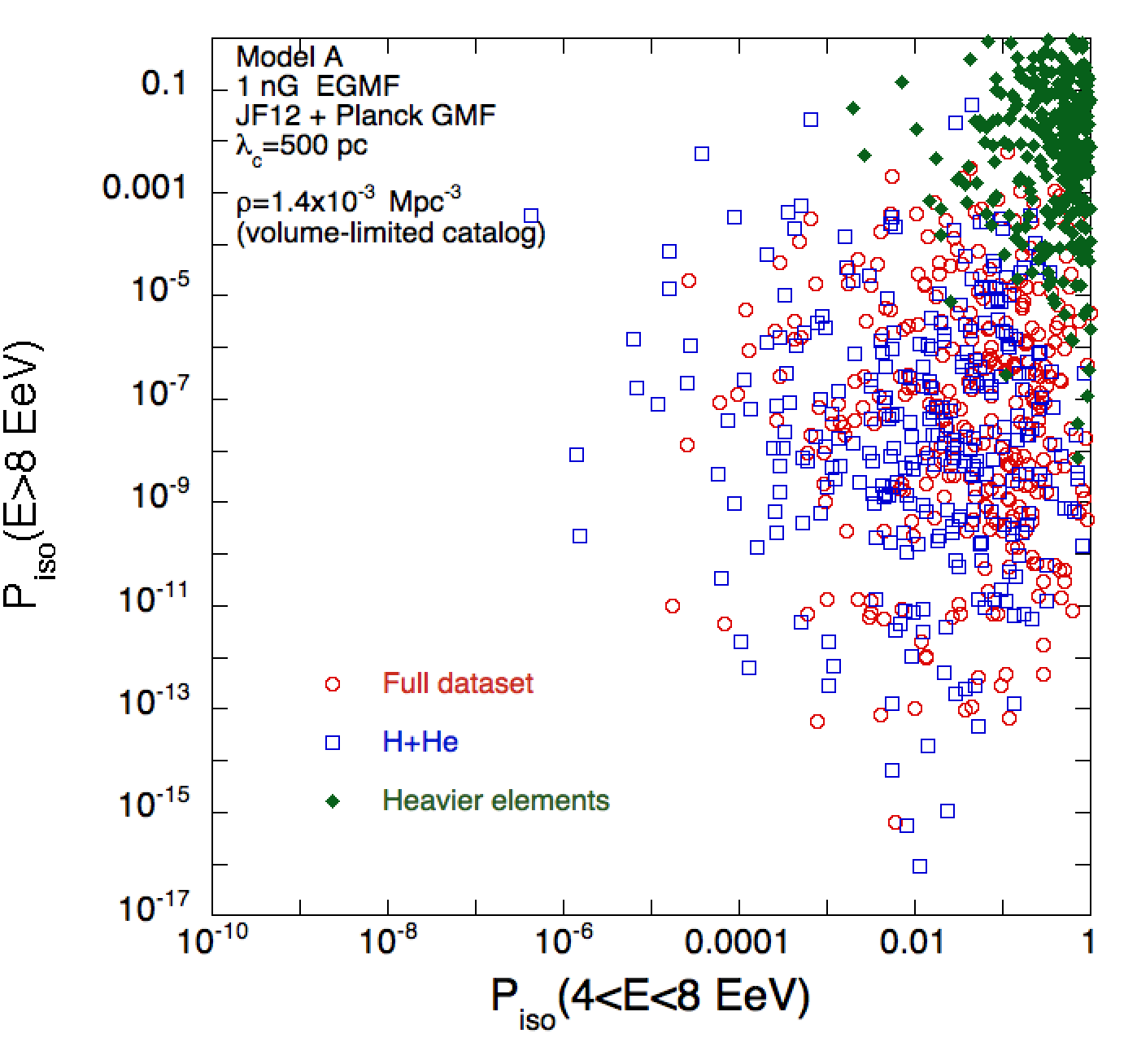}
   \includegraphics[width=7.5cm]{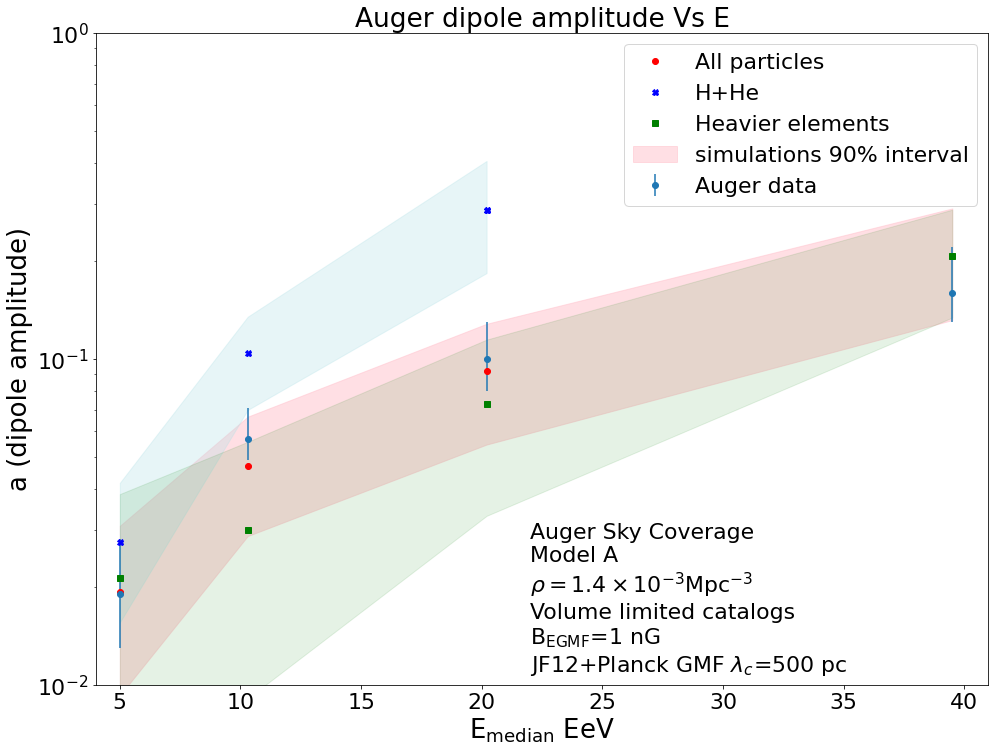}
      \caption{Top : scatter plot of the dipole amplitude reconstructed for 300 datasets based on the baseline volume-limited catalog, source model A, a 1~nG EGMF and the JF12+Planck GMF with $\lambda_{\mathrm{c}}=500$~pc, in the energy bins $4 \le E < 8$~EeV and $E\geq 8$~EeV. The amplitudes obtained for the full datasets are displayed together with those of the isolated H+He and heavier elements sub-components. Center : associated scatter plot of the p-values. Bottom : Associated energy evolution of the dipole amplitude based on the division between four energy bins above 4~EeV (see text).
              }
         \label{FigMapComponents}
\end{figure}

\begin{figure}
   \centering
    \includegraphics[width=7.5cm]{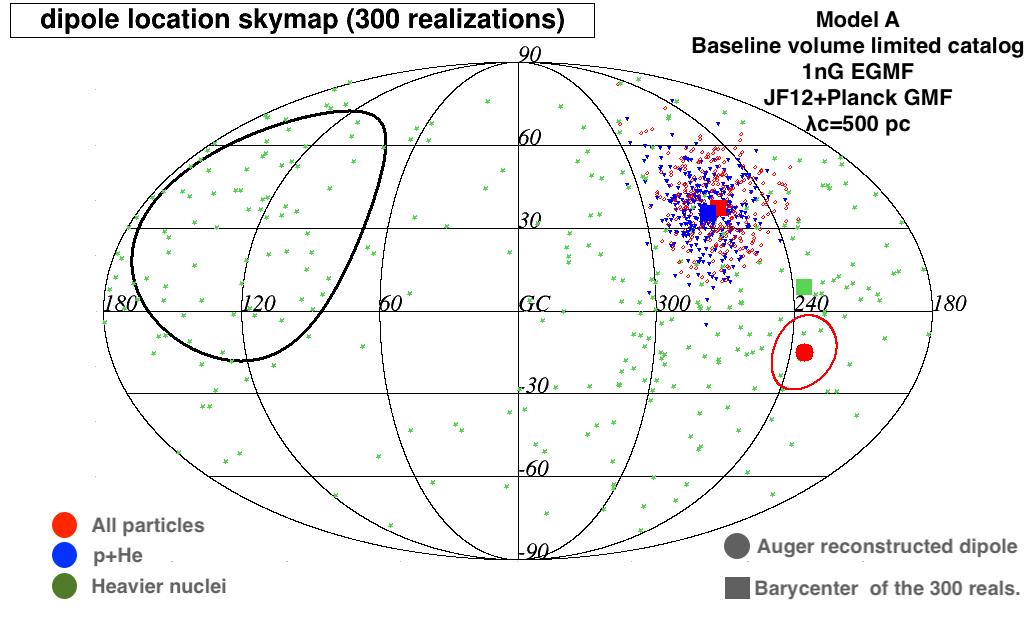}
\caption{Dipole directions reconstructed for the datasets considered in Fig.~\ref{FigMapComponents}.}
  \label{FigMapComponents3}
\end{figure}

\begin{figure}
   \centering
    \includegraphics[width=7.5cm]{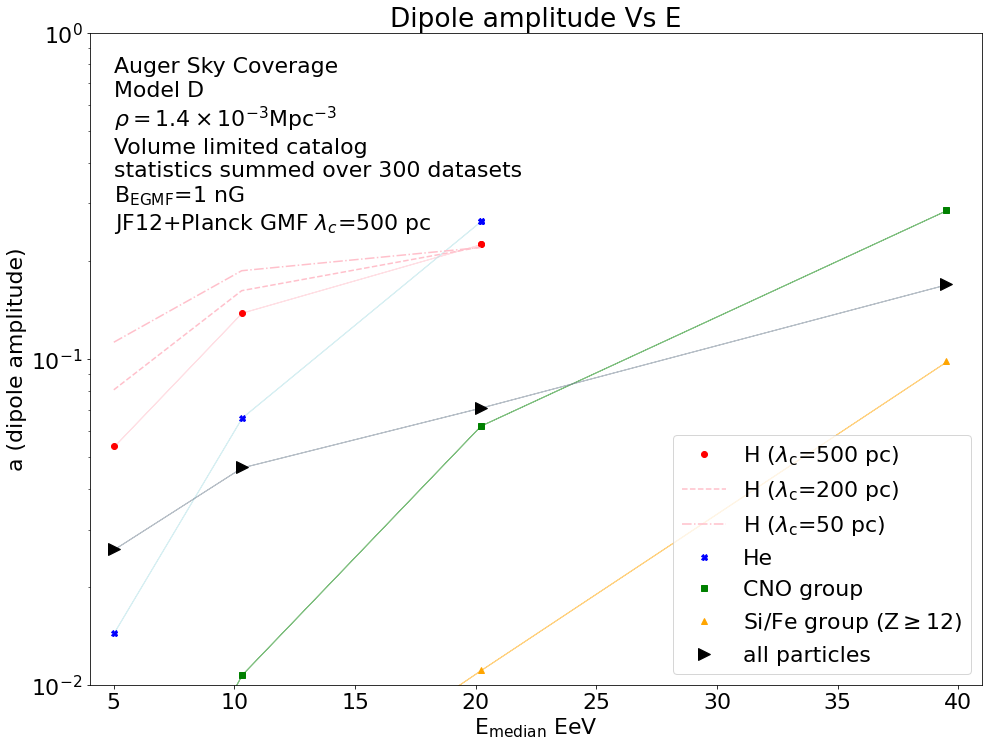}
\caption{Energy evolution of the dipole amplitude in the four energy bins considered in Fig.~\ref{FigDipVsE_compo}, for various nuclear components (see legend) as well as for the all particle dataset built by summing 300 Auger size datasets. The considered scenario is based on source model~D, with the baseline volume-limited catalog, a 1~nG EGMF, and the JF12+Planck GMF model with $\lambda_{\rm c}$=500~pc. In the case of the proton component, the energy evolutions obtained assuming $\lambda_{\rm c}$=50 and 200~pc are also shown for comparison.}
  \label{FigMapComponents2}
\end{figure}

%\subsubsection{Contribution of different groups of nuclear species}
\subsection{Contribution of different groups of nuclear species}
\label{sec:contribspecies}

\subsubsection{Enhancing the anisotropy signal}
\label{sec:enhancing}
Since it now seems established that the UHECR composition is mixed, it is interesting to study the contribution of different groups of nuclear species to the observed anisotropies. This is especially relevant in the context of 'Auger Prime', the upgrade of the Pierre Auger observatory currently under construction \citep{AugerUpgrade2020}, which is expected to provide an improved sensitivity to the UHECR mass composition and thus allow to build separate datasets with larger contributions of either light elements or heavier nuclei. To estimate the value of such a separation for the study of anisotropies, we have extracted from our simulated datasets the 'light component', consisting of H and He only, and the 'heavier component', consisting of all other nuclei (even though the assumption of a perfect identification of the different nuclei in the actual data is probably overly optimistic at this stage).

%To understand the building up of the anisotropy signal, since it seems now established that the UHECR composition is mixed, it is also useful to study the contribution of the groups of nuclear species. This is especially relevant in the context of the currently under construction upgrade of the Pierre Auger observatory \citep{AugerUpgrade2020} which is expected to provide an improved sensitivity to the UHECR mass composition and should allow to separate datasets between light (say protons and He) and heavier nuclei components. We have implemented such a separation (which we have supposed to be perfect) in our datasets in order to understand the weight of light nuclei in the observed anisotropies and also to discuss its relevance in the context of future anisotropy studies related the characterisation of the dipolar or quadrupolar modulations of UHECR arrival direction.

In the top panel of Fig.~\ref{FigMapComponents}, we show the scatter plot of the dipole amplitudes for each component in the energy bins $4\leq E<8$~EeV and $E\geq 8$~EeV for 300 realizations, assuming as above the baseline volumed-limited catalog, source model~A, a 1~nG EGMF, and the JF12+Planck GMF with $\lambda_{\rm c} = 500$~pc. One can see that the isolated light component has a larger dipole value in both energy bins and a predominant contribution to the anisotropy found in the full dataset, while the heavier component is significantly less anisotropic, and tends to wash out the anisotropy signal when all particles are considered together. This was expected, of course, given the much larger deflections in the intervening magnetic fields.

The corresponding scatter plot of the p-values, displayed in the top-right panel of Fig.~\ref{FigMapComponents}, further shows that the heavier component on its own is barely distinguishable from isotropy in these two energy bins. Note, however, that this is no longer the case if one considers the energy bin corresponding to $16\leq E<32$~EeV, and even more so the $E\geq 32$~EeV energy bin. The bottom-left panel shows the energy evolution of the amplitude for the two separate components. The energy evolution of the light component is steeper (in the energy bins where it is present) than that of the heavier (or all-particle) component, this is not only due to the fact the rigidity of the light component is evolving faster with the energy  but also and mostly because the horizon of He nuclei is evolving (shrinking) the fastest in this energy range due to the photo-interactions with far-infrared and then CMB photons which take place at higher energies for protons (if present) and heavier nuclei. 

Finally, Fig.~\ref{FigMapComponents3} shows the reconstructed dipole direction of the light and heavier components for the energy bin where the dipole moment is the most significant (i.e. $E>8$~EeV). As can be seen, the spread in the dipole direction of the heavier component (green dots) is huge, which is in line with the fact that the reconstructed dipole is essentially not significant at this statistics in this energy bin. A significantly lower value of $\lambda_{\mathrm{c}}$ ( $\lesssim$50~pc) would be for instance required for the predictions to be more concentrated around the barycenter. On the other hand, the light component shows a more consistent direction from one realisation to the next, very close to the all-particle dipole direction, which is again expected since this component makes up for the overwhelming part of the anisotropy in this energy bin.

\subsubsection{Rigidities and GZK horizon scales}
\label{sec:Rigzk}
Even though it appears to be rather out of reach at the present time, we investigated what could be observed if an even more precise separation of the nuclear components were available, namely if one could distinguish between 4 components: H, He, intermediate-mass nuclei ($2<Z<12$, i.e. mostly CNO) and heavier elements ($Z\geq 12$). The result is shown in Fig.~\ref{FigMapComponents2}, where we again plot the energy evolution of the dipole amplitude with energy, obtained from the stacking of our 300 datasets into a single high statistics one, for a scenario based on model~D, a 1~nG EGMF and the JF12+Planck GMF with $\lambda_{\mathrm{c}}=500$~pc. We have chosen model~D for this discussion, because its proton component extends up to higher energies than in the case of the other source composition models (namely up to the $16 \leq E <32$~EeV energy bin), and thus reaches higher rigidities than the other elements for this particular model. (NB: for the sake of simplicity, on the plot we keep showing the dipole amplitudes at the same energy for all  nuclear components, namely the median energy of the all-particle spectrum in each energy bin, even though the actual median energy is in fact different for each component.)

As can be seen, the amplitude of the He component is evolving very steeply between the $8 \leq E<16$~EeV and the $16 \leq E<32$~EeV energy bins, which is mostly due to the above-mentioned decrease of the horizon distance. The value of the amplitude predicted for He even slightly exceed that of proton in the $16 \leq E<32$~EeV bin, despite the higher rigidity of the protons. This is because He nuclei have a closer horizon than protons, and in this energy bin, we are looking  into the He GZK suppression, where the quasi isotropic contribution of more distant sources is lowered.

A qualitatively similar phenomenon takes place in the higher energy bin, $E \geq 32$~EeV, for the CNO component: this energy range is now covers the CNO GZK suppression, and the dipole amplitude obtained for this component reaches values comparable to that of He in the previous energy bin. Then again, the same is expected above 50 EeV for the heavier nuclei component. This is a significantly higher energy than the median energy in the $E \geq 32$~EeV energy bin, which is why it is not visible on Fig.~\ref{FigMapComponents2}.

These results further illustrate how useful a fine separation of the different nuclear components can be when studying anisotropies in the case of a low $E_{\max}$ UHECR model. Keeping protons out of the discussion for now, in each energy bin above 8~EeV, the lightest species, which have the largest rigidities, also turn out to have the shortest horizon. Both factors reinforce each other to make the contribution of these species the most anisotropic among the composed nuclei, until they eventually disappear. Thus, isolating the contributions of He and intermediate nuclei would be very interesting in the above-mentioned energy ranges, where their anisotropies are specifically influenced by the evolution of their horizons as a function of energy.

It is interesting to note that such a precise separation of the nuclear components is not so much important to observe intrinsically strong anisotropies at the highest energies, say above 50~EeV, because all the remaining nuclear species at these energies do experience a strong reduction of their horizon scale because of photo-interactions (i.e. they are captured "inside" their GZK cutoff). The main challenge, at such high energies, is thus mostly to accumulate large enough datasets for the underlying strong anisotropies to become statistically significant.

%Note that the constraints that can be obtained from the study of the anisotropies of various elements at various energies are not redundant, precisely because there is a combination of effects due to the evolution of rigidity, and to the evolution of the horizon scale. Now, while we said above that in a given energy bin the lightest composed species have the shortest horizons, it is also true that in a given rigidity bin, the heaviest species have the shortest horizon (see Fig.~7 of \citet{GAP2008}) {\color{red}Ce serait bien d'"expliquer" un peu si on y parvient, par exemple en disant quelque chose comme "because at a given rigidity, the energy is proportional to $Z$, and thus roughly to $A$, while the photo-dissociation cross sections evolve slowlier". Mais il faudrait trouver le bon angle…}. {{\bf \color{red}This implies that the weight of the isotropic contribution due to distant sources should vary when considering different species in the energy range of their  flux cut-off.}}{\color{red}Je ne comprends pas ce que tu veux dire… Evolve? Comment? Manque un bout de phrase ? par ex. "differently for different nuclear components"}. We shall come back to this discussion in Sect.~\ref{sec:conclusion}. 

Note that the specific value of the amplitudes and the associated p-values obtained for the all-particle datasets, or for the light and the heavier components, depend on the relative contributions of individual species within these components. This explains the different steepness of the dipole amplitude evolution seen in Fig.~\ref{FigDipVsE_compo}) for the different composition models. For instance the predominant contribution of CNO in the case of model B explains the steeper increase of the all particles dipole amplitude in the last two energy bins, compared to the other source composition models, all other parameters being fixed. Conversely, the larger contribution of both H and Fe nuclei to the datasets built with source model D explains the flatter energy evolution.

\subsubsection{Influence of the GMF coherence length}

The way the dipole amplitude evolves with rigidity also depends on the coherence length of the GMF, especially in the rigidity range where the UHECRs are resonant with the turbulent field. The cases with $\lambda_{\mathrm{c}}$=50, 200 or 500~pc are different in this respect for the protons in the first three energy bins, as can be seen in Fig.~\ref{FigMapComponents2}. In particular, in the case of the lower coherence length, the dipole amplitude of the proton component is already rather high in the first energy bin, so the evolution from one bin to the next remains moderate. On the contrary, the dipole amplitude is more strongly suppressed in the case of $\lambda_{\mathrm{c}} = 500$˜pc in the lowest energy bin, so the increase of the amplitude with energy is sharper, to reach essentially the same amplitude as for $\lambda_{\mathrm{c}} = 200$˜pc in the  $16 \leq E <32 $~EeV  bin, since the turbulent GMF has much lower effect at such high rigidities.

\begin{figure}
   \centering
   \includegraphics[width=\hsize]{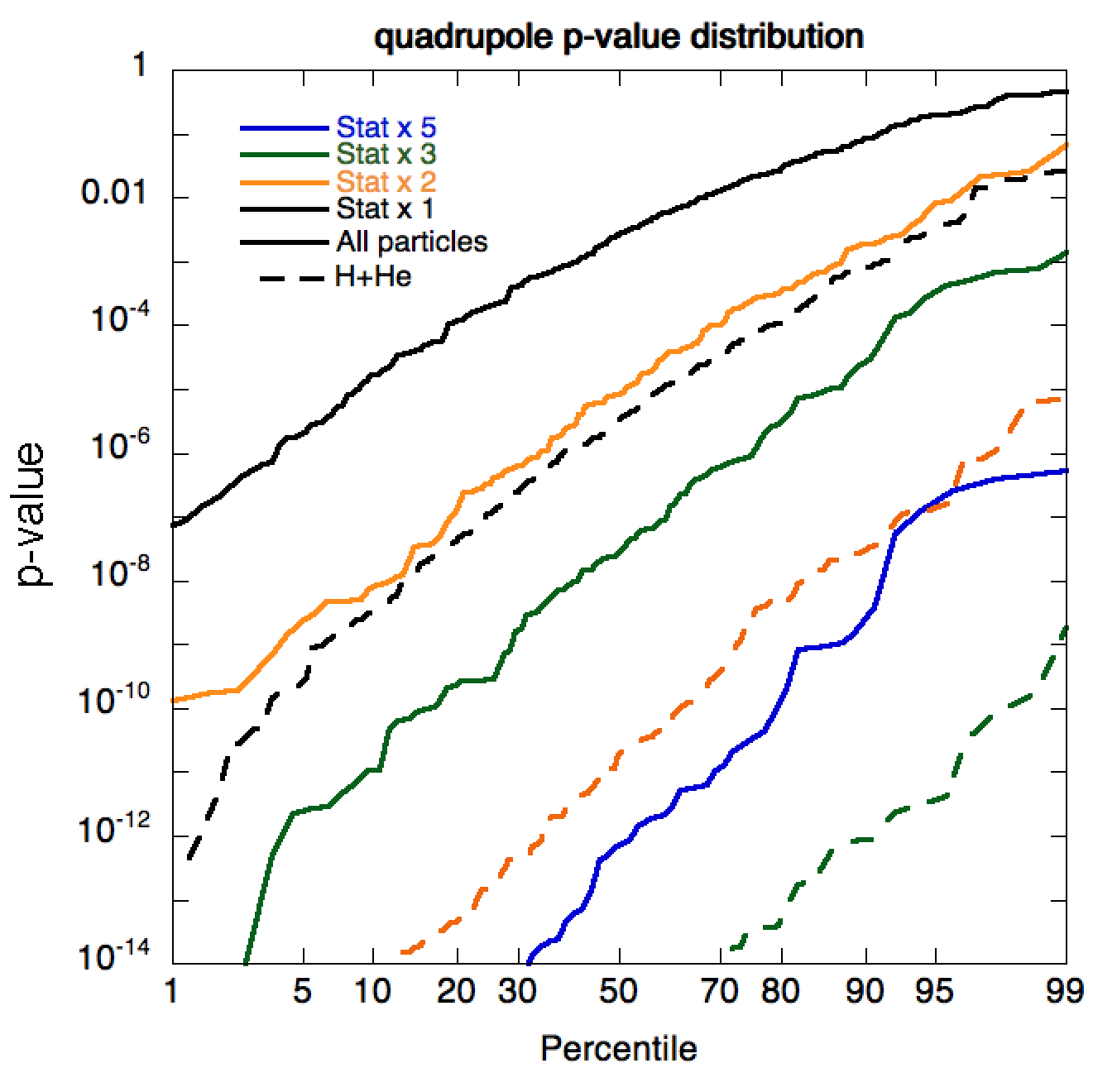}
      \caption{Probability distribution of the p-value of the quadrupolar right ascension modulation obtained for various statistics (in units of the full Auger dataset at the time of the dipole discovery) of the full dataset, considering either the analysis of the full dataset or of the isolated light (H+He) sub-component and the UHECR event with $E\geq8$~EeV. The astrphysical scenario considered is the baseline volume-limited catalog, source model A, a 1~nG EGMF and the GMF JF12+Planck model with $\lambda_{\mathrm{c}}=500$~pc.
              }
         \label{FigQuadComponents}
\end{figure}

\subsubsection{Quadrupole amplitude}

Another promising advantage of being able to separate the light component when studying the large scales anisotropies, relates to the next order of modulation of the UHECR arrival directions, namely the quadrupole. In Fig.~\ref{FigQuadComponents}, we show the cumulative distribution function of the p-value of the right ascension quadrupolar modulation obtained for 300 realisations of the astrophysical scenario discussed in Figs.~\ref{FigQuadScat1} and \ref{FigMapComponents}, for the UHECR events with $E>8$~EeV. Recall our earlier finding that the absence of a significant quadrupole in the Auger dataset at the time of the dipole discovery is not constraining for the models. The p-value distributions are shown for four different statistics, ranging from that of the Auger dataset to a five time larger dataset. The solid lines correspond to the all-particle datasets, while the dashed lines show the quadrupole p-value for the light (H+He) component only. It is striking that the p-values are much lower for the latter component, to the extent that a mere doubling of the statistics (compared to the current Auger dataset) should lead to a significant quadrupole amplitude, if the model assumptions hold. Correspondingly, the quadripole ampliltude appears much larger for the light component than for the all-particle dataset, namely  $\sim 7$\% instead of $\sim 2.6$\%, for the displayed scenario.

Finally, while we do agree with the result of the analysis of \citep{Matteo2018} that the quadrupole amplitude is expected to increase with the energy for most extragalactic UHECR scenarios, the $E>8$~EeV energy bin should be more optimal to find a significant signal (that is rejecting a statistical fluctuation with a high significance), all the more so with an efficient separation of the light component, at least for source models with an energy evolution of the mass similar to those considered here (either models A, B, C or D).

\subsection{Anisotropy of the composition}
\label{sec:anisocompo}
\begin{figure}[ht!]
   \centering
   \includegraphics[width=\hsize]{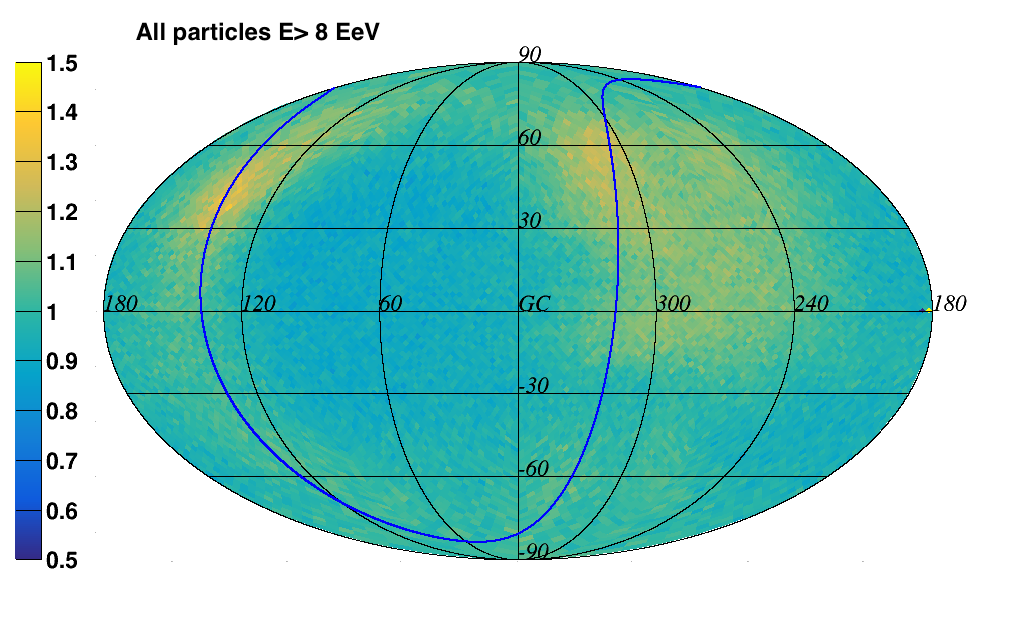}
   \includegraphics[width=\hsize]{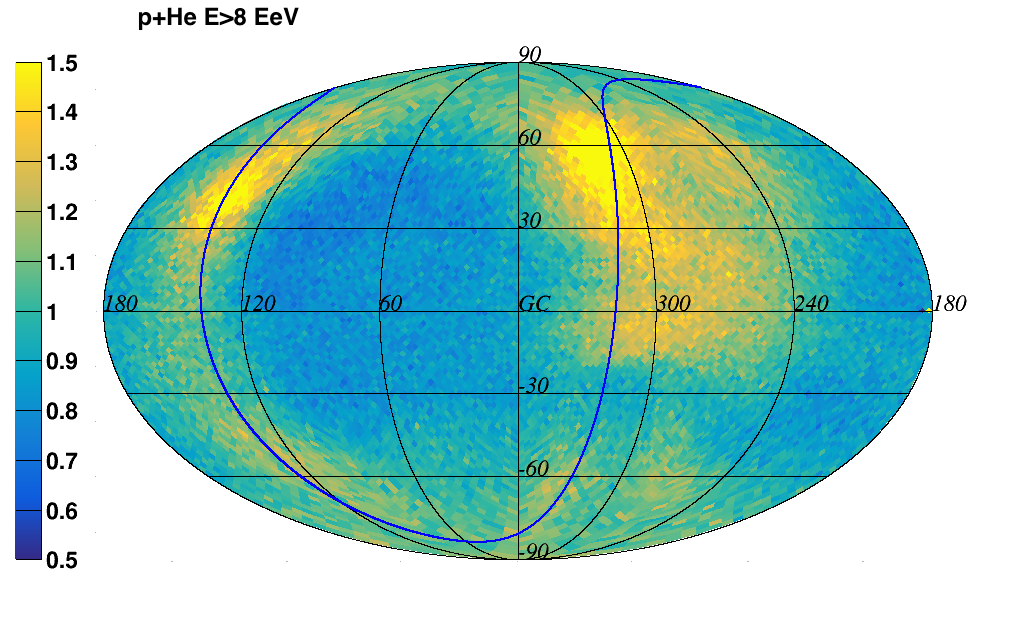}
    \includegraphics[width=\hsize]{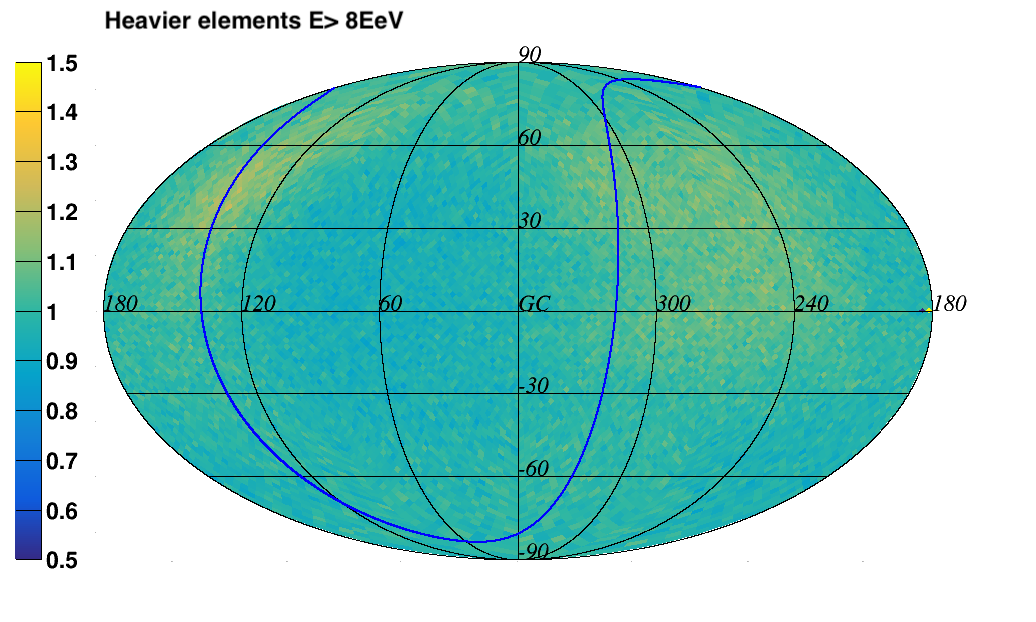}
      \caption{Density skymap obtained by summing 300 datasets assuming a uniform full-sky exposure for the astrophysical scenario assuming the baseline volume-limited catalog, source model A, a 1~nG EGMF, the JF12+Planck model with $\lambda_{\mathrm{c}}=200$~pc. Top panel shows the all particle skymap, while the central and bottom panels show respectively the light (H+He) and heavier elements sub-components skymaps.
              }
         \label{FigMapANICompo}
\end{figure}

The predominance of the light component in the anisotropy signal, in the energy bins where this component is present, comes from the fact that light particles have a lower charge and thus a higher rigidity than heavier nuclei, and is further amplified by the rapid decrease of the energy loss horizon of He nuclei around $\sim 10^{19}$~eV, as we discussed in the previous paragraphs ({\it i.e} Sects.~\ref{sec:enhancing} and~\ref{sec:Rigzk}). The light particles are expected to be less deflected than heavier nuclei and thus their arrival directions should retain more information about the genuine spatial distribution of the sources, especially in the relatively nearby universe. This situation is clearly visible on the different panels of Fig.~\ref{FigMapANICompo}, where we show the density map (with a linear color scale) of the arrival direction of UHECRs above 8 EeV, cumulated over 300 datasets, assuming as above the baseline volume-limited catalog, model A a 1~nG EGMG, and the JF12+Planck GMF with $\lambda_{\mathrm{c}}=200$~pc, in the case of a uniform full-sky coverage. The top-panel shows the all particles sky map, while the central and bottom panels respectively show the light and heavier component sky maps. The sharper constrast of the light component map as well as its dominant contribution to the overall anisotropy are clearly seen. The regions where the event counts are larger than expected from an isotropic distribution of events are indeed those where light nuclei are preferentially found.

This is a general result of all the mixed composition models that we investigated: the “event number anisotropies” revealed by analyses such as the ones we use in this paper are always accompanied by associated “composition anisotropies”. The regions of the sky where the composition is lighter are also the regions where a numerical excess of the event count is found. Note that this should be true also at higher energies (e.g. in the $E>32$~EeV energy bin), even if the light component eventually disappears and we are left with a mix of nuclei going from CNO to Fe. The CNO group nuclei then replaces light particles in the above description, as discussed in Sect.~\ref{sec:Rigzk}.

\begin{figure}
   \centering
   \includegraphics[width=\hsize]{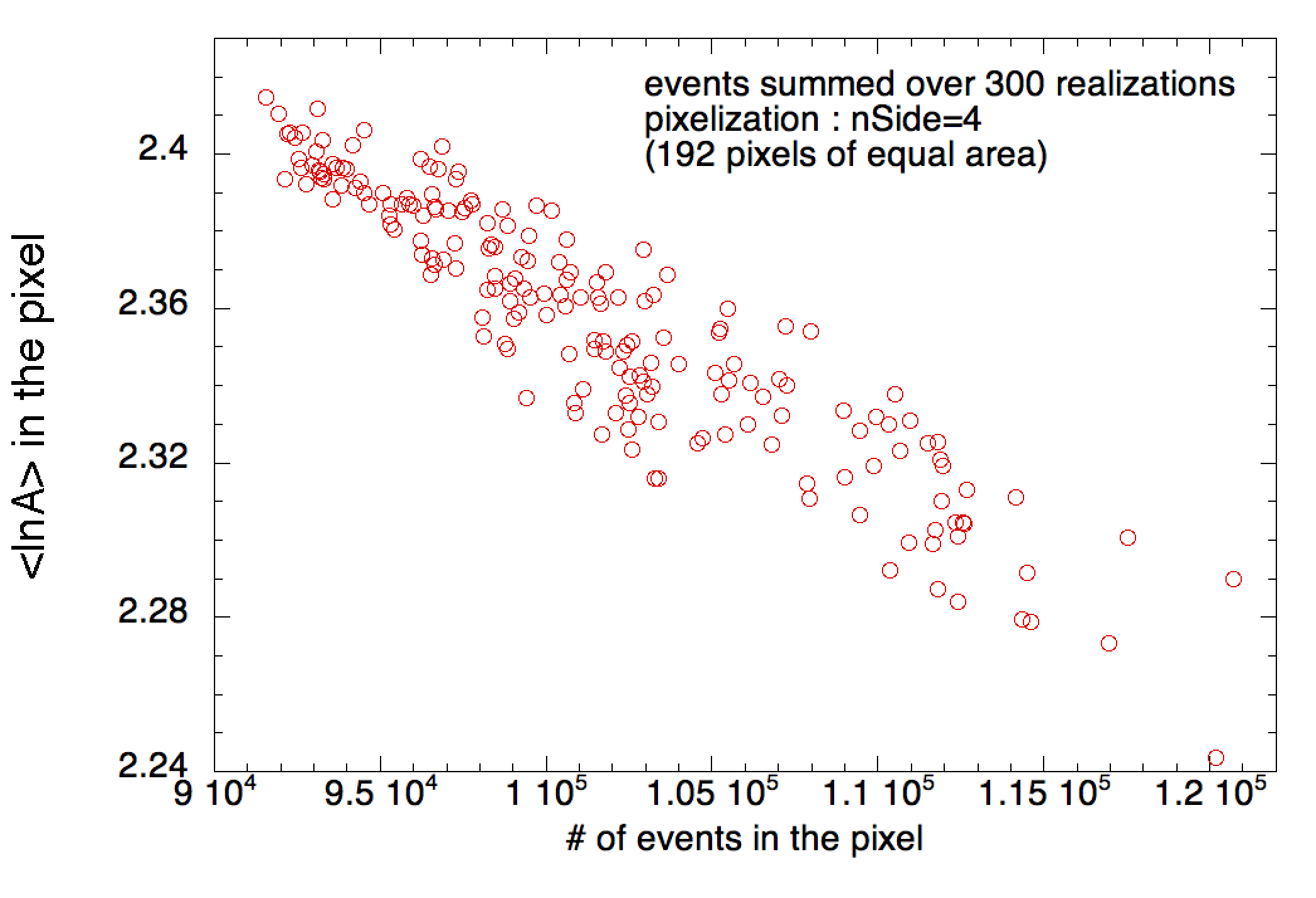}
    \includegraphics[width=\hsize]{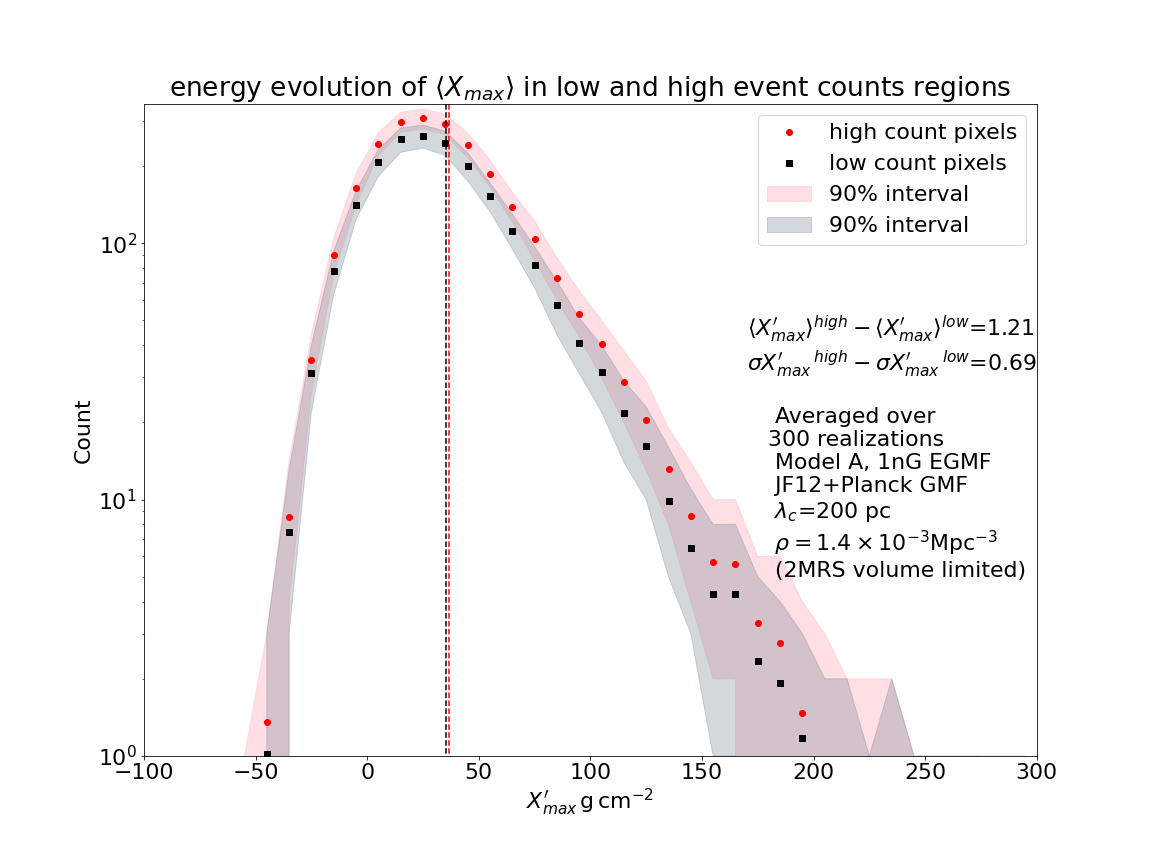}
    \includegraphics[width=\hsize]{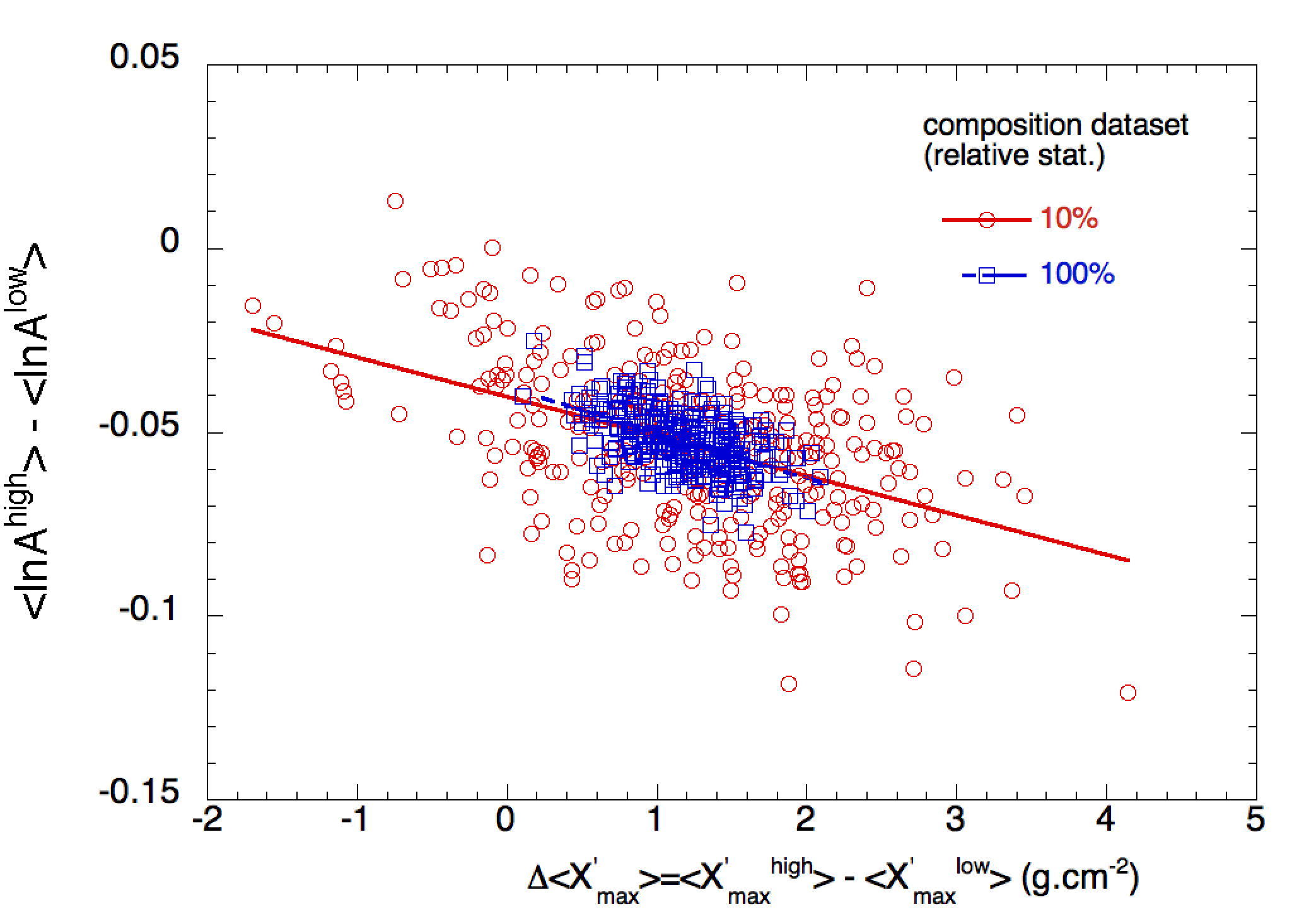}
      \caption{Top : Scatter plot showing the value of $\langle \ln A \rangle$ in a pixel as a function of the number of events in the pixel for the skymap displayed on the top panel of Fig.~\ref{FigMapANICompo} divided into 192 pixels of equal area. Center : average distribution (over 300 datasets) of $X^\prime_{\rm max}$ in the high and low count pixels (see text) for the events with $E\geq8$~EeV, for the scenario considered in Fig.~\ref{FigMapANICompo}, the shaded areas show the interval in which 90\% of the datasets are found. The value of $X_{\rm max}$ is assumed to be measured for 10\% of the events (that is in average for 6500 events per dataset). Bottom : Scatter plot of the difference in $\langle \ln A \rangle$ as a function of $\langle X^\prime_{\rm max} \rangle$ between high and low pixel counts for the 300 datasets corresponding to the scenario considered in Fig.~\ref{FigMapANICompo}. The different red and blue colors shows cases for which the value of $X_{\rm max}$ and $\ln A$ is known respectively for 10\%  and 100\%  of the events.  
              }
         \label{FigNVsLnA}
\end{figure}

To illustrate this effect, we have divided the skymap of the top-panel of Fig.~\ref{FigMapANICompo} into 192 pixels of equal angular size and recorded the event number count and the mean logarithmic mass, $\langle \ln A \rangle$, in each pixel. The resulting scatter plot of $\langle \ln A \rangle$ as a function of the event count is shown on the top panel of Fig.~\ref{FigNVsLnA}. A clear correlation is visible, showing how tightly the event count anisotropy is related to a composition anisotropy.

However, the range of variations in $\langle \ln A \rangle$ is quite limited, of the order of 0.12, and it remains to be explored whether such composition anisotropies can be expected to be measured with more limited statistics (recall that 300 datasets have been merged for this plot).

%The observability of the latter effect is however not demonstrated by this scatter plot since, first, it is build with a statistics cumulated over 300 datasets and second, the difference in $\langle \ln A \rangle$ is quite modest between the high and low number counts pixels.      

For the sake of simplicity, we perform the study in the case of a full-sky exposure, with twice the Auger statistics, i.e. 65000 events above 8~EeV. For each dataset, we still divide the skymap into 192 pixels of equal area (corresponding to the Healpix pixelisation with nSide = 4), and we select the 1/3 ({\it i.e} 64) pixels with the largest event counts as well as the 1/3 pixels with the smallest event counts. For each of these sub-datasets, we determine the value of $\langle \ln A \rangle$ and of $\sigma^2\ln A$, and we register the distribution of the depths of the maximum shower development, $X_{\rm max}$, using CONEX showers \citep{Pierog2006} with the EPOS-LHC hadronic model. We consider alternatively the case when this information is available for all the events, or for only 10\% of the events, taking into account the typical duty cycle of a ground-based fluorescence detector duty cycle (this is a somewhat crude assumption, since in reality the fraction of events for which fluorescence data are available depends on the energy). Note that such an exercise, with an optimisation based on high- and low-count pixels, is relevant only if one may expect that a real signal is present beyond the statistical fluctuations in the event counts, which is presumably the case for the Auger data in the energy bin with $E \geq 8$~EeV, and which is already known to be the case in this energy range for the astrophysical scenarios under consideration (see above and Figs.~\ref{FigCl},~\ref{FigCl_Auger},~\ref{FigDipScat1}). 

%that this exercise and the optimization on high and low count pixels only make sense because an anisotropic signal is already known to exist in this energy range ($E\geq8$~EeV) for this astrophysical scenario,  actually somewhat stronger than in Auger data, on the energy range of interest, as already shown in Figs.~\ref{FigCl},~\ref{FigCl_Auger},~\ref{FigDipScat1}, with different analyses. 

In order to derive a single composition indicator from events with different energies, we use the same procedure as introduced by \citet{Mayotte2021}, defining a standardised shower development depth, $X^\prime_{\rm max}$, obtained from the usual depth $X_{\max}$ by subtracting the mean expected depth of an Fe shower with the same energy. This automatically corrects for the energy dependence of $X_{\max}$, although it of course relies on an a priori knowledge of this energy dependence, derived from an assumed hadronic model. Here, we use the EPOS-LHC model, but we note that contrary to the global normalisation of $X_{\max}$, the elongation rate is not strongly dependent on the hadronic model in the considered energy range.  The result is shown in the central panel of Fig.~\ref{FigNVsLnA}, where we plotted the distribution of $X^{\prime}_{\rm max}$ for the events above 8~EeV averaged over 300 datasets for the low-count and high-count pixels. The dashed lines indicate the mean values of the two distributions, and the shaded area shows the dispersion within the 300 datasets.

As can be seen, the average differences are extremely small, both in terms of $\langle X^\prime_{\rm max} \rangle$ and $\sigma X^\prime_{\rm max}$ (respectively $\sim$ 1.2 and 0.7 $\rm g\,cm^{-2}$), even though the high-count pixels tend to have slightly larger values of $X^\prime_{\rm max}$, as expected from the fact that their larger flux is associated with a lighter composition, according to the above-mentioned correlation (top panel). However, the average distributions are  hiding  stronger differences visible in individual datasets due to statistical fluctuations. The difference, $\Delta\langle X^\prime_{\rm max} \rangle$, of the values of $\langle X^\prime_{\rm max} \rangle$ in the high-count and low-count pixels for each of the realisations, is plotted  on the bottom panel of Fig.~\ref{FigNVsLnA} against the corresponding (but non directly measurable) difference of the values of $\langle \ln A \rangle$. The value of $\Delta\langle X^\prime_{\rm max} \rangle$ can be seen to vary from dataset to dataset, especially for datasets with lower statistics (i.e. when applying a 10\% duty cycle), and occasionally yield negative values or on the other hand differences much larger than the average. This results from the low statistics and the imperfect correlation between the UHECR mass and the corresponding shower $X_{\max}$, due to shower-to-shower fluctuations. Assuming composition datasets of $\sim 6500$~events for this particular scenario, we find at most $\Delta\langle X^\prime_{\rm max} \rangle \sim 4\,\rm g\,cm^{-2}$, reaching a maximum of $\sim 5.5\,\rm g\,cm^{-2}$ for datasets with twice lower statistics (while of course the average differences remain unchanged). As a conclusion, although this composition feature is genuine, it is expected to be quite difficult to observe, even assuming a scenario which seems intrinsically more anisotropic than the Auger data. In other words, under the above very generic model assumptions, it appears unlikely that a significant composition anisotropy could be seen without an associated event-count anisotropy, of larger significance.

Note that the exact values of $\Delta\langle X^\prime_{\rm max} \rangle$ in the above analysis depend on the source composition model, and the expected differences are indeed somewhat larger in the case of model D, because the corresponding composition is more mixed than that of model A. The average differences of $\langle X^\prime_{\rm max} \rangle$ and $\sigma X^\prime_{\rm max}$ between high-count and low-count pixels reach respectively $\sim 2$ and $\sim 1.2$ $\rm g\,cm^{-2}$ in that case, still for $E\geq 8$, the baseline volume-limited catalog, a 1~nG EGMF and the JF12+Planck GMF with $\lambda_{\rm c} = 500$~pc. The numbers become respectively $\sim 2.5$ and  $\sim 1.6$ $\rm g\,cm^{-2}$ if we assume $\lambda_{\rm c} = 200$~pc. Yet, even in that case, the expected differences remain difficult to observe for datasets (with composition information) of accessible size in the near future, and are subject to large statistical (i.e. dataset-to-dataset) fluctuations. 

Finally, it is worth noting that this composition feature does not necessarily require the presence of a significant component of protons and/or He nuclei but can also exist in the case of a mix of heavier nuclei. For instance, in the case of model A the average difference of $\langle X^\prime_{\rm max} \rangle$ values between high-count and low-count pixels actually increases when changing the event selection energy threshold from 8~EeV to 16~EeV, despite the fact that the relative abundance of light nuclei and in particular of protons decreases. Again this is a consequence of the considerations discussed in Sect.~\ref{sec:Rigzk}.

The above discussion is relevant in the context of the recent claim by the Auger collaboration of the possible presence of an anisotropy in the composition of the UHECRs above 5~EeV \citep{Mayotte2021}. Unlike our approach above, the claimed signal is not based on a pre-established event count anisotropy (which could have been the dipole modulation above 8~EeV) but rather on an statistical optimisation (scan) of the portions of the sky "on" and "off" the Galactic plane (scanning for the absolute value of the Galactic latitude defining the "on" and "off" regions) and the energy threshold of the composition dataset for which the difference of $\langle X^\prime_{\rm max} \rangle$ between the "on" and "off" regions is most significant. The scanning procedure yielded a separation of the "on" and "off" regions at a Galactic latitude of $\pm 30^\circ$ and an energy threshold of 5~EeV for the composition datasets. Differences for $\langle X^\prime_{\rm max} \rangle$ and $\sigma X^\prime_{\rm max}$ of respectively $\sim 9$ and $\sim 6$ $\rm g\,cm^{-2}$ were found between of off-plane and on-plane regions, with deeper showers, implying a lighter composition, in the off-plane region. The magnitude of the difference appears much larger than what we found in the above study. If confirmed, this measurement may thus be difficult to associate with the same type of effects as described here, unless the optimised cuts adopted (through a scanning procedure) in the Auger data analysis are essentially favouring an upward fluctuation of a genuine, but much smaller signal (in which case the observed on/off differences should decrease in future data releases).

%investigated this potential signal with the composition anisotropy we have discussed above. 

Note that when performing the same on-plane/off-plane separation in our datasets (rather than separating between high-count and low-count pixels), we still get an event count anisotropy, which is not surprising, given the flux maps of Fig.~\ref{FigMapANICompo}. Quantitatively, we find an excess of the event count ranging from $\sim 1.5 \sigma$ to $\sim 4 \sigma$ depending on the considered datasets. The differences in the values of $\langle X^\prime_{\rm max} \rangle$ and $\sigma X^\prime_{\rm max}$ between the "on" and "off" regions are however $\sim$twice lower than when separating between high count and low count pixels and thus very far from the difference found in Auger data. 

While the authors of this analysis argue that it would be useful to perform in the future this type of analysis with the higher statistics surface detector composition dataset, we also wish to stress that it would be interesting (and according to us more straightforward) to search in the surface detector dataset for a standard event count anisotropy, assuming the same on-plane and off-plane regions and energy threshold as those suggested by the scan on the fluorescence dataset. Note that the absence of such an event count anisotropy at a significant level, when comparing the two regions, would require a rather unnatural coincidence, by which the excess of deep showers in the off-plane region would be compensated by a deficit in shallow showers of approximately the same magnitude, and vice-versa in the on-plane region. This situation, although possible, is definitely not expected in general, since the arrival directions of heavy nuclei (shallow showers) are generically more isotropic than those of light nuclei, and would require either conspiring statistical fluctuations, or a rather fine-tuned combination of a source distribution and a GMF structure (with its specific rigidity-dependent magnification/deflection patterns). Another possibility, of course, would be that the observed feature, if confirmed at such a high amplitude, is revealing a shear conflict between the types of models investigated here and the actual UHECR phenomenology (for instance if different components of UHECRs are building up the flux observed in different energy and/or rigidity bins). In any case, we argue that, based on simple phenomenological arguments, it is rather natural to search for an event count excess associated with a composition anisotropy, whatever the origin of the latter may be (unless it is a mere statistical fluctuation, of course).

%Regarding the magnitude of the on/off difference in the signal observed by Auger, it seems quite unlikely the origin of this claimed composition anisotropy is related to the one we discussed in the above paragraphs, unless claiming that the optimisation (scan) performed on the composition dataset highlighted a fluctuation of an otherwise much smaller (but genuine) difference, in which  case the observed on/off difference should decrease in future data releases.   

\begin{figure*}[ht!]
   \centering
   \includegraphics[width=8.5cm]{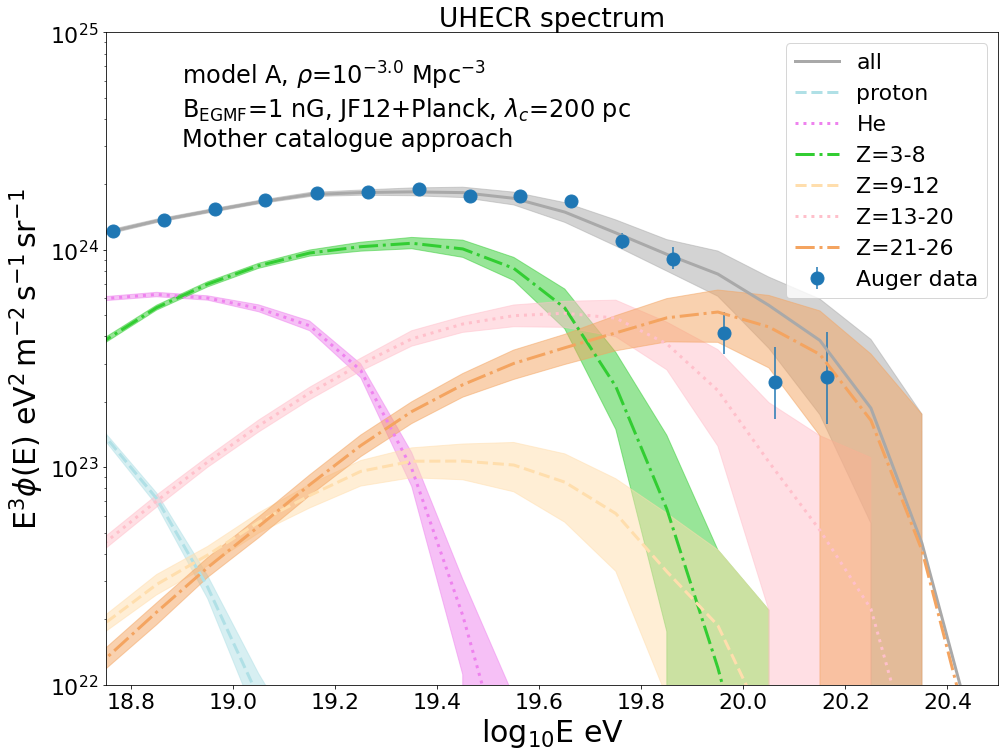}
   \includegraphics[width=8.5cm]{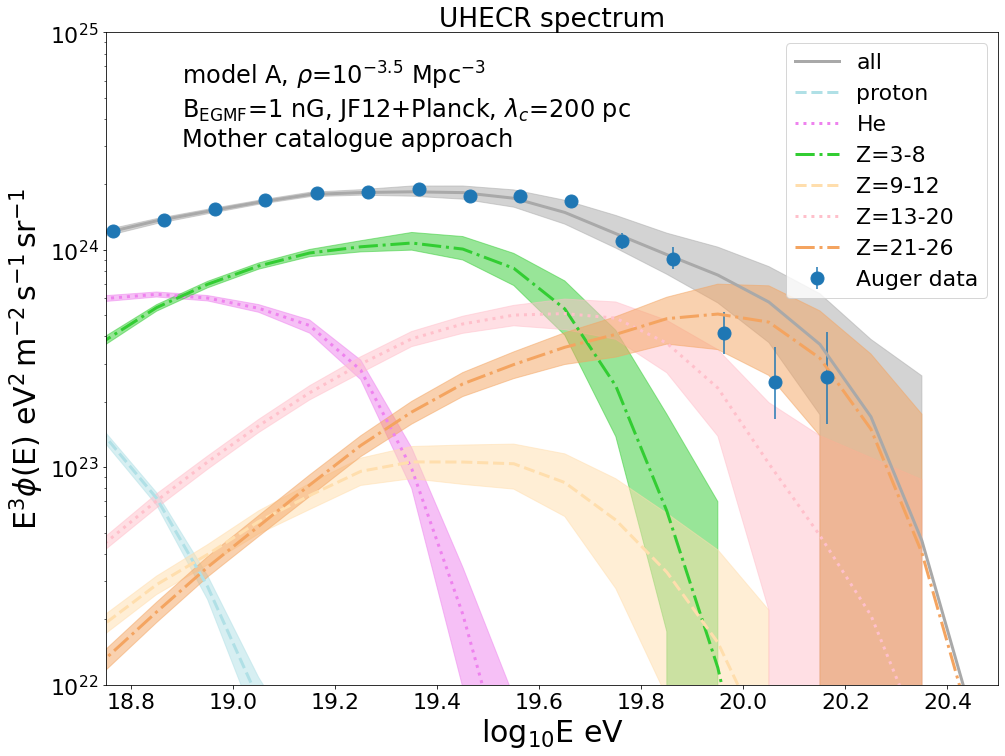}
   \includegraphics[width=8.5cm]{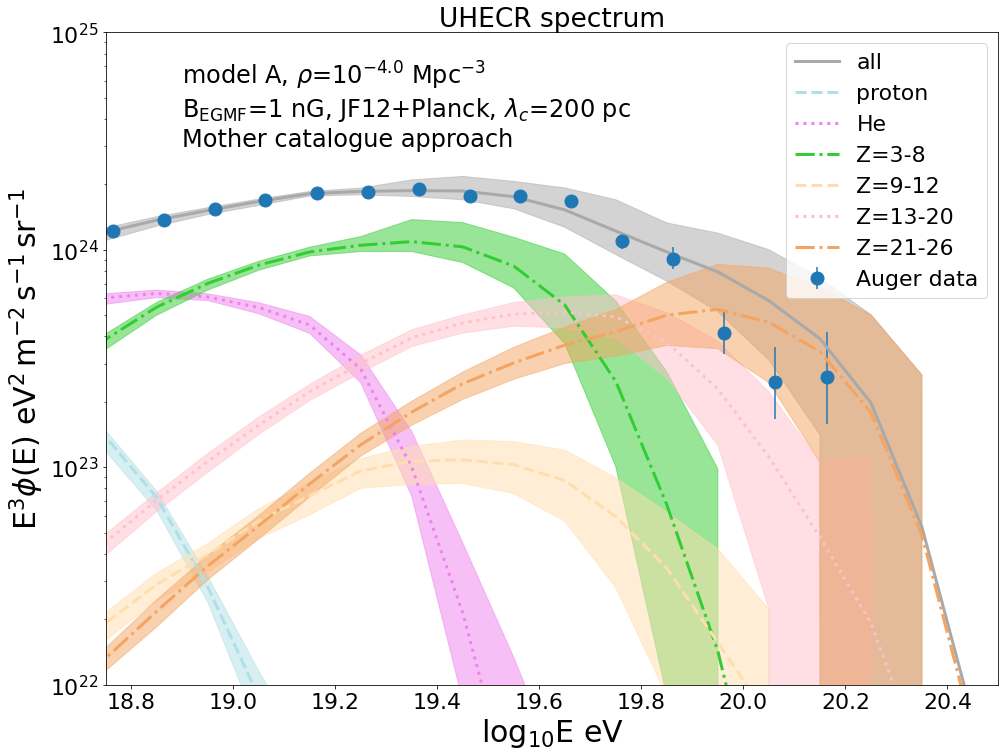}
   \includegraphics[width=8.5cm]{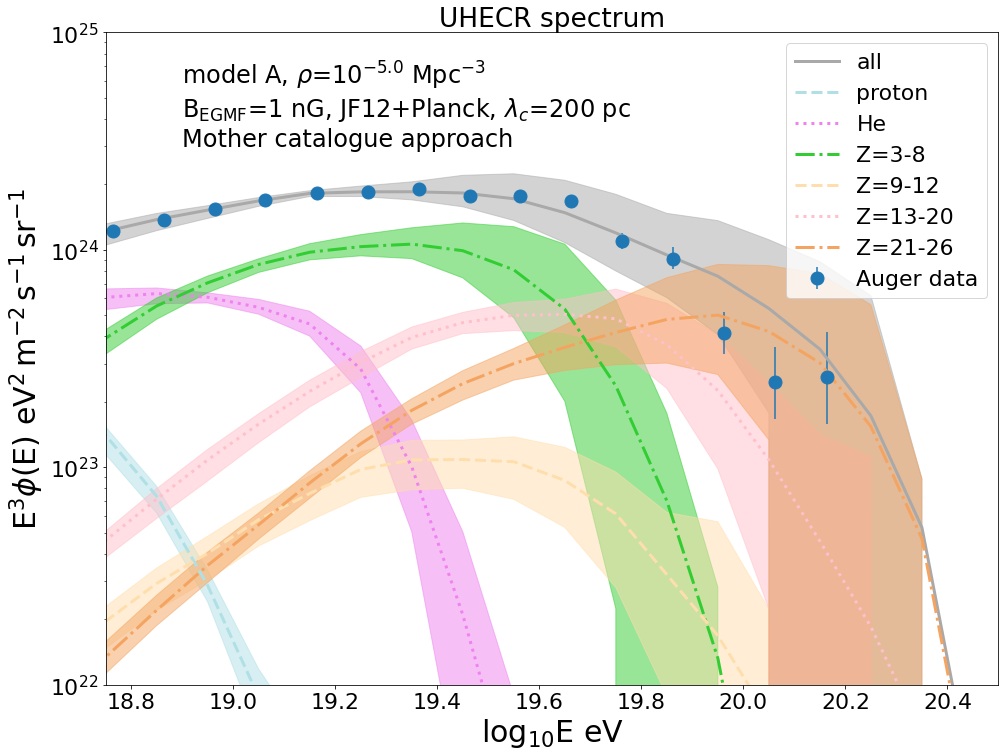}
      \caption{UHECR energy spectra obtained after propagation in the extragalactic and Galactic media, for source model A assuming the different hypotheses on the source density, $10^{-3}$ $\rm Mpc^{-3}$ (top-left), $10^{-3.5}$ $\rm Mpc^{-3}$ (top-right),$10^{-4}$ $\rm Mpc^{-3}$ (bottom-left), $10^{-5}$ $\rm Mpc^{-3}$ (bottom-right), in the mother catalog approach. An EGMF of 1~nG and the "JF12+Planck" GMF model with $\lambda_{\mathrm{c}}=200$~pc are assumed in all cases. The simulated sky coverage is that of the Pierre Auger Observatory. The lines show the mean expected fluxes averaged over 300 realizations and the shaded areas cover the range in which $90\%$ of the realizations are found (energy bin by energy bin). The different colors show different groups of nuclear species, as indicated on each plot.  
      %The scenarios displayed on the bottom-right panel are based on the sampling of an homogeneous and isotropic distribution of source with a  $10^{-4}$ $\rm Mpc^{-3}$ density and the source model $\rm A^\prime$. 
              }
         \label{FigVariance1}
   \end{figure*}

\subsection{Cosmic variance}
\label{sec:cosmicVariance}

We finally turn to the discussion of the cosmic variance that can be expected in the framework of the astrophysical models explored in the present study. 
This allows us to assess the validity domain of some of the statements made in the previous paragraphs, which were obtained for well defined hypotheses on the sources distributions. This is particularly important for the discussion of the observed dipole direction. Indeed, as discussed above, when the UHECR source distribution is assumed to follow that of the brightest galaxies in the 2MRS catalog (with a minimum intrinsic luminosity chosen to result in the intended overall source density), none of the explored astrophysical models appears to be able to reproduce the dipole direction actually reconstructed from the Auger data, whatever the assumptions on the source spectrum and composition, the EGMF strength and the GMF model (within the limits investigated, ensuring a relatively good account of the other UHECR observables, namely the energy spectrum and the evolution of the composition). A natural question is thus to ask whether a different choice of the sources could lead to a better agreement with the data, as far as the dipole direction is concerned, without requiring a complete change of perspective regarding the underlying astrophysical scenario. Indeed, the generic assumption of the explored scenarios is that the UHECR sources have a spatial distribution that follows that of the matter in the universe, traced by the galaxies, but that does not imply that the most luminous galaxies in the K-band should be the best proxy for the location of UHECR sources among all galaxies. And indeed, there is no particular reason why it would be so (nor otherwise), at least as long as the origin of UHECRs remains unknown.

To explore how the reconstructed dipole direction would differ if one were to make different assumptions for the source distribution, we simply sampled the galaxies in the local universe without any prejudice, i.e. by drawing randomly the UHECR sources among the galaxies in our “mother catalog” (see Sect.\ref{sec:sourceDistrib}), defining only an intended source density. For each chosen density, we generated 300 independent realisations of the source distribution, as sub-sampled from the mother catalog, and for each of these realisations we generated a dataset assuming either the Auger exposure, the TA exposure, or a full-sky exposure. In addition, for comparison, we generated source catalogs at the various selected densities by drawing randomly the sources from a uniform (homogeneous and isotropic) distribution.

%We finally turn to the discussion of the impact of the cosmic variance on our predictions which is critical to assess the validity domain of some of the statements made in the previous paragraphs in the context of a well defined hypotheses on the sources distributions. This is in particular important for the discussion of the observed dipole direction which we so far failed to reproduce with our various astrophysical scenarios. We remind the reader that our scenarios are  based on four different main ingredients {\it i.e}, a source  model (spectrum and compostion), the EGMF strength, the GMF model and a hypothesis on the source distribution. So far by considering our volume-limited catalogs we kept, on purpose, the cosmic variance out of the discussion. We will now use catalogs built on the random sampling of the {\it mother catalog} assuming various source densities as explained in Sect.~\ref{sec:sourceDistrib}. For each hypothesis on the source density we generate 300 realizations of the sampling of the mother catalog (resulting in 300 different sub-catalogs all based on the same general assumption that the UHECR sources distribution follows that of galaxies in the Universe) for each of which we generate a (Auger, TA or full-sky exposure) dataset. Likewise, catalogs sampling an homogenous and isotropic distribution of sources assuming various densities were also produced. 

\begin{figure}
   \centering
   \includegraphics[width=8.5cm]{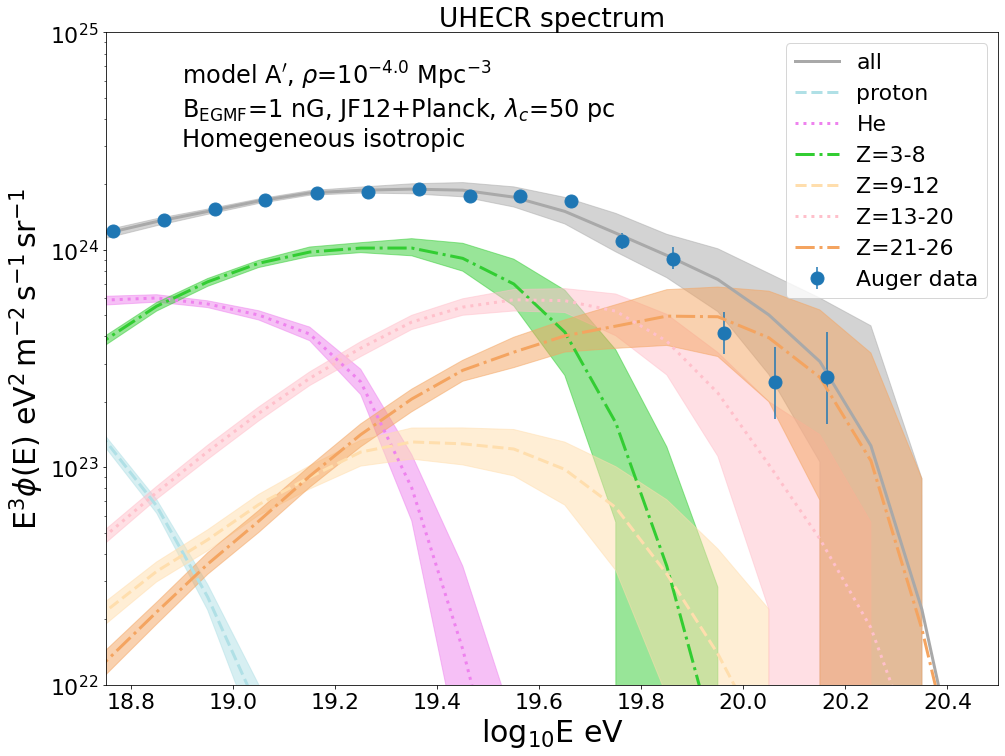}
   \caption{Same as Fig.~\ref{FigVariance1}, the scenarios displayed are now based on the sampling of an homogeneous and isotropic distribution of source with a  $10^{-4}$ $\rm Mpc^{-3}$ density and the source model $\rm A^\prime$. An EGMF of 1~nG and the "JF12+Planck" GMF model with $\lambda_{\mathrm{c}}=50$~pc are assumed.}
   \label{FigVariance1iso}
\end{figure}

\subsubsection{Energy spectrum and composition}
  
The spectra obtained for source model A, in the mother catalog approach, are displayed in Fig.~\ref{FigVariance1} for four different choices of the source density, ranging from $10^{-3}$ $\rm Mpc^{-3}$ to $10^{-5}$ $\rm Mpc^{-3}$. Unsurprisingly, in each energy bin the spread in the flux levels obtained for different realisations of the source distribution increases as the source density decreases, which reflects the corresponding variance in the integrated overdensity of a given realisation as a function of distance, as shown in Fig.~\ref{Fig2MRSDens}. In all cases the spread in the simulated fluxes is minimal around $10^{19}$~eV (which was also true, though less visible for volume-limited catalogs). This is a mere consequence of the fact that the statistics of the simulated datasets is defined by the number of events above 8~EeV, so that this number is by construction the same for all realisations.

Although part of the dispersion in the simulated fluxes is due to statistical fluctuations, associated with the finite size of the datasets, a comparison with Fig.~\ref{FigSpec} shows that the cosmic variance is the most important factor, at least for densities lower than $10^{-3}$ $\rm Mpc^{-3}$.

For comparison, we also show in Fig.~\ref{FigVariance1iso} the range of source spectra obtained in the case when discrete sources are drawn with a density of $10^{-4}$ $\rm Mpc^{-3}$ from an a priori uniform (homogeneous and isotropic) distribution. The source model $\rm A^\prime$ considered in this case is a variation of model A, in which in particular a slightly harder source spectrum is assumed, to better reproduce on average the observed spectrum. The reason why a harder spectral index is needed in this case is that the actual distribution of galaxies has a significant overdensity at small distances from us, which by definition is not present, on average, in source realisations draw from a uniform distribution. The other parameters of model $\rm A^\prime$ are further adjusted so as to result in a similar composition as model A for the propagated UHECRs.

\begin{figure}
   \centering
   \includegraphics[width=8.5cm]{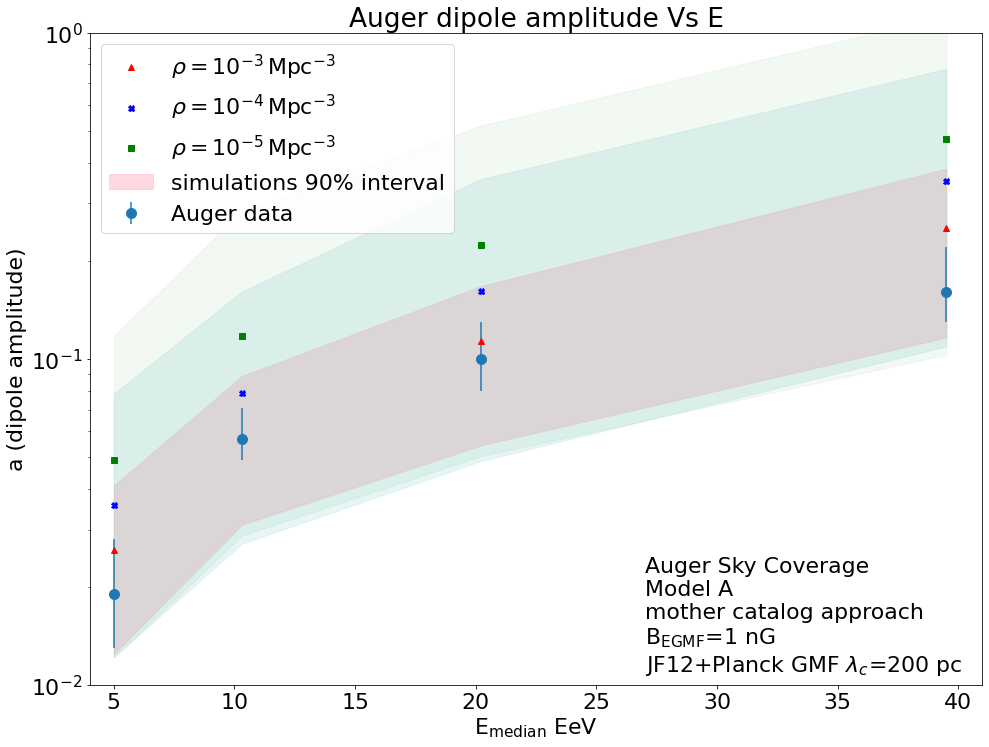}
   \includegraphics[width=8.5cm]{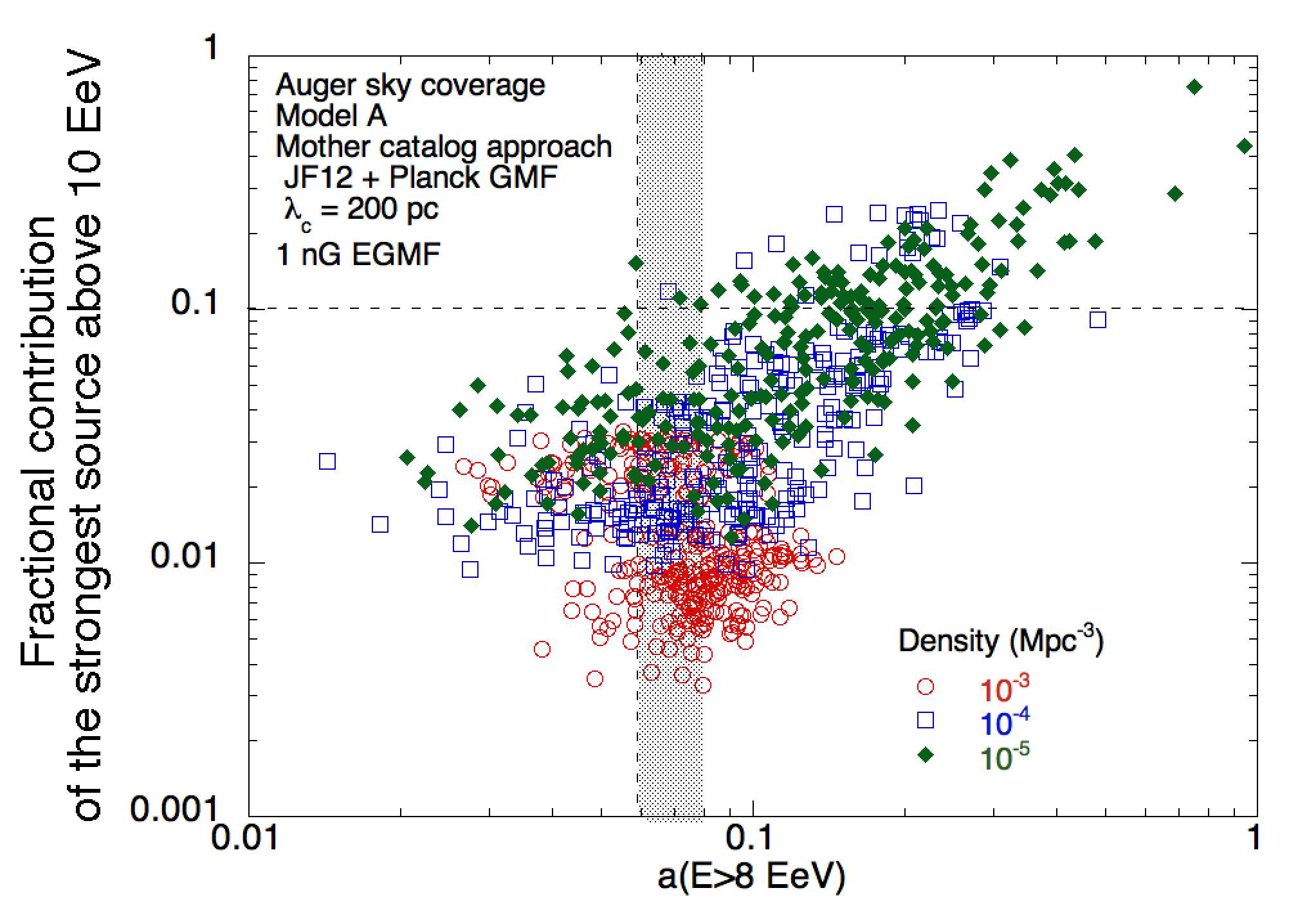}
   \includegraphics[width=8.5cm]{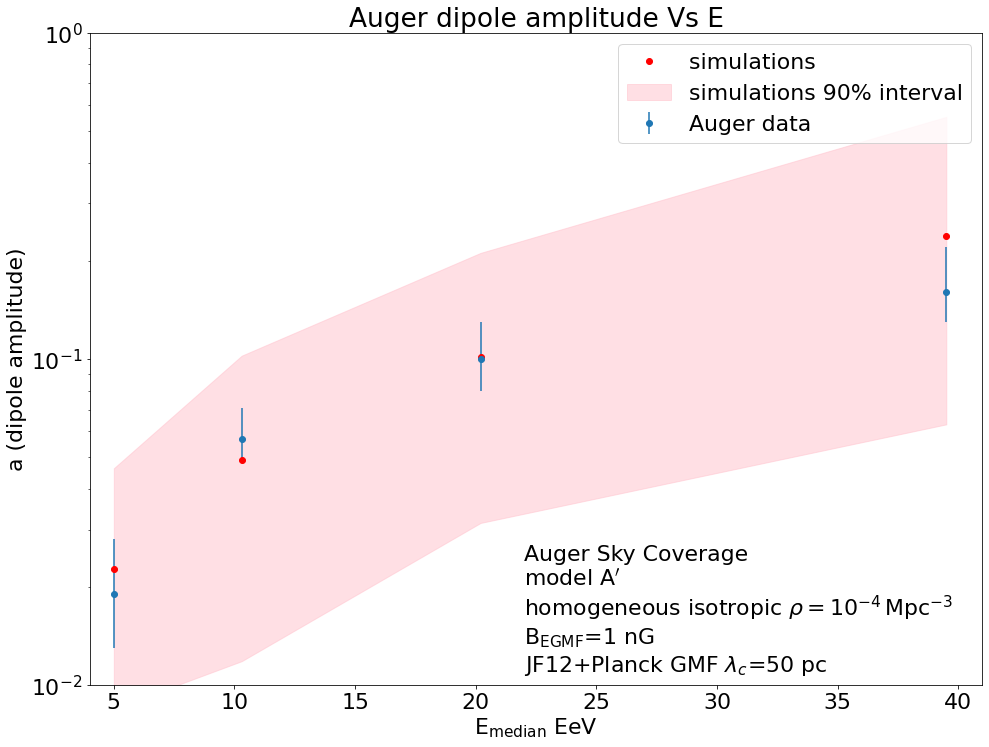}
\caption{Top : Energy evolution of the dipole amplitude predicted for  scenarios based on model A, a 1 nG EGMF and the JF12+Planck GMF with $\lambda_{\mathrm{c}}$=200~pc assuming source densities of $10^{-3}$, $10^{-4}$ and $10^{-5}$ $\rm Mpc^{-3}$, in the mother catalog approach. The mean values and the dispersion of the predictions (90\% interval), estimated over 300 realizations are shown respectively with markers and shaded areas and compared to the values reconstructed with Auger data. Center : Scatter plot of the fractional contribution of the strongest source above 10 EeV to the total flux as a function the dipole amplitude above 8 EeV for the different realizations considered in the top panel. The $1\sigma$ interval estimated with Auger data on the dipole amplitude is shown with a shaded area. Bottom : Same as top panel now assuming source model A$^\prime$, a 1 nG EGMF and the JF12+Planck GMF with $\lambda_{\mathrm{c}}$=50~pc and 300 realizations of an homogeneous and isotropic distribution of sources with a $10^{-4}$ $\rm Mpc^{-3}$  density.  
              }
         \label{FigVariance2}
\end{figure}

\subsubsection{Dipole amplitude and its variation with energy}

The energy evolution of the reconstructed dipole amplitude for the above-mentioned models is displayed on Fig.~\ref{FigVariance2}. The top panel shows the average and the range of results obtained for the various source densities with which we sub-sampled the mother catalog (except the case of a $10^{-3.5}$ $\rm Mpc^{-3}$ density, which is omitted for clarity). One sees that the spread in the expectations considerably increases as the density decreases, as already mentioned for the spectra displayed in Fig.~\ref{FigVariance1}. Arguably, given the spread in the spectra themselves, some realisations could be rejected on this sole basis, since they are essentially incompatible with the actually observed spectrum. Particularly in the case of a $10^{-4}$ $\rm Mpc^{-3}$ or \emph{a fortiori} a $10^{-5}$ $\rm Mpc^{-3}$ source density, some realisations lead to model A simulated spectra which are either too hard or too soft compared to the Auger data. These are realizations which are either particularly overdense or underdense in the local universe, as discussed in Sect.~\ref{sec:sourceDistrib} in the context of Fig.~\ref{Fig2MRSDens}. However, moderate amendments of model A can be proposed in each case, with parameters adjusted to address specifically overdense or underdense realisations of the catalog, as discussed in the previous section to adapt model A into model A$^\prime$. Ideally, one could adjust the source model for each realisation to best match the spectrum and composition measurements. However, we checked that applying such a procedure to a few realisations did not significantly reduce the spread in the reconstructed dipole amplitudes, so we finally kept the parameters of model A in all cases, for the sake of simplicity. As a matter of fact, the spread of the dipole amplitudes turns out to be essentially due to the realisation-to-realisation variability at each given source density, with little dependence on the fine details of the source model.

As expected, Fig.~\ref{FigVariance2} shows that the dipole amplitude increases with decreasing source density: even in the case of realisations that closely follow the density profile of the mother catalog (cf. Fig.~\ref{Fig2MRSDens}), the UHECR flux is shared among a lower number of sources as the density decreases, and the global anisotropy level thus increases (not only the dipole amplitude). The predictions in the case of the $10^{-3}$ $\rm Mpc^{-3}$ density, assuming the JF12+Planck GMF and $\lambda_{\mathrm{c}}$=200~pc, are in good agreement with the values reconstructed from the Auger data. In the case of the lower densities displayed on the figure, the data appear somewhat less “natural”, since they lie in the lower part of the predicted range, which corresponds to realisation which are relatively underdense locally. This observation would however be weakened if a larger value of the GMF coherence length, $\lambda_{\rm c}$, were assumed (and/or a larger value of the EGMF).

Note also that the partial sky coverage of the current UHECR observatories brings additional subtleties to this discussion, if some of the most nearby sources happen to sit outside the direct field of view of a given observatory, especially in the case of a low source density, where individual sources contribute a larger fraction of the total flux. Strong differences between the observations of observatories with different sky exposures (which is the case for Auger and TA) may thus be expected in that case (see {\it e.g} \citet{GAPLP2017}).

The relation between the locally overdense or underdense character of a given realisation and the expected value of the dipole amplitude is illustrated in the central panel of  Fig.~\ref{FigVariance2}, where we show the correlation between the amplitude of the dipole above 8~EeV and the fractional contribution of the strongest source to the total flux above 10~EeV, which is obviously related to the distance of the most nearby sources present in a given realisation of the source distribution. The figure shows a scatter plot where each point corresponds to one particular realisation, with different colors and symbol shapes for different assumed source densities (the same as in the top panel). The correlation is very clear, at least when the fractional contribution of the strongest source exceeds $\sim 5\%$ (which is rarely the case for the $10^{-3}$ $\rm Mpc^{-3}$ density case).

As can be seen, the realisations that show strong fractional contributions in the low density cases usually lead to dipole amplitudes which are too strong to be compatible with the Auger data. An increased value of $\lambda_{\mathrm{c}}$ (or of the EGMF) would however somewhat move the cloud of points in the figure “to the left”, i.e. result in a global shift to lower dipole amplitudes, making the average realisations more compatible with the amplitude measured by Auger in the case of a $10^{-4}$ $\rm Mpc^{-3}$ density (as already noted above).

An important conclusion of these results is that the amplitude of the dipole, by itself, does not bring strong constraints on the actual UHECR source density, when one takes into account the cosmic variance and the limited knowledge on the other physical free parameters (for instance the cosmic magnetic fields). Let us also indicate here that, even though the cases displayed here assume standard candle sources, we have also considered different types of intrinsic luminosity distributions (either power law or broken power law distributed over 2 orders of magnitude in UHECR integrated luminosity), which we do not show here. As expected, they lead to a somewhat larger dispersion of the predictions, but without changing significantly the above findings.

In the case of a $10^{-3}$ $\rm Mpc^{-3}$ source density, a GMF coherence length of $\lambda_{\rm c}$=200~pc is large enough for the average predicted amplitudes to match the data (assuming a 1~nG EGMF). This is a lower value than that used for our baseline volume-limited catalog with a similar density. This is due to the already mentioned fact that this catalog has a somewhat larger local overdensity of galaxies than the mother catalog itself (and thus than average realisations produced in the mother catalog approach). Moreover, in the volume-limited approach, the source luminosities were assumed to follow the 2MRS luminosity distribution function, whereas we now use standard candle sources. In the case of model~D, a value of $\lambda_{\mathrm{c}}$=500~pc is still required to better match the data: given its larger proton abundance, a value of $\lambda_{\mathrm{c}}$=200~pc would result in a too large dipole amplitude for most realisations, in particular in the $4\leq \rm E <8$~EeV energy bin. Note that the tension mentioned in Sect.~\ref{sec:compo} between the data and the model~D expectations regarding the quadrupole amplitude (assuming the baseline volume-limited catalog) is significantly reduced when one relaxes the assumption on the actual source distribution. Indeed, when one accounts for the cosmic variance, $\sim 50\%$ of the realisations with a $10^{-3}$ $\rm Mpc^{-3}$ density appear to be within the $\pm 1\sigma$ interval derived from the Auger data.

Likewise, valuable information about the actual UHECR source distribution can hardly be deduced from the observed energy evolution of the dipole amplitude, since distributions very different from that of the galaxies can reproduce the data, as shown for the uniform case on the bottom panel of Fig.~\ref{FigVariance2}. One sees that the wide interval of predictions obtained with a $10^{-4}$ $\rm Mpc^{-3}$ source density does enclose the Auger data, which lie close to the predicted average when the JF12+Planck GMF is used with $\lambda_{\mathrm{c}}$=50~pc. This required lower value of $\lambda_{\mathrm{c}}$, compared with the previous cases, is due to the fact that the anisotropy level expected for discrete sources draw from an underlying uniform distribution is on average significantly lower than that expected for sources randomly drawn out of an intrinsically anisotropic galaxy catalog (see {\it e.g} \citet{Harari2015} for an earlier discussion). However, such a value of $\lambda_{\mathrm{c}}$ is by no means disfavoured by the currently existing constraints on the GMF.

\begin{figure*}[ht!]
   \centering
   \includegraphics[width=8.5cm]{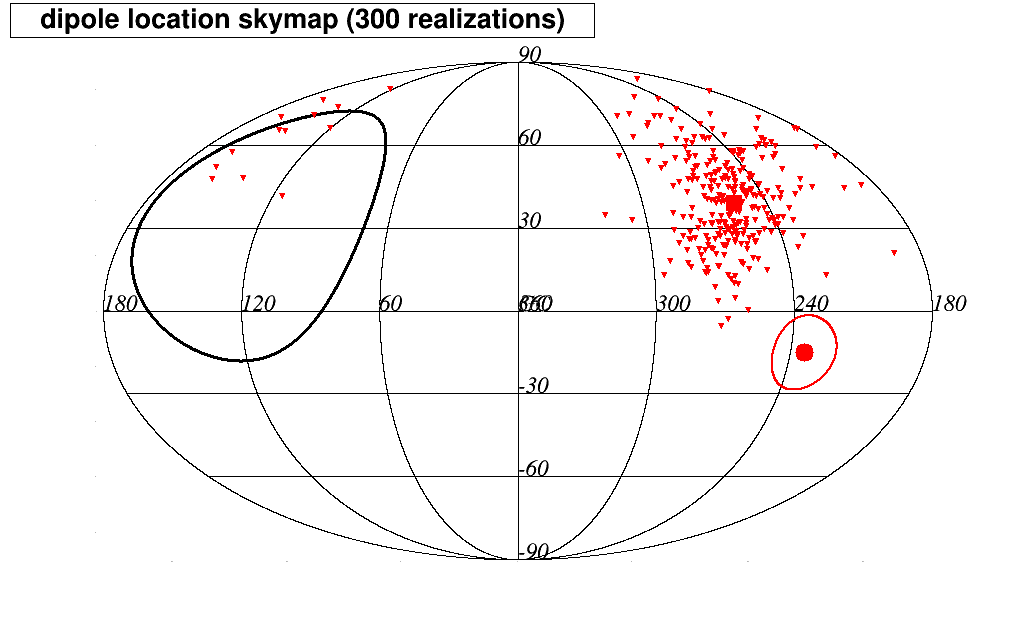}
   \includegraphics[width=8.5cm]{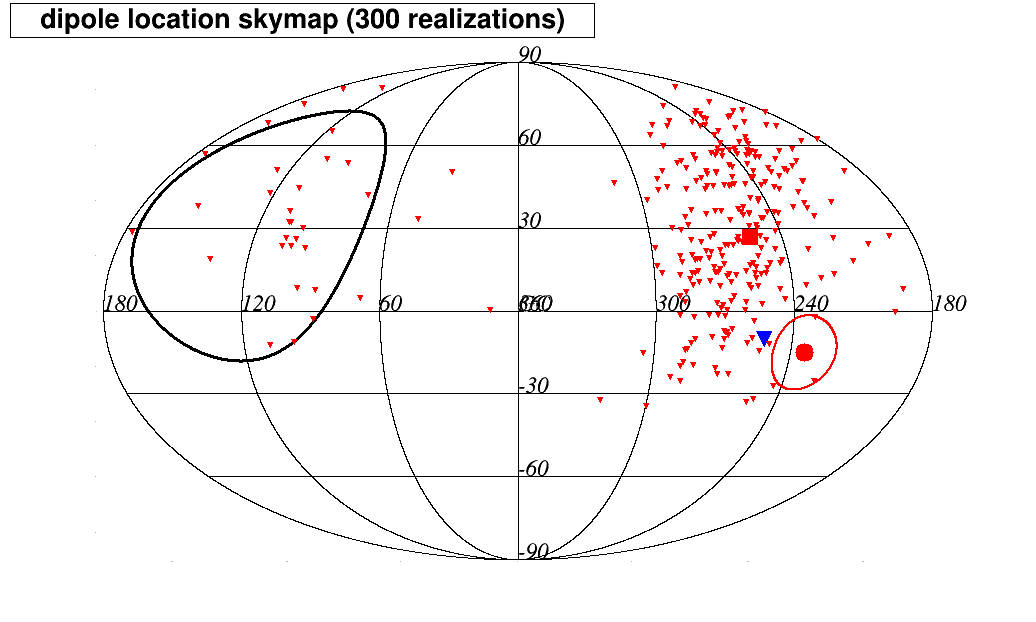}
   \includegraphics[width=8.5cm]{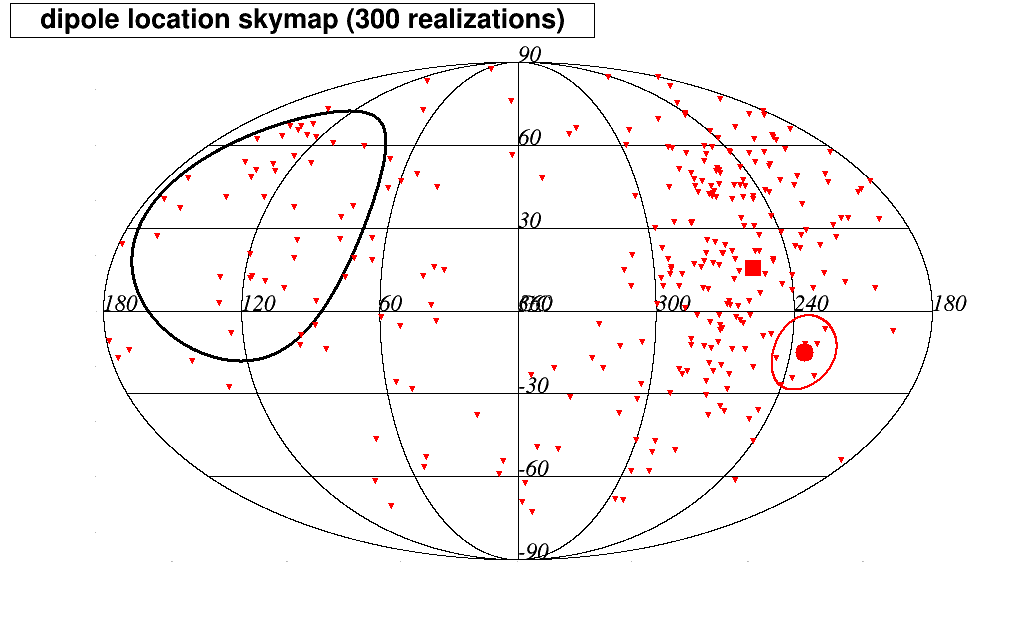}
   \includegraphics[width=8.5cm]{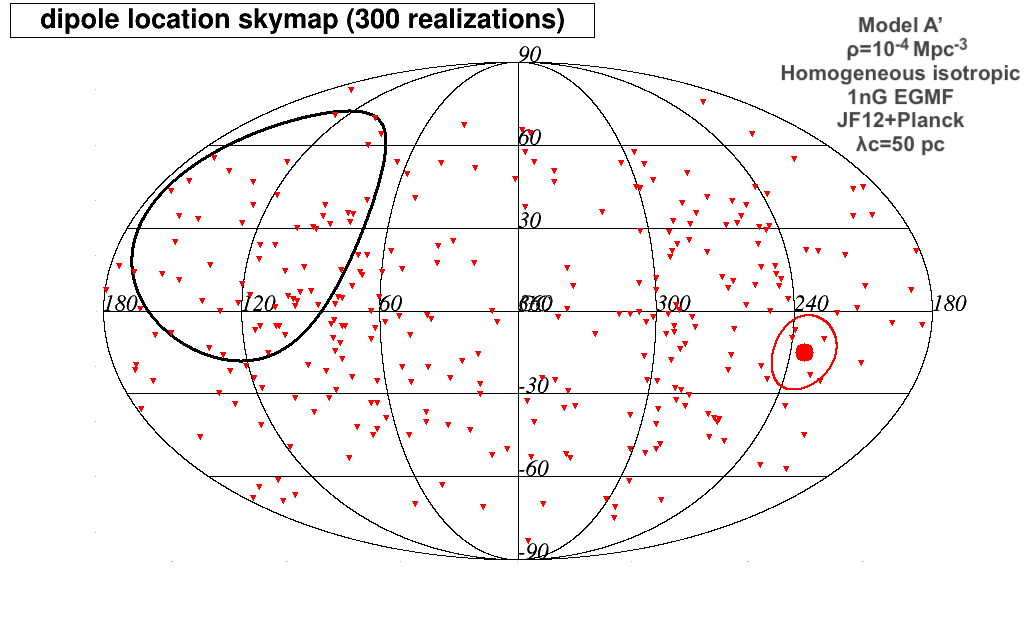}

      \caption{Reconstructed direction of the dipole of the UHECR events with $E\geq8$~EeV for each of the 300 computed realizations. The scenarios considered are based on source model A, a 1~nG EGMF, the "JF12+Planck" GMF model with $\lambda_{\mathrm{c}}=200$~pc and different hypotheses on the source density, $10^{-3}$ $\rm Mpc^{-3}$ (top-left), $10^{-4}$ $\rm Mpc^{-3}$ (top-right) and $10^{-5}$ $\rm Mpc^{-3}$ (bottom-left) in the mother catalog approach. The large square shows the barycenter of the reconstructed dipole. In the  $10^{-4}$ $\rm Mpc^{-3}$ case, the large blue full-triangle shows the dipole direction of the selected realization (see text). Bottom-right : the scenario displayed is based on source model A$^\prime$, a 1~nG EGMF, the JF12+Planck GMF with $\lambda_{\mathrm{c}}$=50~pc and an underlying homogeneous and isotropic distribution of sources with a $10^{-4}$ $\rm Mpc^{-3}$  density. 
              }
         \label{FigVariance3}
   \end{figure*}

\begin{figure}
   \centering
   \includegraphics[width=8.5cm]{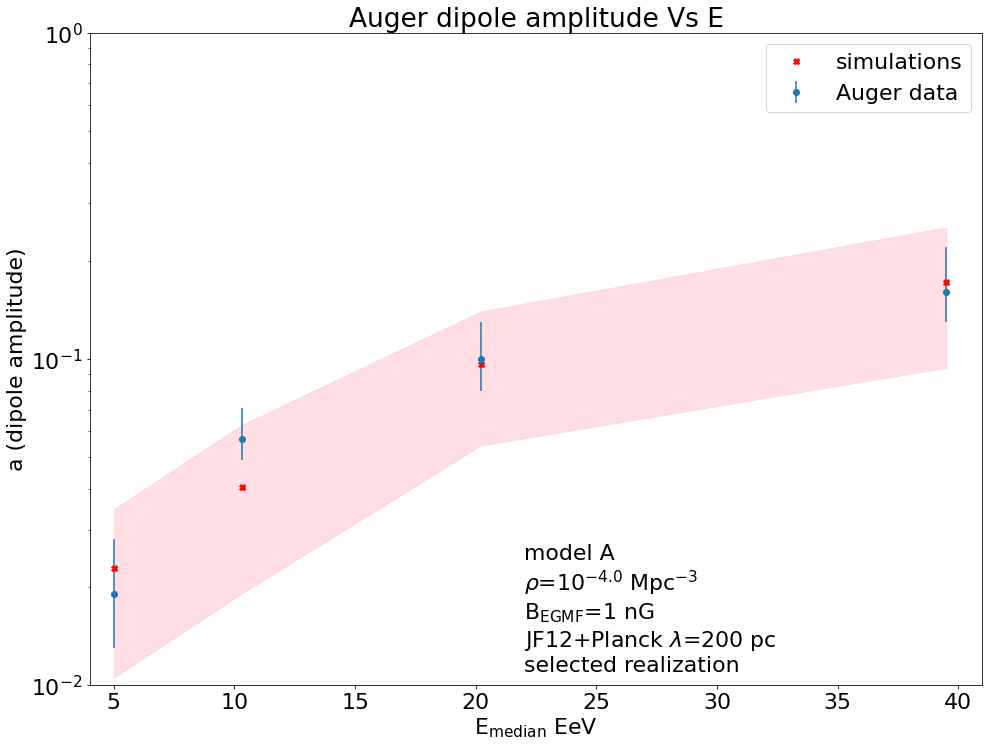}
   \includegraphics[width=8.5cm]{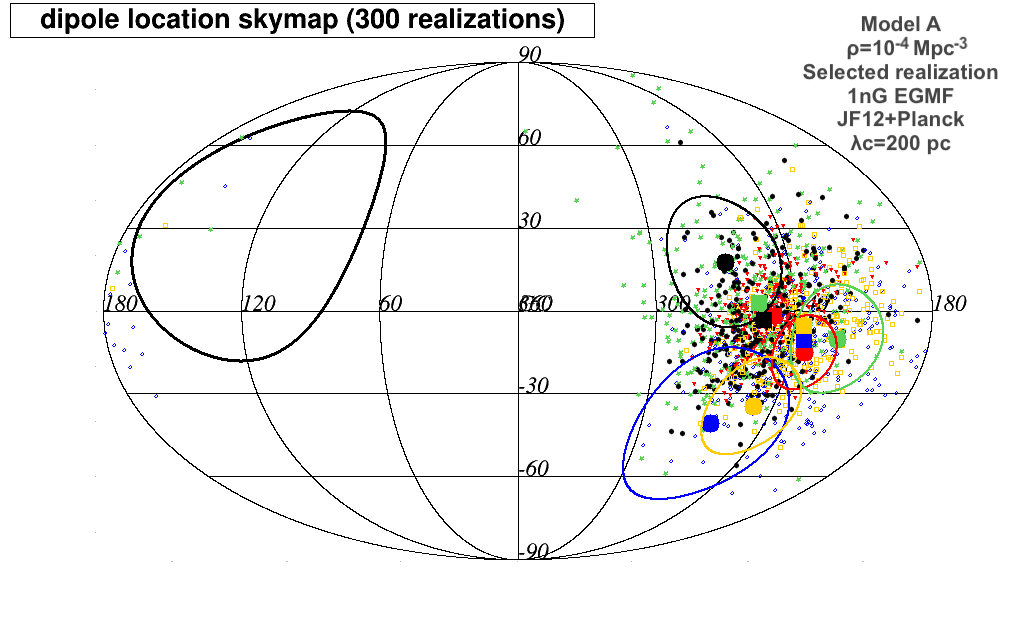}
\caption{Top : Energy evolution of the dipole amplitude predicted for model A, assuming a 1 nG EGMF and the JF12+Planck GMF with $\lambda_{\mathrm{c}}$=200~pc for a selected realization of the mother catalog sampling for a $10^{-4}$ $\rm Mpc^{-3}$ density (see text). The mean values   and the dispersion of the predictions (90\% interval) estimated over the 300 datasets are shown respectively with markers and shaded areas and compared to the values reconstructed with Auger data. Bottom : Skymaps in galactic coordinates showing the reconstructed directions of the dipoles for each of the 300 datasets in five different energy bins (see the color code legend in Fig.~\ref{FigLocation1}). The barycenters of the 300 datasets are shown by large squared markers. The predictions can be compared to the directions reconstructed by Auger shown by full circles and their $1\sigma$ error ellipses (large filled circles).
              }
         \label{FigVariance4}
   \end{figure}

Let us note that accounting for the cosmic variance as well as for the possibility that the actual distribution of UHECR sources departs from that of the galaxies also leads to a larger spread in the predictions concerning the slope of the energy evolution of the dipole amplitude, for a given source composition scenario, compared to what was discussed in Sect.~\ref{sec:compo} and shown in Fig.~\ref{FigDipVsE_compo}. In particular, the incompatibility of model B with the data in this respect is alleviated, although this model remains the least favoured of the four source composition models that we investigated.

\subsubsection{Dipole direction}

As shown in Sect.~\ref{sec:direction}, the aspect of the Auger dipole anisotropy analysis that is the most difficult to reproduce, assuming that the distribution of UHECR sources follows that of the galaxies in the Universe, is the direction in which the dipole is reconstructed, in particular in the $E\geq8$~EeV energy bin, where the isotropy of the UHECR arrival directions is the most significantly rejected (see also \cite{No2019,No2020,Ding2021}). We now examine whether this remains true if one takes the cosmic variance into account.

To this end, we have applied the Auger analysis on all the datasets simulated in the mother catalog approach, and obtained the dipole direction for each realisation of the source distribution already shown in Fig.~\ref{FigVariance2}. The results are displayed in Fig.~\ref{FigVariance3} for the energy bin $E\geq8$~EeV, and compared to the dipole direction reconstructed from the Auger data.

Unsurprisingly, when sampling the mother catalog, the variance of the reconstructed dipole direction increases as the density decreases. The variance is low for the largest density assumed, $10^{-3}$ $\rm Mpc^{-3}$, for which the reconstructed dipole direction remains incompatible with the data for our GMF model assumptions. As the assumed density decreases, the spread of the reconstructed dipole directions increases. While the bulk of them remains in the same broad region of the sky, rather far away from the Auger dipole direction, a handful of realisations out the 300 simulated ones for each set of parameters becomes compatible with then data, in the case of a $10^{-4}$ or $10^{-5}$~$\rm Mpc^{-3}$ source density. However the number of compatible realisations remains as marginal as in a scenario where the sources would be drawn randomly out of an underlying uniform (homogeneous and isotropic) distribution, which is displayed in the bottom-right panel of Fig.~\ref{FigVariance3}, for a $10^{-4}$~$\rm Mpc^{-3}$ source density.

Nevertheless, as an illustration, we focused on one such compatible realisation in the $10^{-4}$~$\rm Mpc^{-3}$ density case, shown by a blue triangle on the top right panel of Fig.~\ref{FigVariance3}. This realisation was selected because it appeared to be close to the data from the point of view of both the reconstructed dipole amplitude and the direction of the dipole above 8~EeV. For this a posteriori selected realisation of the source distribution, we produced 300 different Auger-like datasets, differing from each other only by the statistical fluctuations. We then considered the energy evolution of the resulting dipole amplitude, reconstructed in the four energy bins, as well as the dipole directions reconstructed for each of these 300 datasets. The results are shown in Fig.~\ref{FigVariance4}.

The fact that the agreement with the data appears to be good for both observables suggest that the specificity of the initial dataset (regarding its resemblance with the Auger data), was not due to a mere statistical fluctuation in the building up of the dataset, but to a statistical fluctuation of the source drawing procedure itself. It should thus be kept in mind that this realisation is certainly not typical of what can be expected for this class of models. Furthermore, this particular realisation turns out to predict a right ascension quadrupole modulation that exceeds the observations for most of the resulting datasets. Overall, our simulations show that, under the general astrophysical assumptions explored in the present paper, reproducing the observed characteristics of the dipole modulation measured by Auger is not achieved more easily by with UHECR sources following the distribution the galaxies in the Universe than with a biased version of this scenario (as was considered in Sect.~\ref{sec:bias}), or even than with an underlying homogeneous and isotropic distribution of sources, at least with the assumed GMF models, which are those usually assumed by most authors in the domain and beyond.

Finally, we also studied the impact of specific nearby sources on the observed energy evolution of the amplitude and the direction of the dipole by forcing the presence of these sources in the catalogs produced by sub-sampling the mother catalog. This exercise (whose results will be further discussed in a compagnon paper addressing more specifically the question of anisotropies at intermediate angular scale and higher energy) was conducted for eight nearby sources often mentioned among UHECR source candidates : CenA, NGC253, M82, M87, FornaxA, M104, Circinus, NGC1068. Quite generally, none of these forced inclusions in our catalog appeared to reconcile the model predictions regarding the dipole direction with the observations. Indeed, this outcome is not surprising given the results discussed in Sect.~\ref{sec:distance} regarding the contribution of a relatively limited subset of sources, be they nearby sources, to the dipole direction reconstructed from the entire dataset.

\section{Summary and discussion}
\label{sec:conclusion}

In this paper, we have investigated the general anisotropy features at large angular scale, which can be expected for UHECR source models belonging to a wide range of scenarios, characterised by different values of both astrophysical and physical parameters, while accounting reasonably well for the observed UHECR energy spectrum and the evolution of its composition with energy. The main common assumption of these scenarios is that the distribution of the UHECRs sources in the universe follows on average the distribution of galaxies, with possible biases either towards high-luminosity galaxies, or towards galaxies outside the largest galaxy clusters. Other parameters, notably the UHECR source composition and energy spectrum, the source density, the typical strength of the extragalactic magnetic field, the structure of the Galactic magnetic fields and the coherence length of its turbulent component, have been varied within typically accepted ranges, and both statistical fluctuations and the cosmic variance have been taken into account.

The main, general conclusion, is that: i) it appears relatively easy to reproduce the first-order (dipole) anisotropy observed in the Auger data, as well as its evolution as a function of energy, ii) this general agreement can be obtained with different sets of assumptions on the astrophysical and physical parameters, and thus cannot be used, at the present stage, to derive strong constraints on the UHECR source scenarios and draw model-independent constraints on the various parameters individually, and iii) the actual direction of the dipole modulation reconstructed from the Auger data appears highly non natural in essentially all scenarios investigated, and calls for a reconsideration of their main assumptions, either regarding the source distribution itself or the assumed magnetic field configuration, especially in the Galaxy.

More specifically, we found that the amplitude of the dipole modulation above 8~EeV, {\it i.e} the portion of the Auger dataset where it allows to reject the isotropic hypothesis with the highest significance, could be reproduced with different astrophysical scenarios with various combinations of parameters, and that the finer details of the dipole amplitude energy evolution, obtained by dividing the Auger dataset into 4 energy bins above 4~EeV, also turned out not be very constraining for extragalactic scenarios. This energy evolution is sensitive to the UHECR source composition model (see Sects.~\ref{sec:compo} and~\ref{sec:contribspecies}), but the cosmic variance makes it difficult to disentangle from the dependence on source distribution (presence or absence of a local overdensity of sources) or on the cosmic magnetic fields. In sum, the degeneracy between the multiple model parameters influencing the dipole amplitude and its energy evolution prevents one from drawing any strong conclusion regarding any of these parameters in particular, at least with the current data and the present knowledge regarding cosmic magnetic fields.

The picture dramatically changes when it comes to the discussion of the dipole direction. We found that the direction of the dipole reconstructed from the Auger data could not be reproduced (except in the highest energy bin containing UHECRs with a reconstructed energy $E\geq 32$~EeV) with a model where the UHECR source distribution is assumed to follow the distribution of galaxies in the universe, for all of the investigated source composition models, and all combinations of EGMF amd GMF parameters explored. This discrepancy is particularly significant in the $E\geq8$~EeV or $8\leq E<16$~EeV energy bins. Even when taking into account the cosmic variance, only a handful of simulations succeeded at reproducing the observed direction, with no larger probability than for discrete sources drawn from an underlying homogeneous and isotropic distribution.

At this stage, it is not clear whether this discrepancy points towards a genuine difference between the actual UHECR source distribution and that of the standard galaxies, or a large misunderstanding of the cosmic magnetic fields, which the most commonly used GMF models would fail to describe accurately. The role of the GMF in shaping the observed anisotropies, for a given UHECR scenario, is indeed important, as can be seen from the large differences between the reconstructed dipole directions obtained with the JF12+Planck model or the Sun+Planck model (see Fig.~\ref{FigLocation1}). Even when restricting to the framework of the JF12 GMF model, changing some of the physical ingredients of the regular field model within the range allowed by the observational constraints can lead to considerable modification of the deflection patterns, as shown by \citet{Unger2017, Unger2019}, even at rigidities of 10˜EV (see Fig.˜2 of the first reference), which is larger than the UHECR median rigidity at all energies (cf. Figs.~\ref{FigRigid} and~\ref{FigDipDist}). Thus, the uncertainties on the GMF intensity and configuration is a major obstacle for the understanding of the anisotropy signal and its use to constrain the origin of UHECRs and the main astrophysical parameters of their potential sources.

As a consequence, the perspective of using UHECR anisotropy observations to constrain at the same time the nature of the UHECR sources, their density and distribution, as well as the intensity and structure of the extragalactic magnetic fields in the local universe and the local group, appears out of reach without better constrained GMF models. For instance, unfortunately, the thorough study of \cite{Sigl2004} in the case of a proton-dominated composition at the highest energies cannot be straightforwardly extended to the mixed composition case within low $E_{\max}$ scenarios. In such a context, a good knowledge of the GMF, but also independent observational constraints and/or progresses in the theoretical modeling of the EGMF intensity and structure, including galaxy groups and clusters and the immediate environment of the Galaxy, will be required to facilitate the interpretation of the UHECR sky anisotropy, in a way that will provide reliable information about the UHECR origin.

On the other hand, the progress achieved in the recent years concerning the measurement of the UHECR composition as well as the perspectives offered by the 'Auger Prime' upgrade of the Pierre Auger observatory currently under construction \citep{AugerUpgrade2020}, open very interesting possibilities for the discovery of anisotropy signals, as discussed in Sect.~\ref{sec:contribspecies}. If the separation of nuclear component is confirmed to be reliable in practice, we have shown that the possibility of isolating the contribution of elements or groups of elements in the energy range where they experience a shortening of their energy loss horizon ({\it i.e} in their GZK suppression feature) can be very valuable, since this is the range in which the anisotropy will be the least "polluted" by the (quasi-isotropic) contribution of more distant sources. This strategy is complementary to that aiming at accumulating a much larger statistics above, say, $5-6\,10^{19}$~eV, where the background from distant sources will strongly decrease, even without separating between nuclear components. This should be accessible with the proposed ground-based \citep{GRAND2020, GCOS2021} and space-based \citep{Bertaina2019, POEMMA2021} observatories targeting UHECRs.

It is also important to distinguish between two different goals of anisotropy studies: i) the demonstration that the UHECR sky is not isotropic, and ii) the collection of information that will help identify the sources. The first goal may now be considered to be achieved. However, the energy range and type of observations that allow to exclude an isotropic distribution with the highest significance \citep{Lemoine2009} may not be those which will provide the most valuable information about the underlying astrophysical model. For this, focusing on regions where only a few sources provide the dominant contribution appears particularly promising, especially if one can, in parallel, obtain more reliable information about the magnetic fields.

Likewise, we considered the link between the anisotropy revealed by an excess of events in some directions, and the existence a “composition anisotropy”, characterised for instance by different mean masses in different regions of the sky. We have shown that, for all our models and parameter sets, on average a lighter composition goes along with an excess of events, and that a composition anisotropy appears much more difficult to establish firmly than the even count anisotropy (see Sect.~\ref{sec:anisocompo}). In this respect, we found that the difference in average mass of the events in or out of an extended region around the Galactic plane recently reported by Auger \citep{Mayotte2021}, if confirmed with larger statistics, will most probably be of a different origin and could thus point towards a different type of UHECR source scenario (for instance with an additional component, possibly Galactic).

Coming back to the energy evolution of the dipole amplitude, the fact that it can be reproduced quite naturally in the framework of the investigated models can be regarded as an additional support in favour of the general assumption of an extragalactic UHECR source distribution. However, the apparent discrepancy with the dipole direction forces us to remain open to alternative scenarios, for instance scenarios in which a one extragalactic source provides a dominant contribution to the UHECR flux (see e.g, \cite{Mollerach2019}), or even the most extreme case of a Galactic scenario \citep{Calvez2010}. Such scenarios, however, could only be envisaged with major revisions of the most common ideas about the magnetic field inside and in the vicinity of the Galaxy, with an extended turbulent halo that could significantly deflect, or even partially confine the UHECRs.

Concerning the direction of the reconstructed dipole, we studied its evolution with energy, and found that, whenever the statistics is sufficient to overcome the statistical fluctuations of the dataset, the directions obtained in different energy bins are not very different (see Sect.~\ref{sec:direction}). This appears somewhat in tension with the current angular distance between the dipole directions reconstructed from the Auger data above 8~EeV and above 32~EeV. If this trends is confirmed with larger statistics, it may question the assumption of a unique origin of all the UHECRs above the ankle. More generally, the drift of the reconstructed dipole direction with energy may be turned into a criterion for single (dominant) component models.

Finally, we also studied the quadrupole modulation of our simulated datasets, and found that the current lack of observation of a significant quadrupole modulation is in tension with scenarios combining a relatively weak EGMF ($\lesssim 1$~nG homogeneous turbulent field) and a coherence length of the GMF $\lambda_{\mathrm{c}}\lesssim 100$~pc, at least when assuming that the distribution of UHECR sources follows that of the galaxies (even when introducing some bias, as we did in Sect.~\ref{sec:bias}). The lack of significant signal in larger future (all particles) datasets, would provide additional information, and further restrict the allowed range in the parameter space combining the cosmic magnetic fields and the actual distribution of UHECR sources, as already emphasized by \citet{Matteo2018}. Such meaningful constraints would however require a significant increase in the statistics and would benefit from the possibility to isolate the light UHECR component, as suggested by Fig.~\ref{FigQuadComponents}.

In the second paper of this series, we shall investigate the intermediate scale anisotropies associated with the same astrophysical scenarios, notably at the highest energies, and examine the correlations of the UHECR arrival directions with specific sources or catalogs of sources. Following the approach of the present paper, we will discuss their implications for the UHECR source models, in the light of the above-mentioned uncertainties on the various physical parameters.

\begin{acknowledgements}
We wish to thank Noémie Globus for helpful comments and suggestions about the present manuscript and for her contribution to previous related works. We thank Katia Ferri\`ere and Tess Jaffe for interesting exchanges about the Galactic magnetic field and its modeling, Daniel Pomarède for valuable discussions about the structure of the nearby universe and Yehuda Hoffman for giving us permission to use the simulated density field from \citet{LSSS2018} in our calculations.

     % Part of this work was supported by the George Abitbol fundation.
      
\end{acknowledgements}

% WARNING
%-------------------------------------------------------------------
% Please note that we have included the references to the file https://www.overleaf.com/project/5e7f0ca575abcc0001249c60aa.dem in
% order to compile it, but we ask you to:
%
% - use BibTeX with the regular commands:
%   \bibliographystyle{aa} % style aa.bst
%   \bibliography{Yourfile} % your references Yourfile.bib
%
% - join the .bib files when you upload your source files
%-------------------------------------------------------------------

\end{document}